\documentclass[useAMS,usenatbib]{mn2e}
\usepackage{xspace}
\usepackage{amssymb}
\usepackage{graphicx} 
\usepackage[T1]{fontenc}
\newcommand{\sz}{$\sigma_z$\xspace}

\newcommand{\kms}{km~s$^{-1}$\xspace} 
\newcommand{\msun}{M$_{\odot}$\xspace}

\title{Dissecting simulated disc galaxies I:  the structure of mono-age populations}

\author[M. Martig, I. Minchev and C. Flynn]{ Marie Martig$^{1,2}$\thanks{E-mail: marie.martig@gmail.com}, Ivan Minchev$^{3}$ and Chris Flynn$^{1}$\\
$^1$Centre for Astrophysics \& Supercomputing, Swinburne University of Technology, P.O. Box 218, Hawthorn, VIC 3122, Australia \\
$^2$Max-Planck-Institut f\"{u}r Astronomie, K\"{o}nigstuhl 17, 69177 Heidelberg, Germany\\
$^{3}$Leibniz-Institut f\"{u}r Astrophysik Potsdam (AIP), An der Sternwarte 16, 14482, Potsdam, Germany
}
\begin{document}
\date{Accepted 2014 May 16. Received 2014 May 7; in original form 2013 September 12}
\maketitle
\begin{abstract}
We study seven simulated disc galaxies, three with a quiescent merger history, and four with mergers in their last 9 Gyr of evolution. We compare their structure at $z=0$ by decomposing them into ``mono-age populations'' (MAPs) of stars within 500 Myr age bins. All studied galaxies undergo a phase of merging activity at high redshift, so that stars older than 9 Gyr are found in a centrally concentrated component, while younger stars are mostly found in discs.
We find that most MAPs have simple exponential radial and vertical density profiles, with a scale-height that typically increases with age. Because a large range of merger histories can create populations with simple structures, this suggests that the simplicity of the structure of mono-abundance populations observed in the Milky Way by \cite{Bovy2012b,Bovy2012c} is not necessarily a direct indicator of a quiescent history for the Milky Way. Similarly, the anti-correlation between scale-length and scale-height does not necessarily imply a merger-free history. However, mergers produce discontinuities between thin and thick disc components, and jumps in the age-velocity relation. The absence of a structural discontinuity between thin and thick disc observed in the Milky Way would seem to be a good indicator that no merger with a mass ratio larger than 1:15--1:10 occurred in the last 9 Gyr. Mergers at higher redshift might nevertheless be necessary to produce the thickest, hottest components of the Milky Way's disc.
\end{abstract}

\begin{keywords}
galaxies: formation - galaxies:structure -  galaxies: kinematics and dynamics - methods: numerical
\end{keywords}

\section{Introduction}

The stellar disc of the Milky Way, as well as of many nearby galaxies, is traditionally viewed as consisting of two distinct components: a thin and a thick disc \citep[e.g.,][]{Burstein1979, Tsikoudi1979,Gilmore1983, Dalcanton2002, Yoachim2006, Comeron2011}. The Milky Way's thick disc not only has a larger scale-height than the thin disc \citep{Robin1996,Chen2001,Juric2008}, but it is also older, more metal-poor, and enhanced in alpha elements \citep{Gilmore1995,Fuhrmann1998,Prochaska2000,Bensby2005,Reddy2006}. Its stars have a higher velocity dispersion, and lag behind in rotation compared to the thin disc \citep{Soubiran2003,Allende2006,Bond2010}.

Many formation mechanisms have been proposed for the thick disc, such as being formed thick (following gas rich mergers, as in \citealp{Brook2004}, or via clumpy disc instabilities as in \citealp{Bournaud2009}), thickened by minor mergers after it is formed \citep{Quinn1993, Villalobos2008} or built from accreted stars \citep{Abadi2003}. Some authors have also argued that radial migration could create a thick disc \citep{Schonrich2009a,Schonrich2009b, Loebman2011}, although  \citet{Minchev2012b} recently showed that radial migration alone is not efficient at heating discs.
Discriminating between these formation mechanisms is a complex issue; it has been suggested that different scenarios could be constrained by the shape of the discs \citep{Bournaud2009}, and the distribution of their stars' eccentricities \citep{Sales2009,DiMatteo2011} or the presence of an excess of stars at high altitudes \citep{Qu2011}.

It is worth noting that there may not be a clear thin-thick disc dichotomy in the Milky Way \citep{Norris1987, Nemec1991, Nemec1993,Ryan1993}. Using SEGUE  spectroscopic survey data,
\cite{Bovy2012a} have shown that the thin and thick disc are not two separate components, but that instead a continuity of properties is found when slicing the Milky Way disc into populations of similar chemical abundances. They also find that each of these mono-abundance populations has a very simple spatial and kinematical structure \citep{Bovy2012b,Bovy2012c}. There is a clear anti-correlation between the scale-height and scale-length of these populations, so that the youngest ones are also the coldest and the most radially extended (assuming that age can be matched with chemical abundance, as has been shown to be very likely by \citealp{Haywood2013} for a sample of solar neighbourhood stars, and by \citealp{Stinson2013} for a simulated galaxy). Bovy et al. argue this is clear evidence for inside-out formation, and evidence against mergers, which would have created distinct components instead of the observed smooth continuum.
Similarly, \cite{Bovy2012c} also found a continuity of vertical velocity dispersion (\sz) as a function of chemical abundance, with no obvious gap in the kinematics of old and young stars.

In the light of these new observational results, numerical simulations have also been sliced into mono-abundance or mono-age populations \citep{Bird2013, Stinson2013} and show interesting similarities with the structure of the Milky Way.  Both studies find that mono-age populations are simple structures, and confirm that the anti-correlation between  scale-height and scale-length arises as a consequence of inside-out formation. \cite{Stinson2013} perform a direct comparison with the Bovy et al. results, and show that the mono-abundance groups correspond roughly to mono-age groups. They are unable to reproduce the thickness of the oldest stars in the Milky Way, and also do not find a continuity between thin and thick disc, but rather a clear bimodality between the two components.

In this paper, we aim at extending this work by analysing a sample of seven simulated galaxies with different merger histories. We aim at understanding which observed properties are generic features of disc evolution, and which ones are a clear indicator of a merger-free history. This paper is the first of a series; Paper II \citep{Martig2014b} is focused on the age-velocity relation (i.e. the relation between the ages of stars and their vertical velocity dispersion) and a study of disc heating in simulations.
Throughout these papers, our goal is not to find a simulated galaxy that exactly matches all properties of the Milky Way, but rather to explore different types of formation histories and study their consequences for the structure of discs at $z=0$.

We present our simulations in Section 2, and discuss in Section 3 the simplicity (or otherwise) of the spatial and kinematical structure of mono-age populations. In Section 3, we discuss the conditions for an anti-correlation between scale-height and scale-length, and in Section 5 the conditions for a dichotomy between thin and thick components. Finally, we discuss in Section 6 which properties are imprinted at birth or are due to subsequent evolution. We also present constraints on the merger history of the Milky Way, and discuss the limitations of our approach. We conclude and summarize in Section 7.

\section{Simulations and analysis}
\begin{figure*}
\centering 
\includegraphics[width=0.138\textwidth]{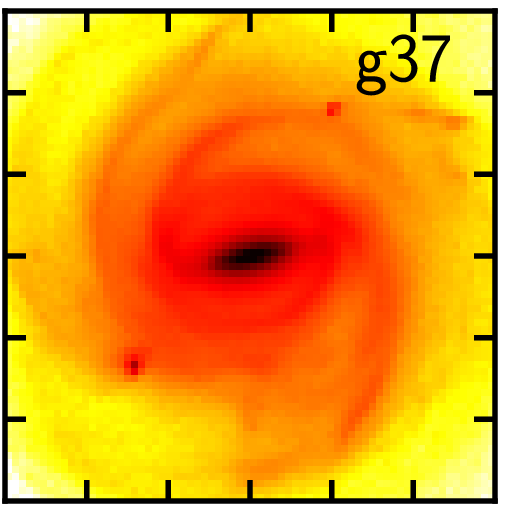}
\includegraphics[width=0.138\textwidth]{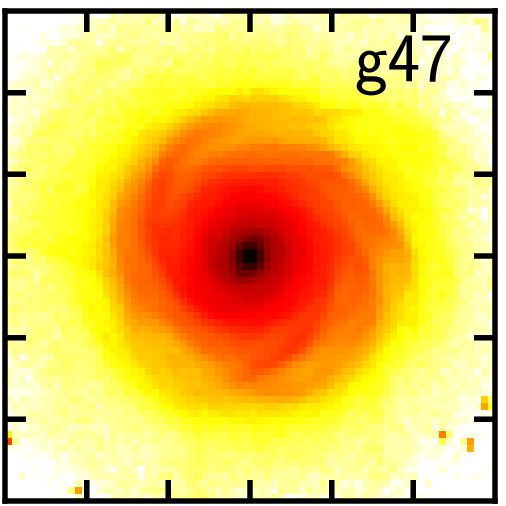}
\includegraphics[width=0.138\textwidth]{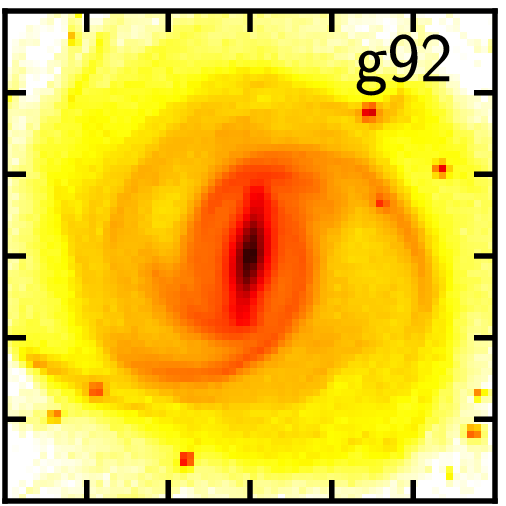}
\includegraphics[width=0.138\textwidth]{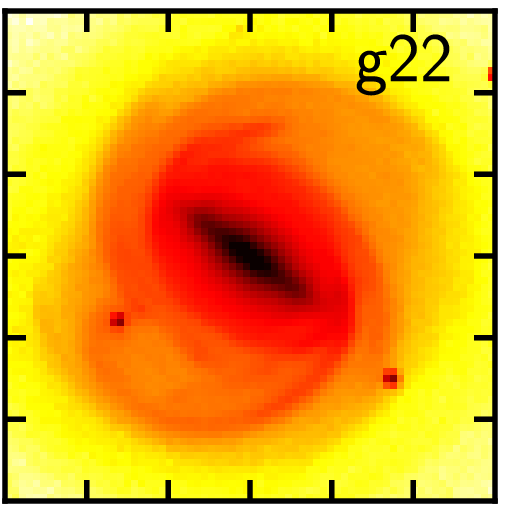}
\includegraphics[width=0.138\textwidth]{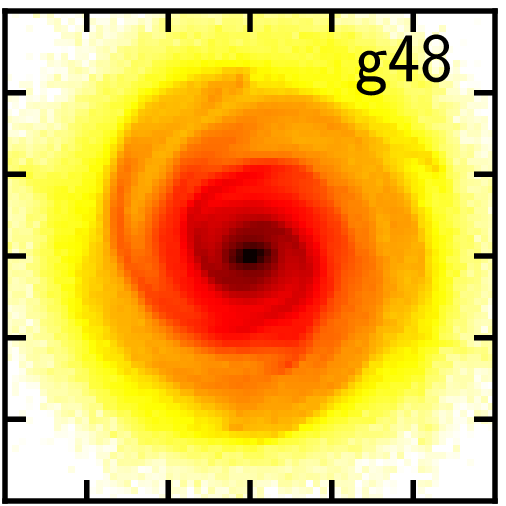}
\includegraphics[width=0.138\textwidth]{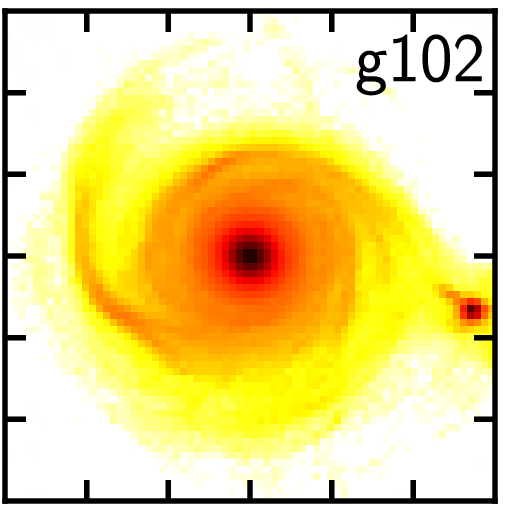}
\includegraphics[width=0.138\textwidth]{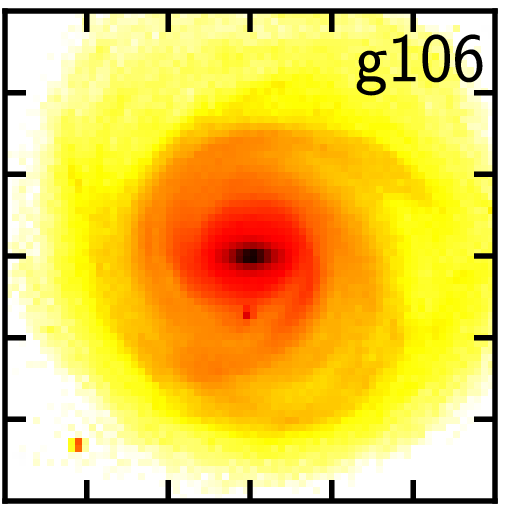}

\includegraphics[width=0.138\textwidth]{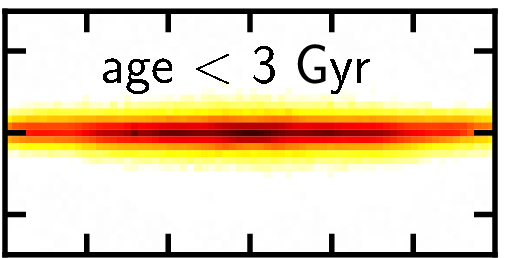}
\includegraphics[width=0.138\textwidth]{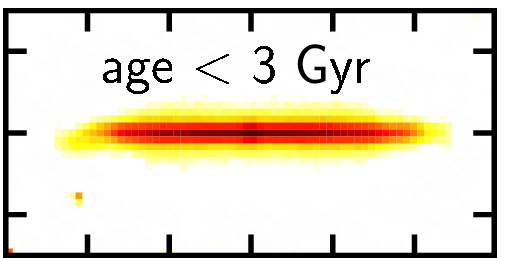}
\includegraphics[width=0.138\textwidth]{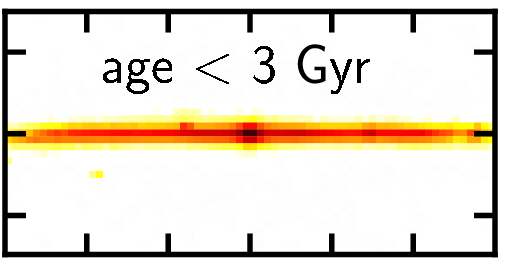}
\includegraphics[width=0.138\textwidth]{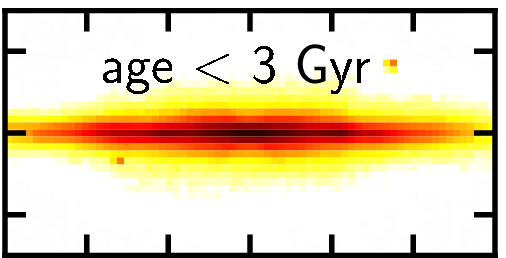}
\includegraphics[width=0.138\textwidth]{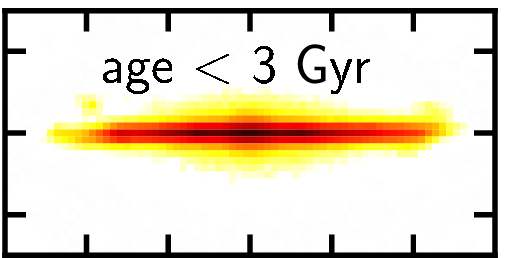}
\includegraphics[width=0.138\textwidth]{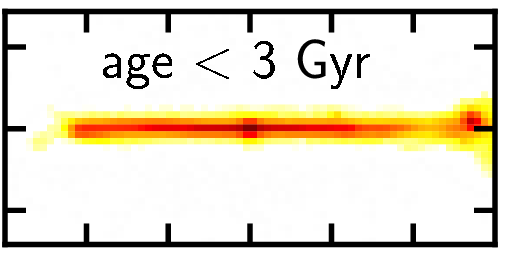}
\includegraphics[width=0.138\textwidth]{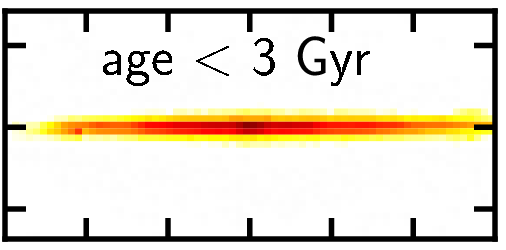}

\includegraphics[width=0.138\textwidth]{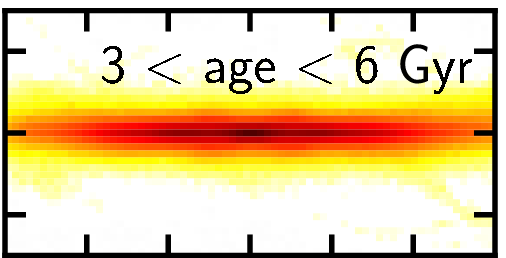}
\includegraphics[width=0.138\textwidth]{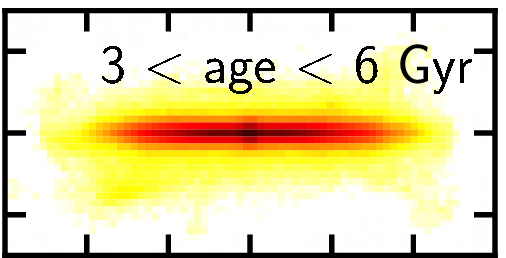}
\includegraphics[width=0.138\textwidth]{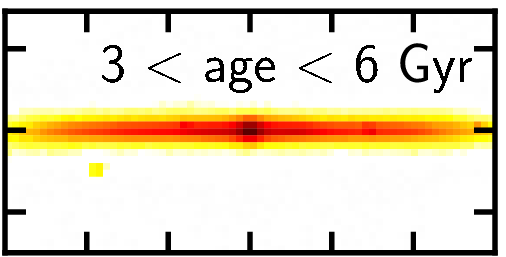}
\includegraphics[width=0.138\textwidth]{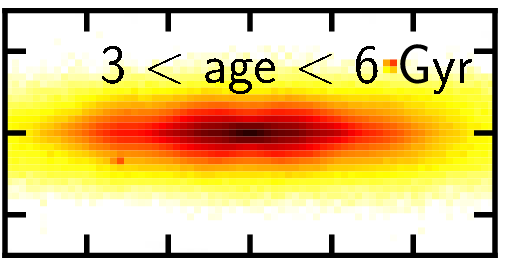}
\includegraphics[width=0.138\textwidth]{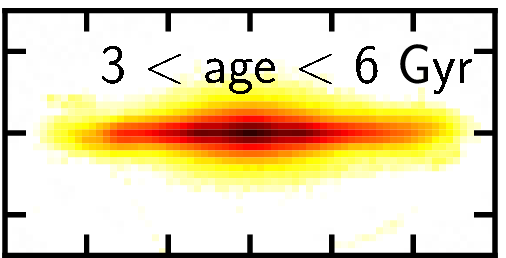}
\includegraphics[width=0.138\textwidth]{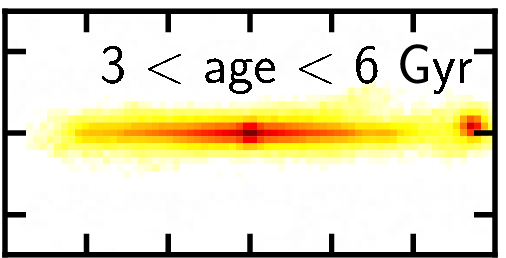}
\includegraphics[width=0.138\textwidth]{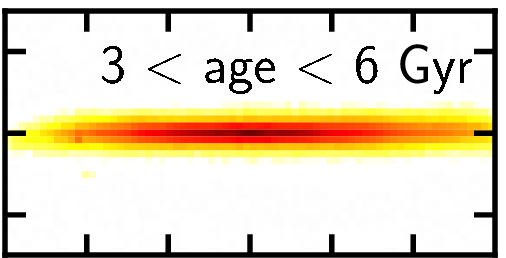}

\includegraphics[width=0.138\textwidth]{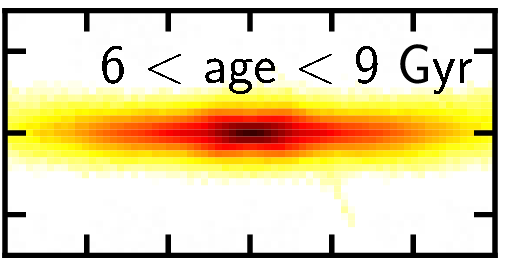}
\includegraphics[width=0.138\textwidth]{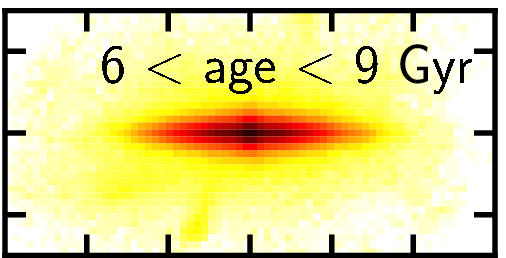}
\includegraphics[width=0.138\textwidth]{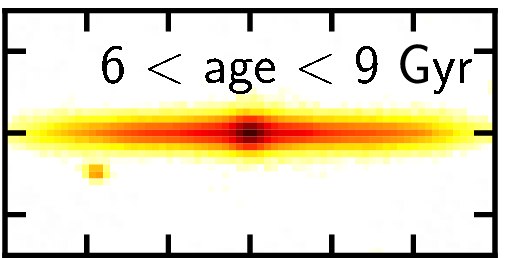}
\includegraphics[width=0.138\textwidth]{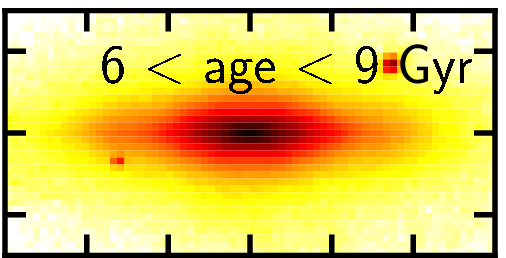}
\includegraphics[width=0.138\textwidth]{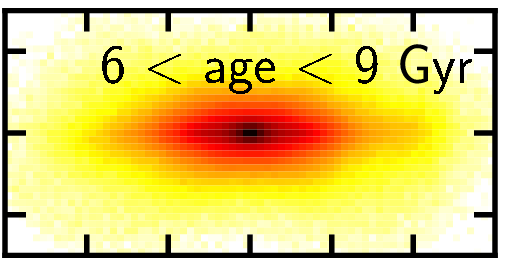}
\includegraphics[width=0.138\textwidth]{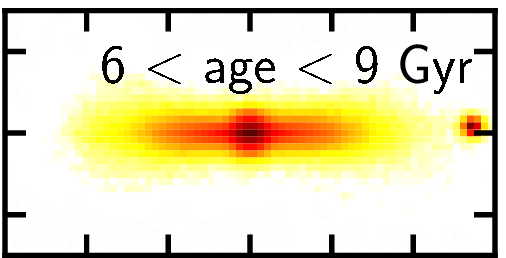}
\includegraphics[width=0.138\textwidth]{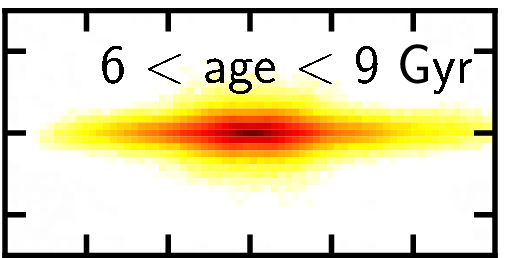}

\includegraphics[width=0.138\textwidth]{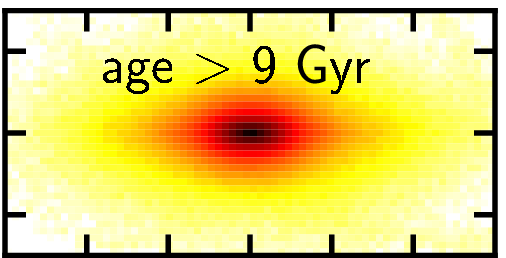}
\includegraphics[width=0.138\textwidth]{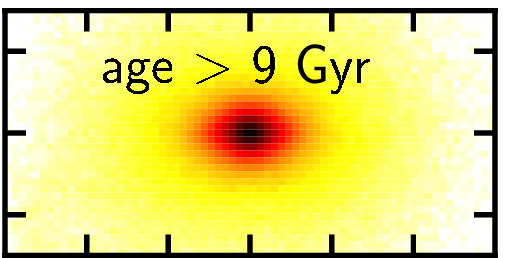}
\includegraphics[width=0.138\textwidth]{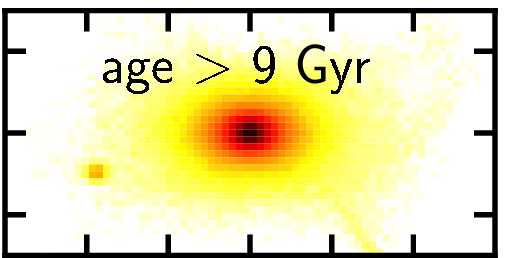}
\includegraphics[width=0.138\textwidth]{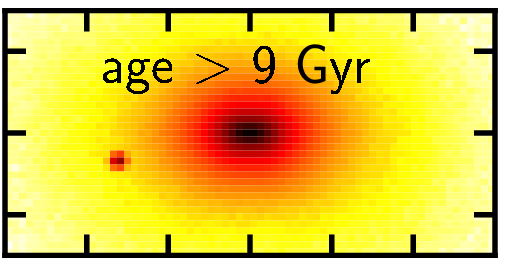}
\includegraphics[width=0.138\textwidth]{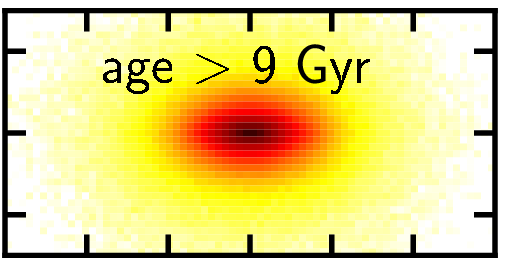}
\includegraphics[width=0.138\textwidth]{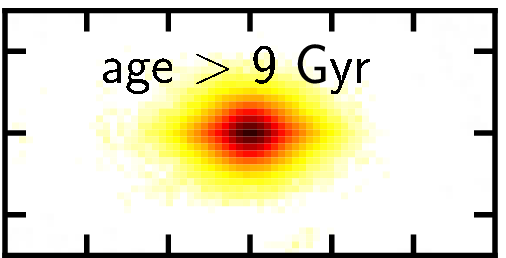}
\includegraphics[width=0.138\textwidth]{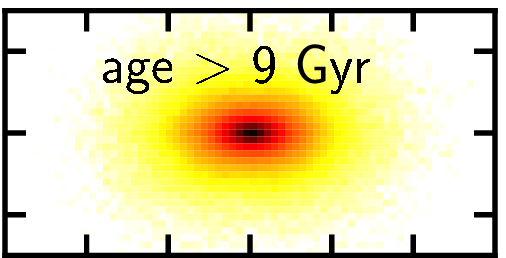}

\caption{The stellar mass distribution for the selected galaxies at $z=0$. The top row shows the face-on distribution for all stars (in 60 kpc $\times$ 60 kpc panels), the four bottom rows show edge-on projections for stars in different age bins (in 60 kpc $\times$ 30 kpc panels, with the same colour scale for all images). }
\label{fig:images}
\end{figure*}

From the sample of 33 galaxies presented in \cite{Martig2012}, we select seven disc galaxies with a range of merger histories, from a galaxy with an extremely quiescent history in the last 8--9 Gyr to a galaxy undergoing a late 1:4 merger. In this Section, we present the simulation technique and the galaxies, and explain how we analyse their structure.

\subsection{Simulation technique}

The galaxies are drawn from a larger sample of 33 simulated galaxies, described in \cite{Martig2012}. We refer the reader to  \cite{Martig2012} for a detailed description of the technique and the sample, and only describe here the main features.

The galaxies are simulated using a zoom-in technique in two steps (the technical details are discussed in \citealt{Martig2009}). The first step consists in running a large dark-matter only simulation with the adaptive mesh refinement code RAMSES \citep{Teyssier2002}, and extracting merger and accretion histories for interesting dark matter haloes. We target haloes with a mass between  $2.7 \times 10^{11} $ and $2 \times 10^{12}$ M$_{\odot}$, and that are in relatively isolated environments. For these target haloes, not only do we record the time, mass, velocity and spin of incoming satellites, but we also keep track of diffuse dark matter accretion, for instance along filaments.

The second step is to re-simulate each merger and accretion history. For that we use the Particle-Mesh code described in \cite{Bournaud2002,Bournaud2003}, with gas dynamics modelled by a sticky-particle algorithm. Our box size is 800 kpc, the spatial resolution is 150 pc\footnote{See Paper II as well as Appendix A of \cite{Martig2012} for resolution tests}, the mass resolution is  $1.5\times 10^4$~M$_{\sun}$ for gas particles, $7.5\times 10^4$~M$_{\sun}$ for star particles ($1.5\times 10^4$~M$_{\sun}$ for star particles formed from gas during the simulation), and  $3\times 10^5$~M$_{\sun}$ for dark matter particles.   For the galaxies studied here, there are typically  2 to 6 million star particles in the discs at $z=0$.

Star formation is modelled by a Schmidt law with an exponent of 1.5 \citep{Kennicutt1998}, using a gas density threshold of 0.03 M$_{\sun}$pc$^{-3}$ (i.e., 1 H cm$^{-3}$). Stellar mass loss is taken into account (see \citealp{Martig2010}), as well as kinetic feedback from supernovae explosions, where 20\% of the energy of the supernovae is distributed to neighbouring gas particles.

Each re-simulation starts at $z=5$ with a seed galaxy, and follows its evolution down to $z=0$ with mergers and diffuse accretion as prescribed by the cosmological simulation. Each halo that was tracked in the cosmological simulation is replaced with a galaxy made of a dark matter halo, a stellar disc and bulge, and a gas disc. The stellar mass of the galaxies is chosen to obey the scaling relations of \cite{Moster2010}, and their gas fraction evolves with redshift. Diffuse accretion is modelled by assuming that gas follows dark matter, in an amount given by the cosmic baryon fraction. Each diffuse dark matter particle of the initial simulation is then replaced with a new blob of gas and dark matter particles.

The main advantage of this re-simulation technique is the relatively low computing time, permitting us to compile statistically interesting samples of galaxies. The 33 galaxies introduced in \cite{Martig2012} span a large range of bulge fractions at $z=0$, from nearly bulgeless to bulge-dominated systems. \cite{Martig2012} studied how this final morphology compared to the morphology at earlier redshift, and discussed how the bulge fraction is connected with the merger and gas accretion history. In the sample, the fraction of barred galaxies at $z=0$ and higher redshift is realistic when compared to  observations \citep{Kraljic2012}, and a detailed chemical modelling of one of the galaxies has shown an interesting match to many features found in the Milky Way \citep{Minchev2013}. Note also that disc dynamics and radial migration properties have been compared with isolated galaxy simulations performed with a Tree-SPH code, and good agreement was found \citep{Minchev2012b}.

\subsection{Overview of the selected galaxies}

\begin{table}
 \caption{Properties of the simulated galaxies at $z=0$: stellar mass within the optical radius, bulge fraction as measured with GALFIT \citep{Peng2002}, optical radius ($R_{25}$ measured in $g$-band), disc scale-length ($R_d$) obtained from an exponential fit to the stellar mass distribution. The last column shows a brief description of their formation history (mergers are denoted by a mass ratio)}
 \label{tab-sample}
 \begin{tabular}{@{}lccccc}
  \hline
  Name & M$_{*}$ (10$^{10}$\msun)& B/T & $R_{25}$ (kpc)& $R_{d}$ (kpc)& History\\
  \hline
 g37 & 12.0  & 0.13 & 35.1 & 7.5 & quiet\\
 g47 & 8.6 & 0.11 & 22.2& 4.2 & fly-by \\
 g92 & 4.4  & 0.04 & 28.9 & 5.1 & quiet \\
 \hline
g22 & 19.1 &0.23 & 32.6 & 5.0 & 1:10\\
g48 & 10.8 & 0.07&  24.7& 3.8 & 1:4\\
g102 & 3.3 &0.48 & 21.9 & 5.0 & 1:15\\
g106 & 4.3  &0.22 & 24.6& 5.1 & 1:5 \\
  \hline
 \end{tabular}
\end{table}

\begin{figure*}
\centering 
\includegraphics[width=0.245\textwidth]{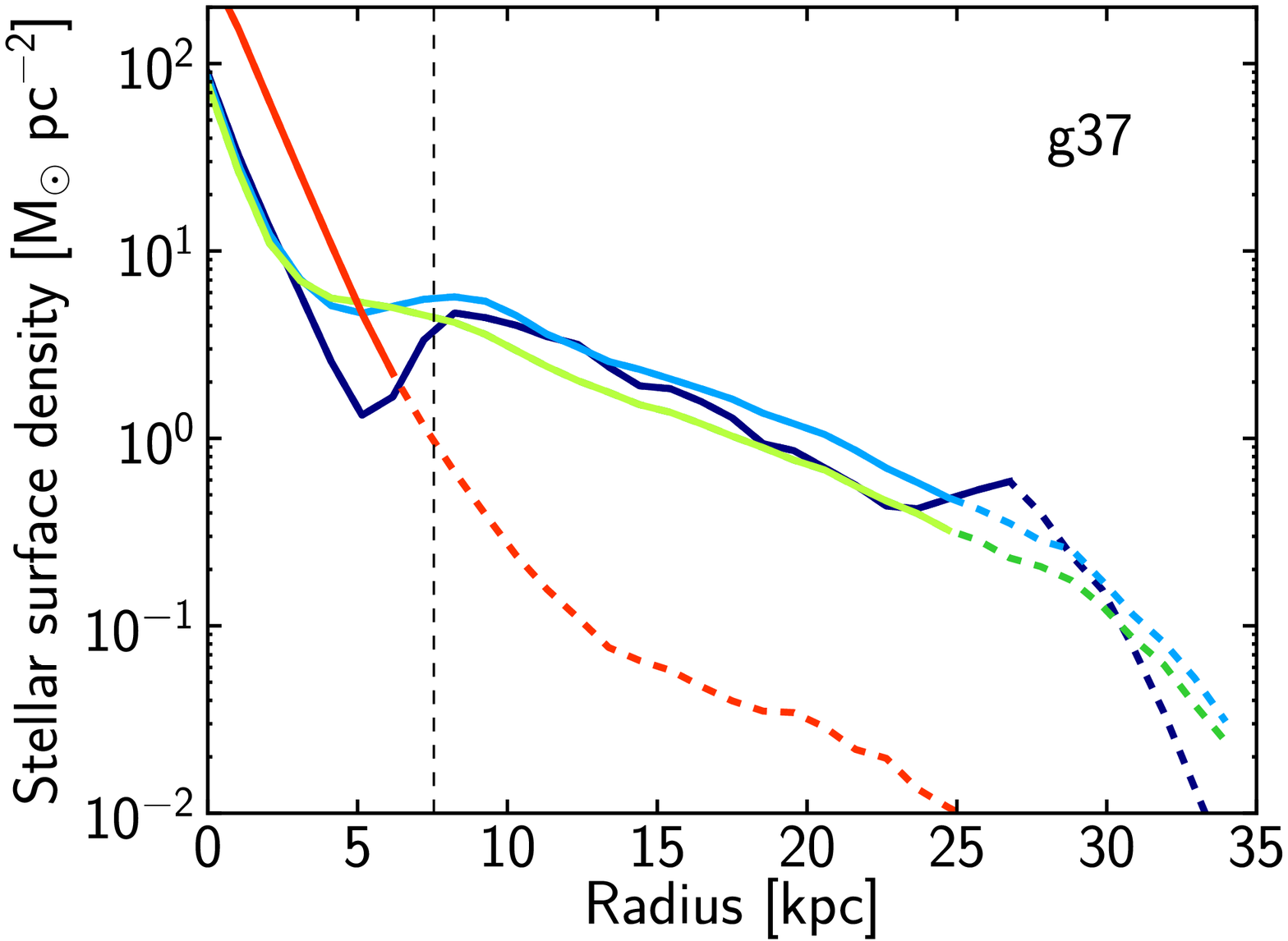}
\includegraphics[width=0.245\textwidth]{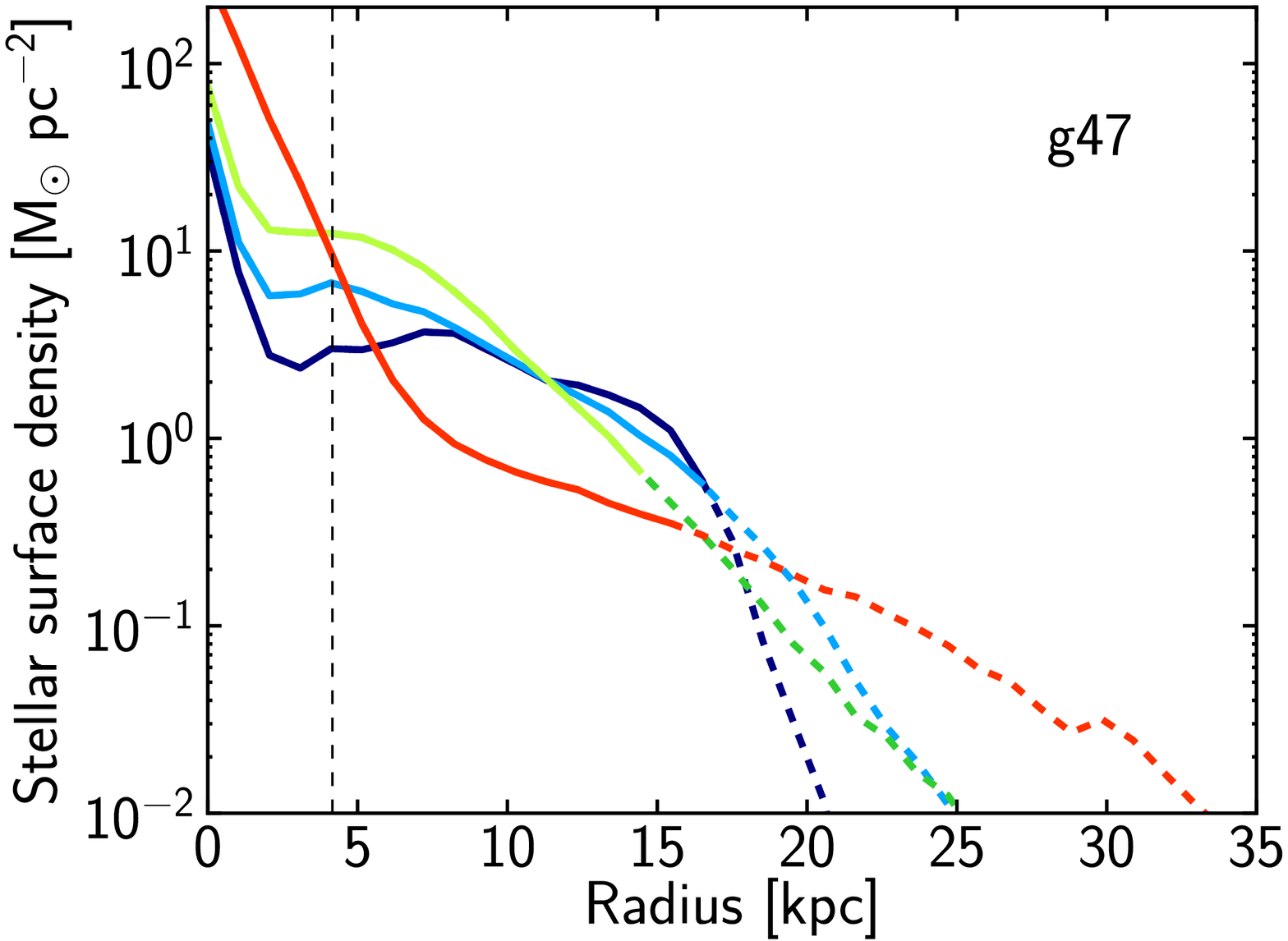}
\includegraphics[width=0.245\textwidth]{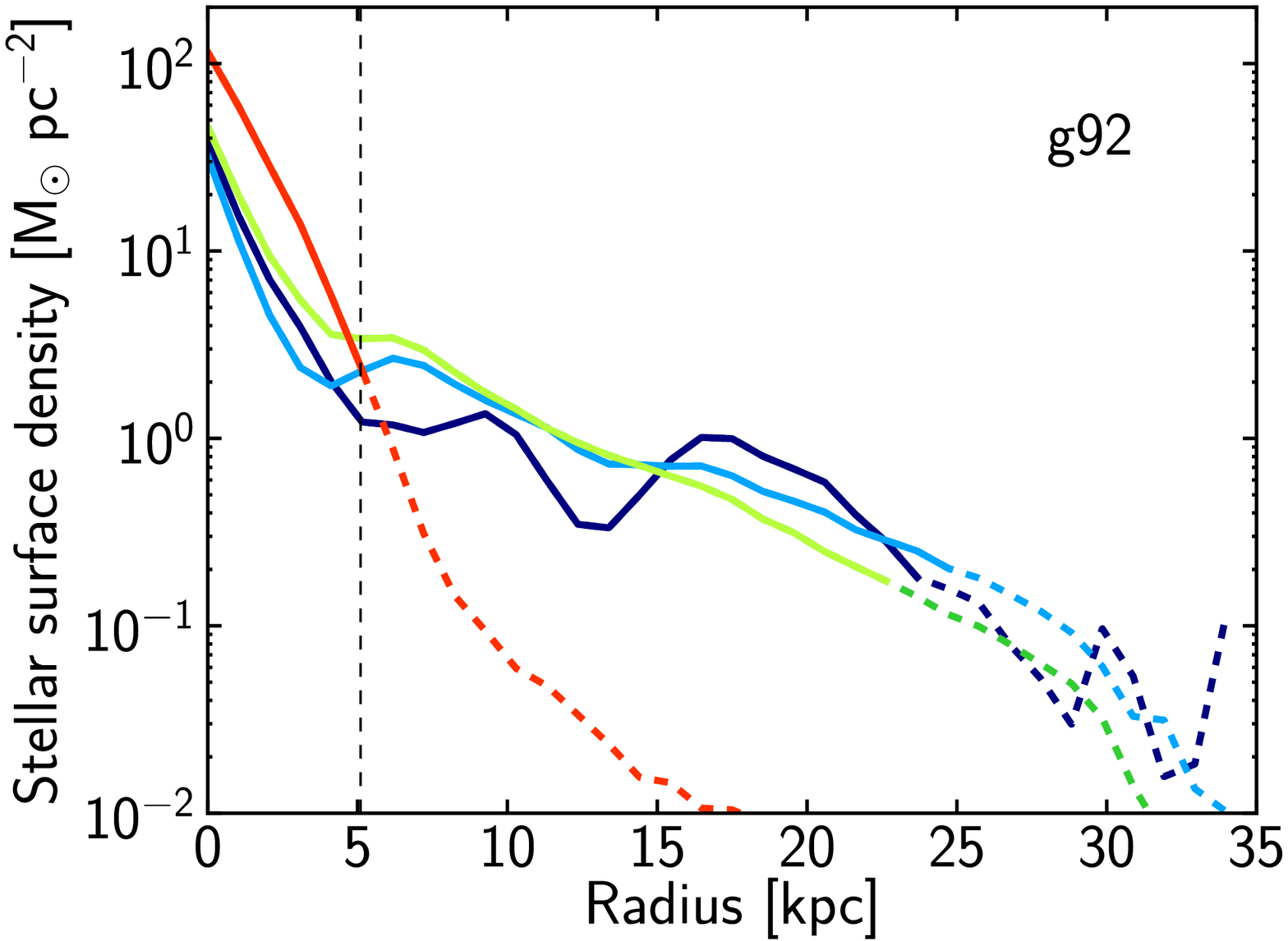}
\includegraphics[width=0.245\textwidth]{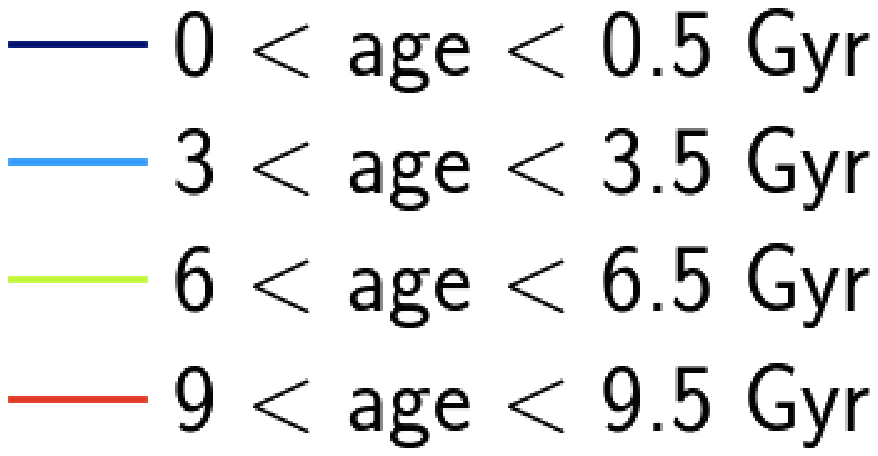}
\includegraphics[width=0.245\textwidth]{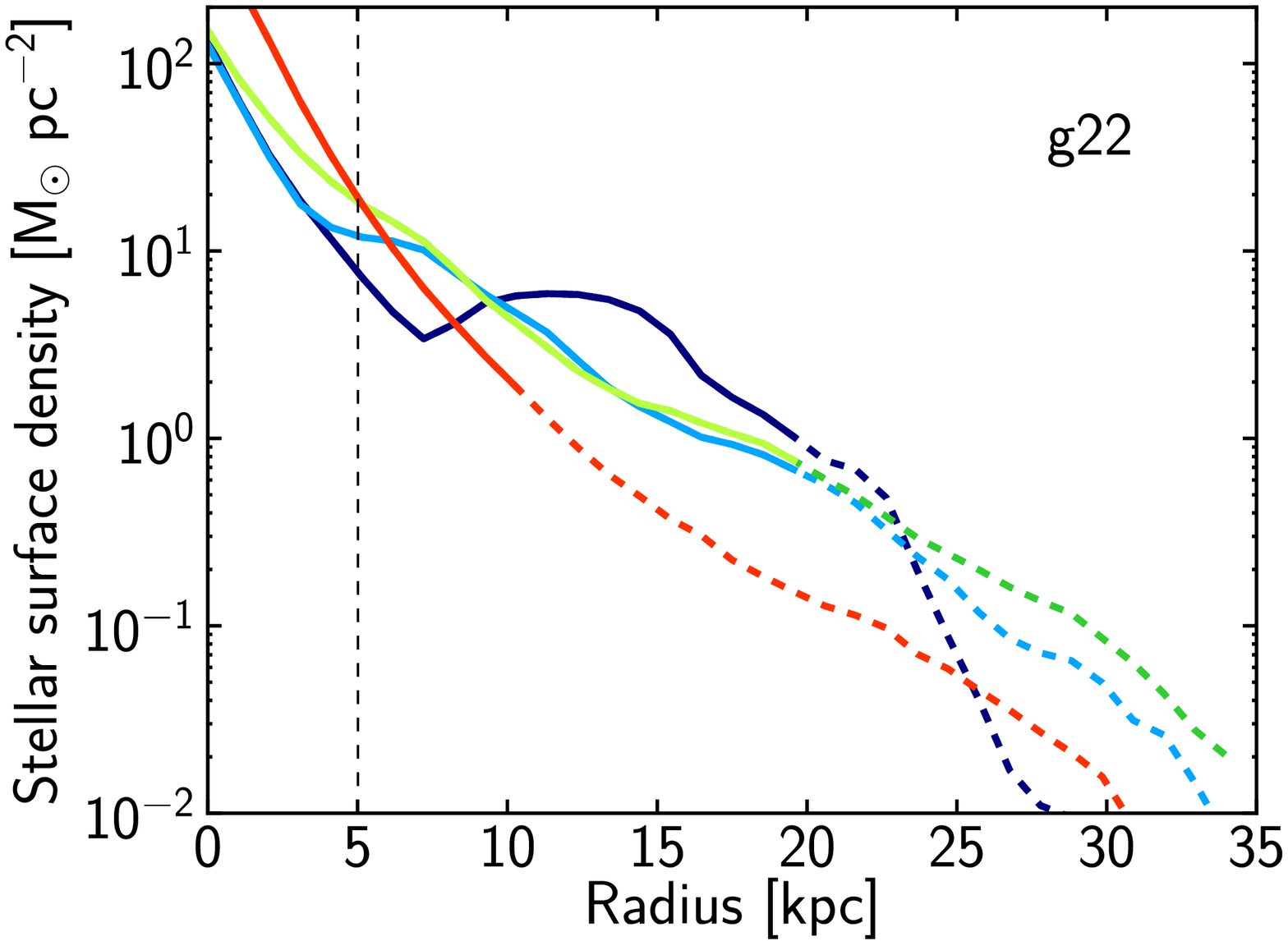}
\includegraphics[width=0.245\textwidth]{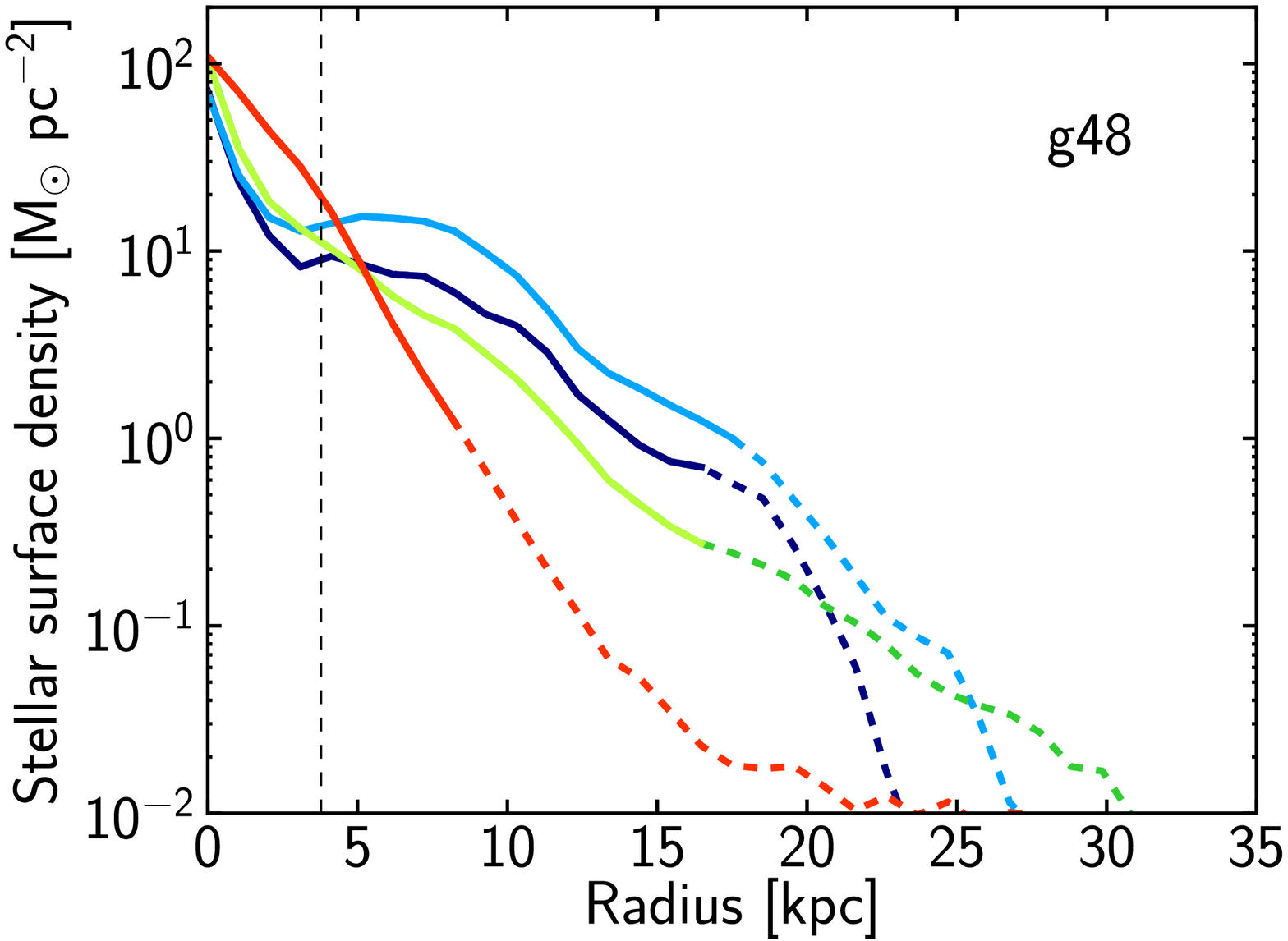}
\includegraphics[width=0.245\textwidth]{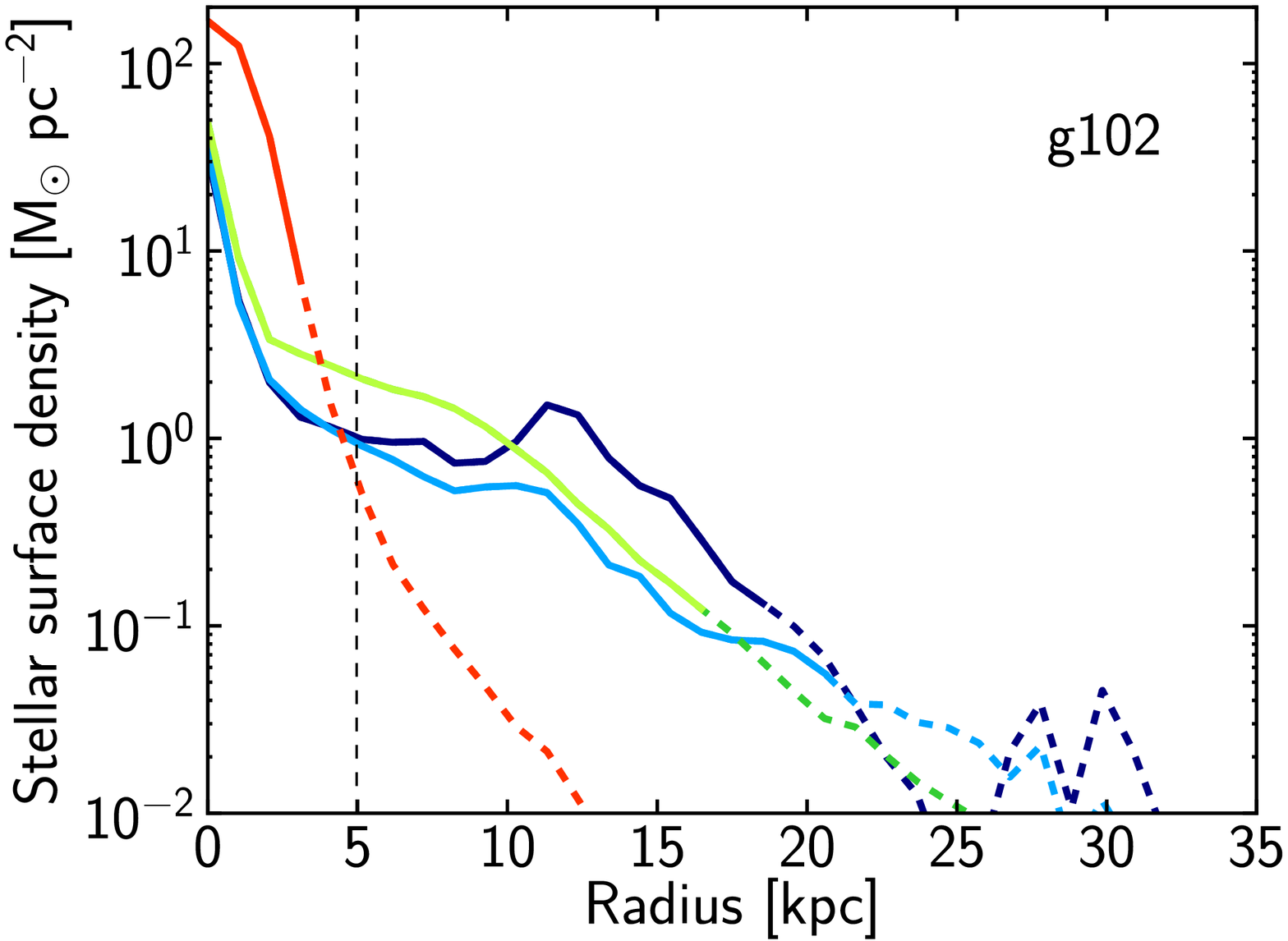}
\includegraphics[width=0.245\textwidth]{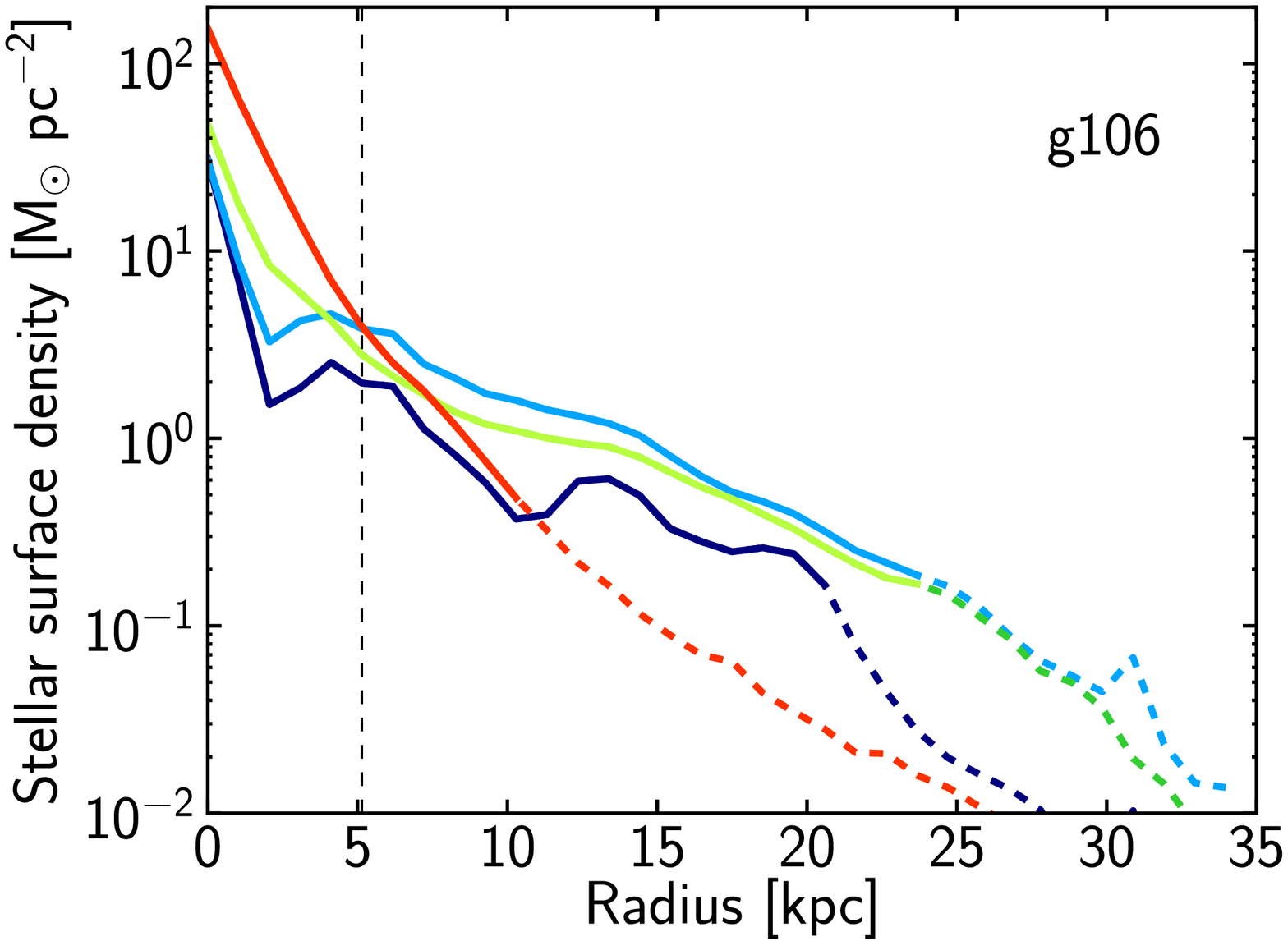}
\caption{Radial profiles of the stellar surface density for 4 different mono-age populations in each simulated galaxy. The solid lines correspond to the radial range up to $R_{95}$ (radius containing 95\% of stars of a given population). The vertical dashed line marks the value of $R_d$ (the scale-length for the total mass distribution), used as the minimum radius for the exponential fits to the profiles. The outer radius for the fits is either  $R_{95}$  or  $R_d$ + 3 kpc if $R_{95}$ is smaller than $R_d$ (as is the case for some of the oldest populations). On the top row are the three galaxies with a quiescent history, the bottom row shows galaxies with a more active merger history.}
\label{fig:radial_profiles}
\end{figure*}
From the initial sample of 33 galaxies, we select seven disc galaxies for this study. Choosing a reduced number of galaxies allows a deeper investigation of each,  while being representative of the diversity of histories and morphologies found in the larger sample. Two have a very quiescent merger history (g92 and g37), one has a quiescent history apart from a close fly-by at t$\sim$9 Gyr (g47), while the other four have various types of mergers after $z=1$ (g22, g48, g102, g106). The stellar mass, bulge fraction, optical radius and  disc scale-length are shown for each galaxy in Table \ref{tab-sample}. Face-on views of their total stellar mass distribution at $z=0$ are shown in Figure \ref{fig:images}, together with edge-on views of stars in different age bins.

All seven galaxies start their evolution with a phase of merger activity which contributes significantly to the buildup of a hot stellar component, as seen in the bottom row of Figure \ref{fig:images} for stars older than 9 Gyr (i.e. formed earlier than $z \sim 1.5$). disc growth starts after the merging phase, and Figure \ref{fig:images} shows that for all galaxies, older disc stars are in a thicker and more concentrated component than the youngest stars. There are nevertheless differences between galaxies in the shape of these components, since these galaxies have different histories for $z<1$.
\begin{description}
\item \textbf{g37} and \textbf{g92} have quiescent histories: over the last 9 Gyr of evolution ($z \lesssim 1.5$) all interactions have mass ratios smaller than 1:50\footnote{Throughout this paper, merger mass ratios are expressed in terms of relative stellar mass, and $t$ is the time since the Big Bang}. 
\item \textbf{g47} is also mostly quiescent, but undergoes a 1:15 interaction with a satellite at $t\sim$8.5--9 Gyr \textbf{($z \sim 0.5$)}, the satellite does not merge with g47 but does significantly disturb its morphology
\item \textbf{g22}  undergoes a 1:10 merger at $t=10$ Gyr \textbf{($z \sim 0.3$)}
\item \textbf{g48} undergoes a 1:4 merger at $t=8$ Gyr \textbf{($z \sim 0.6$)}
\item \textbf{g102} undergoes a 1:15 merger at $t=8.5$ Gyr \textbf{($z \sim 0.5$)}
\item \textbf{g106} undergoes a 1:5 merger at $t=7$ Gyr \textbf{($z \sim 0.7$)}
\end{description}

\subsection{Analysis}

We first determine the global disc scale-length ($R_d$) from an exponential fit\footnote{All fits shown in this paper are performed using the least squares method implemented in SciPy.} to the total stellar mass distribution for $|z|<5$ kpc. The fit is performed between the radii containing 50\% and 95\% of stars within a radius of 30 kpc. This will be used when referring to quantities measured at 2 or 3 $R_d$. The values of $R_d$ for each galaxy are shown in Table \ref{tab-sample}. 

Overall, these scale-lengths are in a reasonable range compared to observed local galaxies: \cite{Fathi2010} measured an average $r$-band scale-length of 5.7$\pm$1.9 kpc for massive galaxies in the SDSS. They are also quite similar to M31, for which the scale-length is between 5 and 7 kpc depending on the band in which it is measured (see a compilation of values in \citealp{Yin2009}). 
However, the simulated galaxies have flatter density profiles than the Milky Way, for which  the scale-length is typically estimated to lie in the 2--3 kpc range (see e.g. \citealp{Freudenreich1998,Robin2003,McMillan2011, Bovy2013}).

We slice the stars at $z=0$ in our galaxies into 500 Myr age bins from 0 to 11 Gyr, which we will then refer to as "mono-age populations" (MAPs). We have tested other age intervals, and find that wider age bins tend to induce errors on the measured properties of the populations, especially in the case of mergers (which induce rapid changes). 

We also remove counter-rotating stars from the selection, as a way to minimize the contamination of our sample by halo stars. These retrograde stars represent 10--40\% of the oldest stars, but none of the young populations (except in the central regions, which we do not study in this paper). Among the oldest stars there remains a contamination by halo stars. As a consequence, we overestimate the velocity dispersion for oldest stars compared to a pure disc sample. These old populations are however not the main focus of this paper (because we are interested in the structure of discs, we mostly focus on stars that are less than 9 Gyr old \footnote{Note however than in the Milky Way some disc stars are older than 10 Gyr, which limits the possible comparisons between simulations and observations for these very old stars}), and this procedure does not affect our conclusions.

The high mass resolution of the simulations allows for large numbers of stellar particles in each MAP. For instance, a 2 kpc-wide cylinder centred around a galactocentric distance of 2$R_d$ (typical of what is used in this study) contains between 100 and 80,000 stellar particles of a given age bin (depending on the simulated galaxy and the age bin --- the smallest numbers of particles are usually found for the oldest populations), with typical values above 5,000 stellar particles.

\section{The simplicity of mono-age populations}

For the Milky Way, \cite{Bovy2012b,Bovy2012c} find that mono-abundance populations have remarkably simple properties, at least in the radial and vertical range probed by the SEGUE data. The spatial structure of each population is well fitted by a single exponential in the radial and vertical directions, and the velocity dispersion is close to constant with height above the disc, i.e. each population is closely isothermal.

In this Section, we investigate if such simple populations are a common feature of our simulated galactic discs, and whether this simplicity requires a quiescent merger history.

\subsection{Radial density profiles} \label{sec:Rcarac}

\begin{figure*}
\centering 
\includegraphics[width=0.245\textwidth]{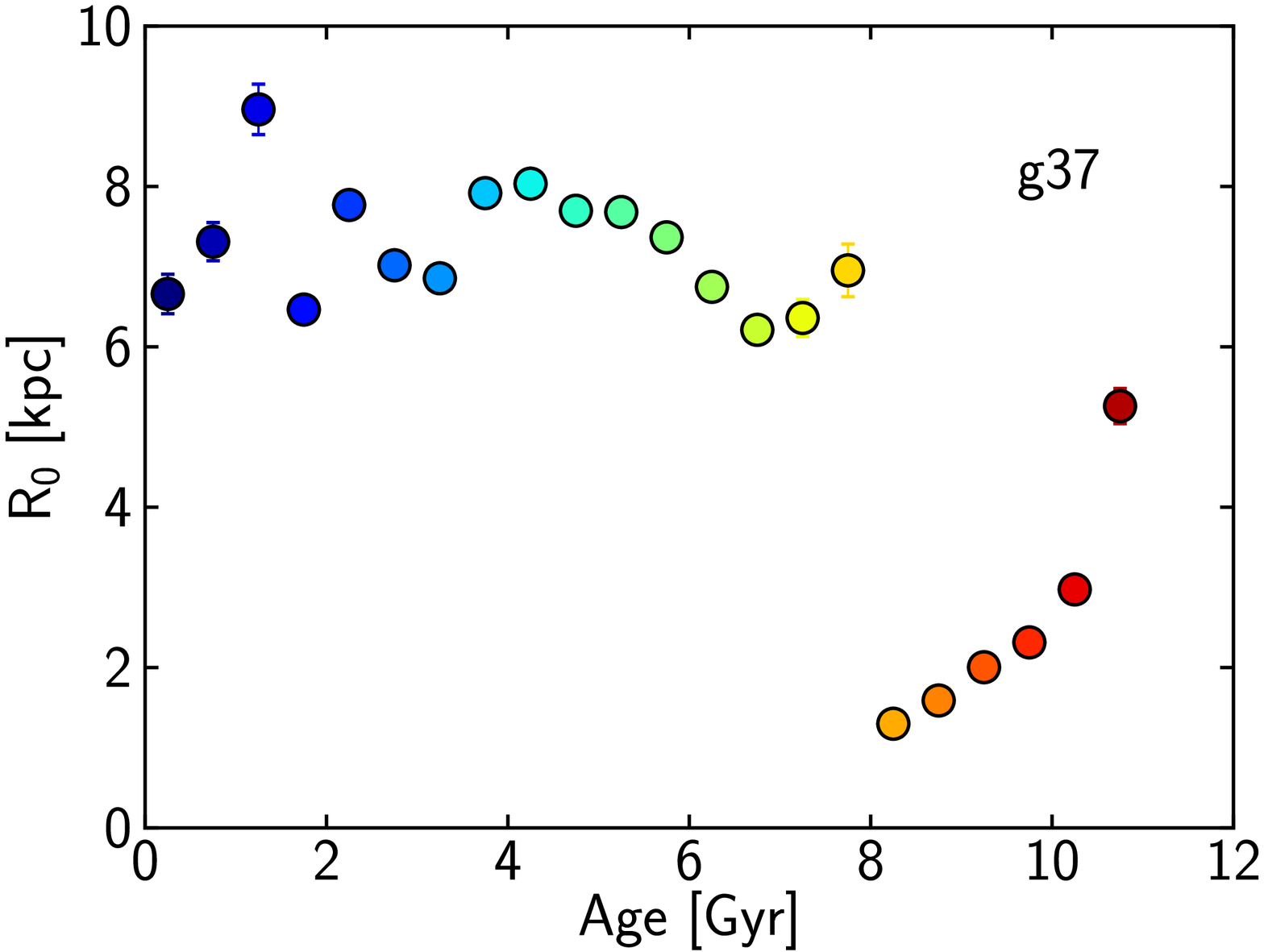}
\includegraphics[width=0.245\textwidth]{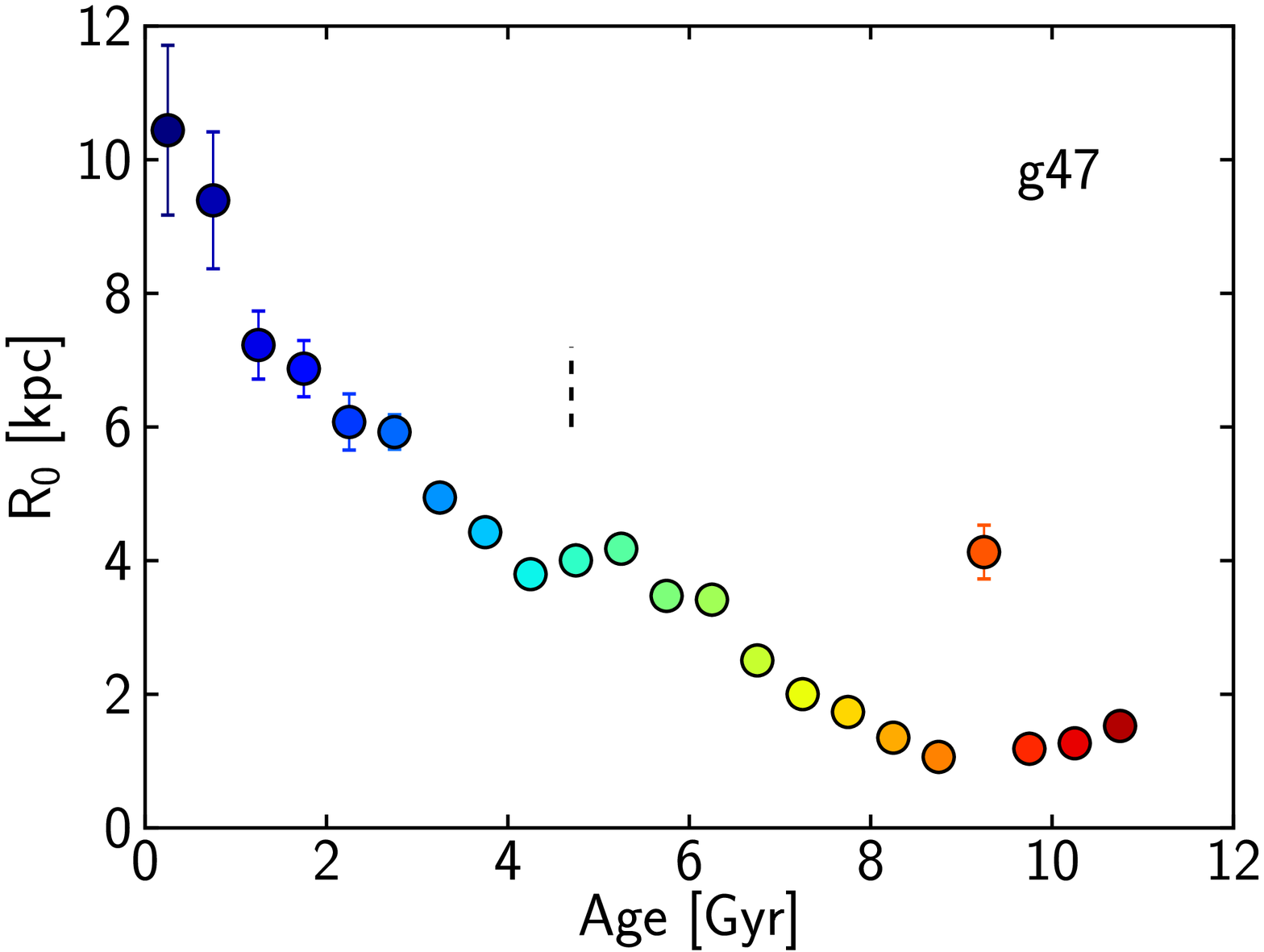}
\includegraphics[width=0.245\textwidth]{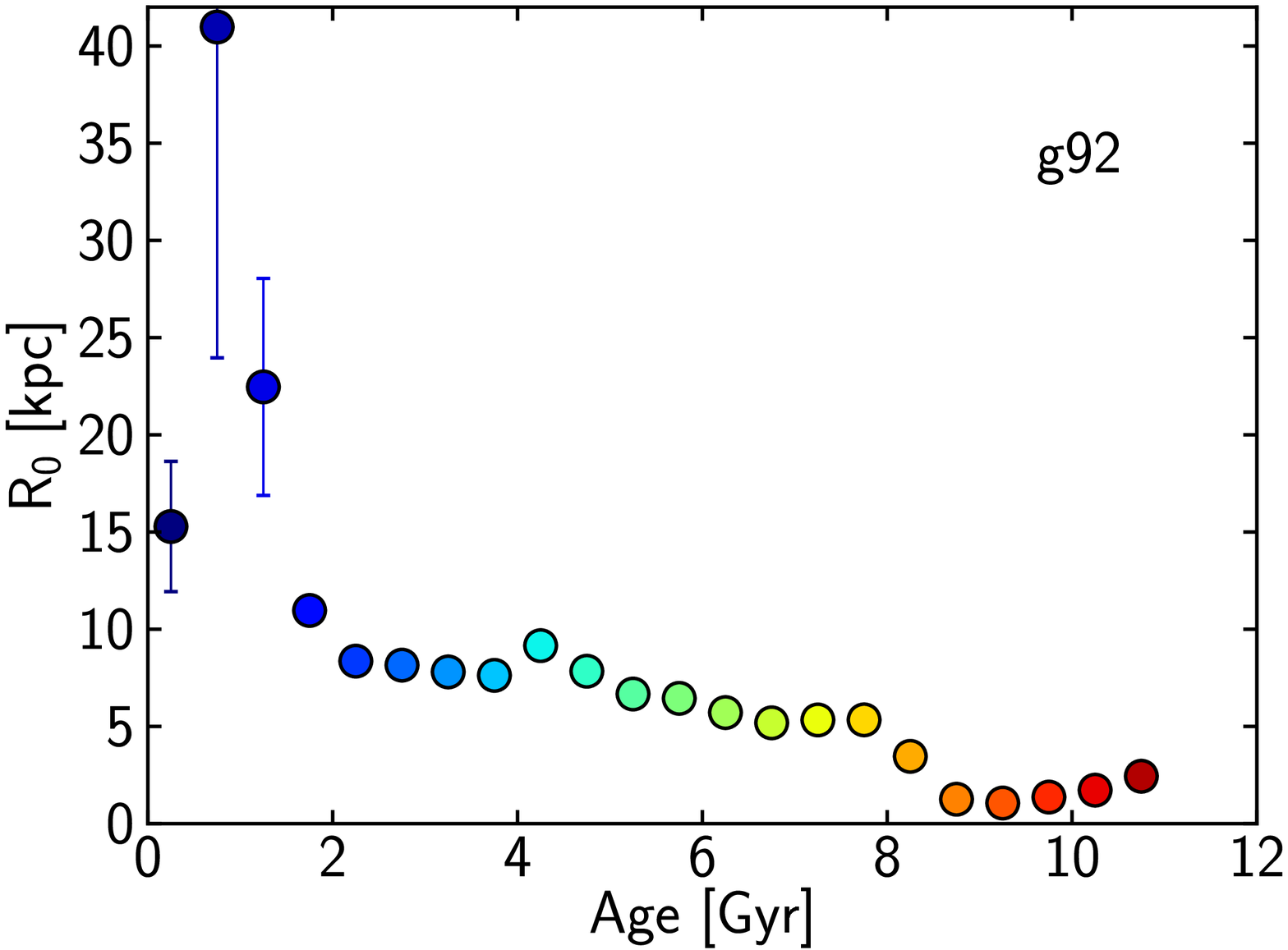}
\includegraphics[width=0.245\textwidth]{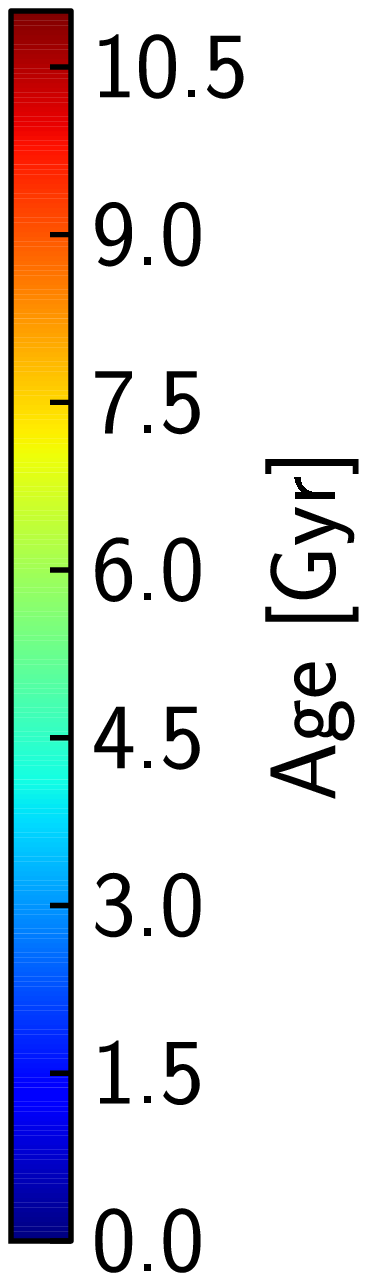}
\includegraphics[width=0.245\textwidth]{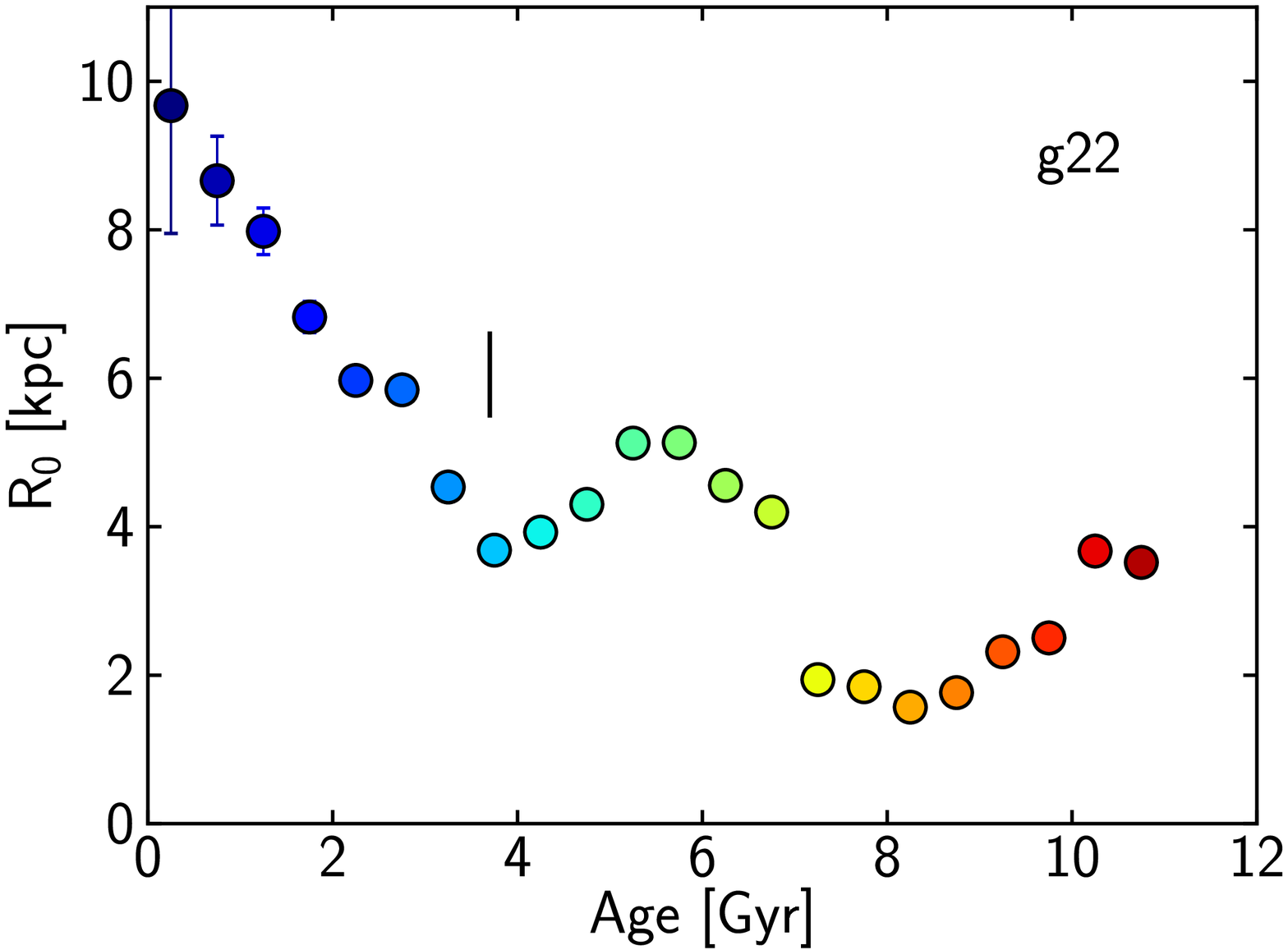}
\includegraphics[width=0.245\textwidth]{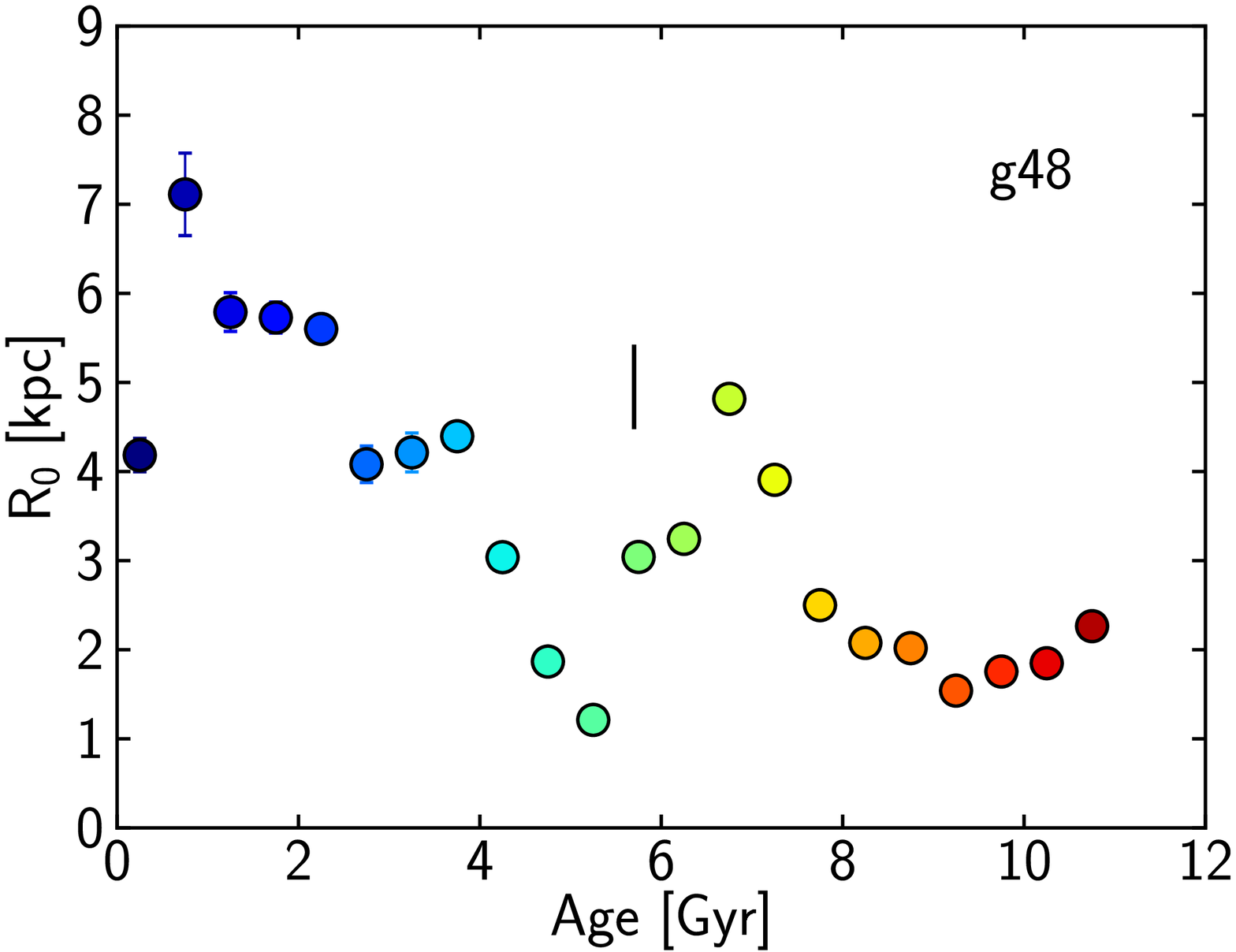}
\includegraphics[width=0.245\textwidth]{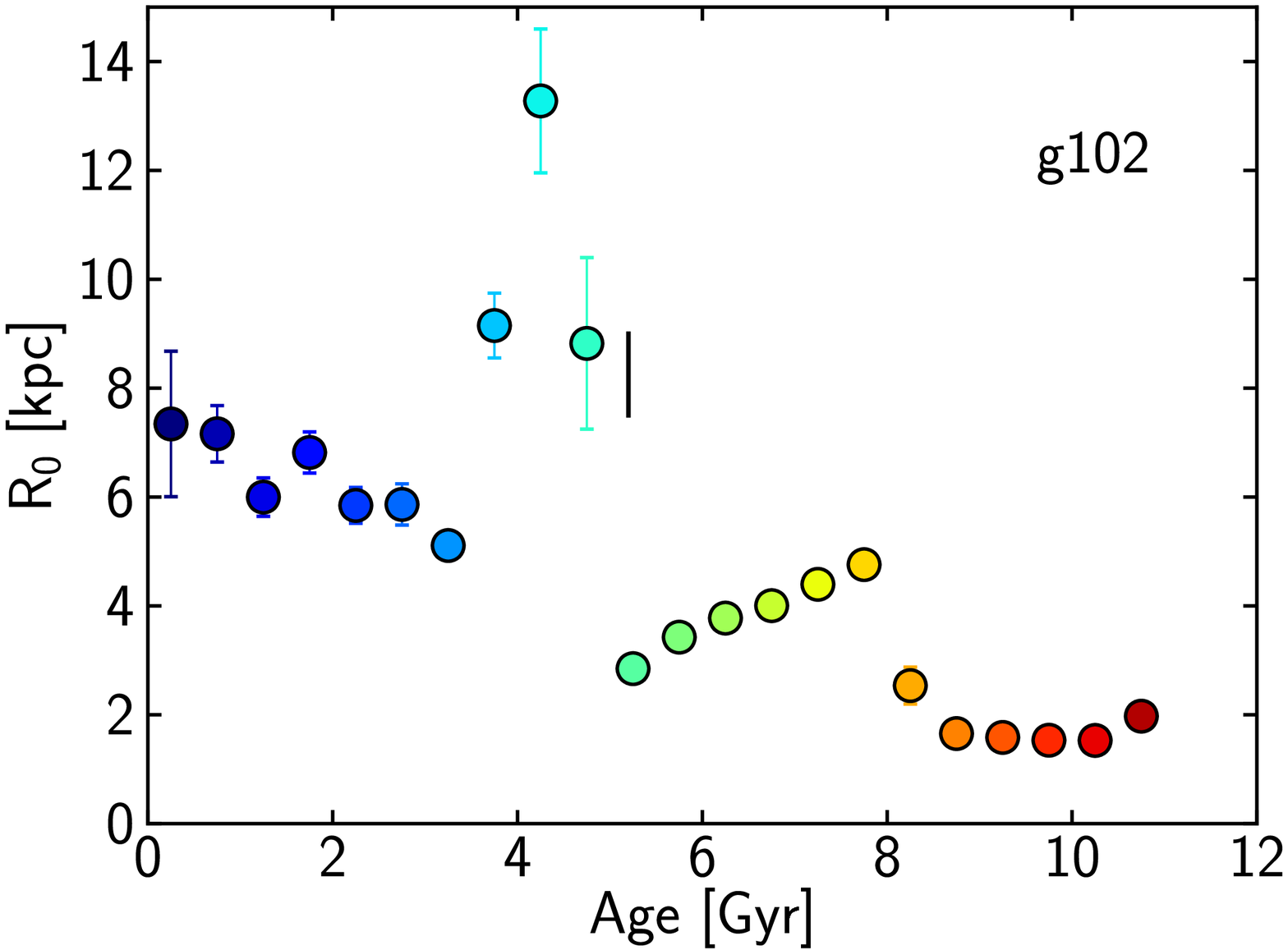}
\includegraphics[width=0.245\textwidth]{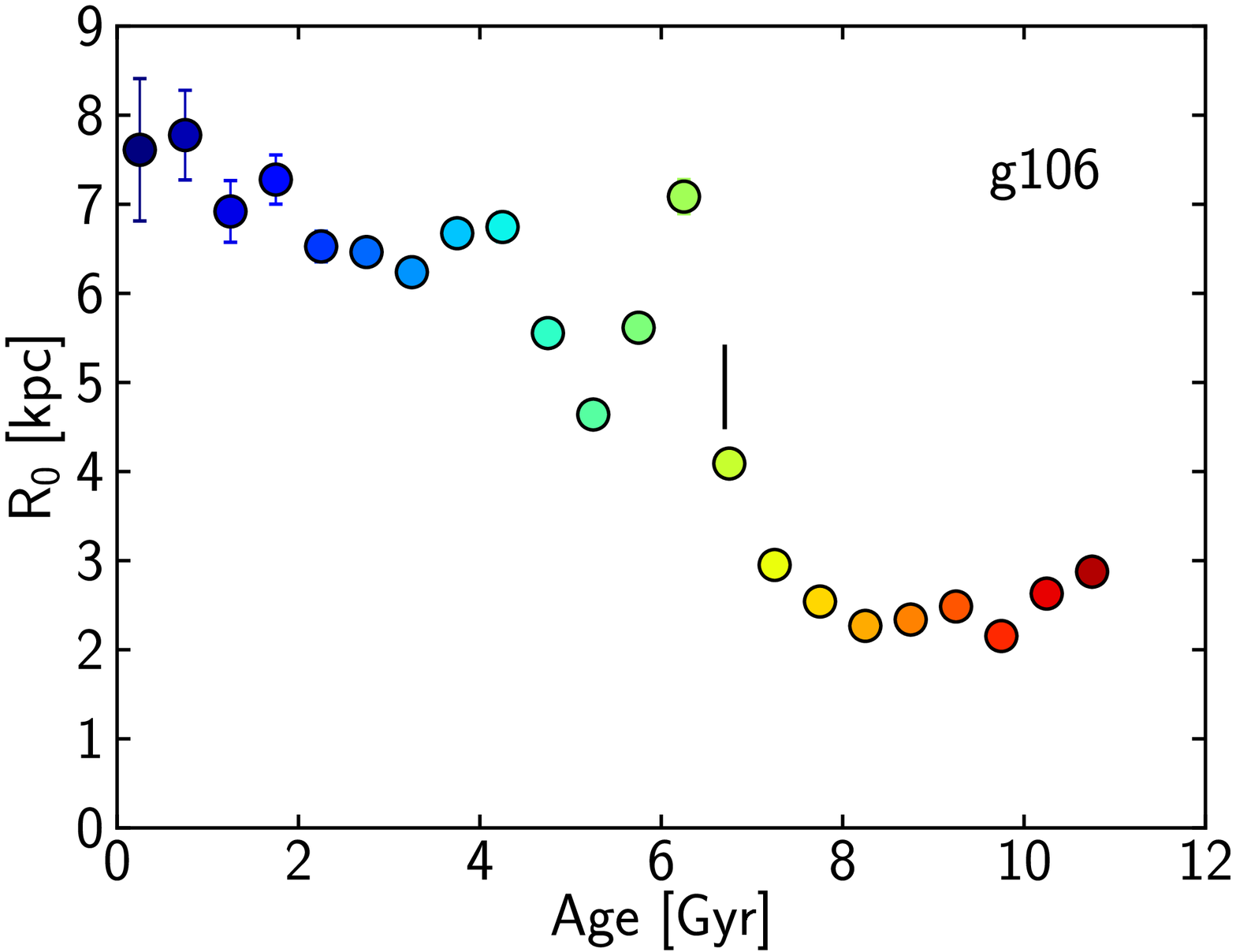}
\caption{Scale-length as a function of age for mono-age populations in the 7 simulated galaxies. The scale-lengths are obtained from exponential fits to the surface density profiles, as described in the text and Figure \ref{fig:radial_profiles} (the error bars represent the 1-$\sigma$ range obtained from the fit). The small vertical lines mark the time of coalescence for the last merger undergone by each galaxy (the dashed line for g47 marks the end of a fly-by). The colour code reflects the age of each population, and is here redundant with the $x$-axis, but is useful for comparisons with further Figures, which will use the same colour code.}
\label{fig:R0}
\end{figure*}
We compute the radial surface density profile for all MAPs (for stars with $|z|<5$~kpc). We show in Figure \ref{fig:radial_profiles} examples of such profiles, for four MAPs in each simulated galaxy. 

In all cases the oldest populations (age $\gtrsim 9$ Gyr) have clearly distinct density profiles. They are the most centrally concentrated populations (as also seen in Figure \ref{fig:images}), and contribute the most to the mass distribution in the bulge/inner halo. Intermediate age and young populations show a clear disc component with, in most cases, smooth and regular profiles, except for the youngest stars. These young stars tend to have a more disturbed morphology, with bumps at the locations of the rings, spiral arms or clumps where they were born, and they have not yet  had the time to migrate and create a smooth profile\footnote{the irregularities in density profiles for young stars are not a statistical effect of a small number of such stars}.

We perform an exponential fit for each density profile (more exactly, a linear fit to the logarithm of the density profile). The inner radius of the fit is chosen to be  $R_d$ to exclude the central regions of a galaxy (in Figure \ref{fig:radial_profiles} that inner radius is marked by the vertical dashed line). The outer radius for the fit is either the radius containing 95\% of the population ($R_{95}$)  or  $R_d$ + 3 kpc if $R_{95}$ is smaller than $R_d$ (as is the case for some of the oldest populations). This outer cut avoids including parts of the profile that are beyond the truncation radius.

The exponential fits are a reasonable approximation to most profiles, but note that in some cases very different results would be obtained if the fit were only performed on a more limited radial range (for instance, a very steep slope would be found for the youngest stars in g106 if the fit were limited to the 5--10 kpc region --- see Figure  \ref{fig:radial_profiles}).

We show in Figure \ref{fig:R0} how the scale-length derived from the fits ($R_0$) varies with the age of the populations. Some of these scale-lengths are very high, corresponding to relatively flat density profiles. Note that while \cite{Bovy2012b} only show scale-lengths up to 4.5 kpc for mono-abundance populations in the Milky Way (see their Figures 4 and 5), some of these populations do  have much larger scale-lengths, although with large errors (Jo Bovy, private communication).

Overall, there is a general trend of decreasing $R_0$ with increasing age, along with significant differences between galaxies. The galaxy with the smoothest variations of $R_0$ with age is g47 (one of our fly-by cases), with $R_0$ varying from 1--2 kpc for the old stars to $\sim 10$ kpc for the youngest stars\footnote{these large variations of $R_0$ do not translate into very different photometric scale-lengths: the scale-lengths measured on mock images only vary from 5 kpc in $K$ band to 6 kpc in $B$ band}. The fly-by shows up as a bump (for ages around 5 Gyr) as the disc scale-length increases and then decreases again. We also find a mostly regular structure for g92, with a  strong increase of $R_0$ for stars younger than 1.5 Gyr. The last of our quiescent galaxies, g37, does not show strong variations of $R_0$ with age (as is also clearly apparent from the profiles in Figure \ref{fig:radial_profiles}).

Mergers tend to produce more abrupt variations in the time evolution of $R_0$ (except for g22, which is actually quite similar to g47 in terms of $R_0$ evolution). They produce torques that strongly affect the radial distribution of gas and stars in galaxies; they both redistribute the old stars and change the spatial distribution of newly formed stars.
For instance, in the case of g48 the striking V-shaped feature in Figure \ref{fig:R0} is associated with the major merger: the large $R_0$ values seen in this galaxy for ages of 6.5--7 Gyr correspond mostly to stars formed in situ and undergoing strong radial extension under the effect of the merger,  and the smallest $R_0$ corresponds to star formation being limited to the inner few kpc of the galaxy soon after coalescence.

In summary, we find that the radial surface density profiles of our galaxies are in most cases successfully fitted by simple exponentials. The oldest and youngest stars are often an exception to that rule. We find a general trend of increasing scale-lengths for younger stars, but with large variations from one galaxy to another.

\subsection{Vertical density profiles}
\begin{figure}
\centering 
\includegraphics[width=0.3\textwidth]{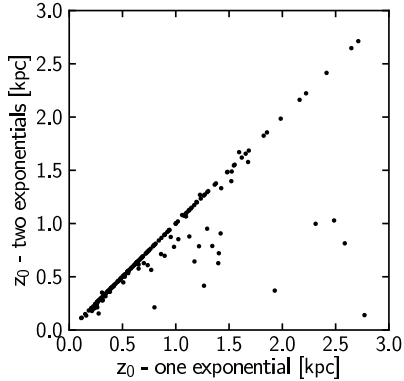}
\caption{Testing the quality of single exponential fits to the vertical density profiles of each mono-age population in each simulated galaxy. We fit the profiles both with a single exponential and the sum of two exponential representing a thin and thick components. We compare the scale-height for the dominant (most massive) component of the two exponential fit with the scale-height for the single component fit. Points are shown for all seven galaxies and all mono-age populations sampled at four different radii (only shown are results obtained for reasonable fits). We find that a single exponential is a good fit to most vertical density profiles, at the exception of some of the oldest populations (the outliers on this Figure), for which a single component fit overestimates the scale-height.}
\label{fig:z0_fits}
\end{figure}

To measure the vertical density profiles of our galaxies, we divide each MAP into 2 kpc-wide concentric radial bins. In each radial bin, we examine the vertical range encompassing 95 \% of stars in that MAP (limited to 5 kpc). Over that vertical range, we then compute the vertical density profile using 50 bins.

Each vertical density profile is fitted both with an exponential and a sech$^2$ profile. For the exponential fits, we only consider stars 200 pc above the plane: in some cases the density profiles show signs of flattening at small $z$ which does not allow a good overall exponential fit. Depending on the galaxy, the age of the populations and the radius considered, either the exponential or sech$^2$ fit can be the best match to the profile, and we do not find one of these functions to be the best in all cases. For simplicity, we only present here the results obtained with the exponential fits, but our conclusions would be unchanged if we had used the sech$^2$ fits results.

In nearly all cases a single exponential is enough to provide a good fit to the density profiles. However, we also tested fitting the profiles with the sum of two exponentials representing a thin and thick component. In many cases the double exponential fits converge towards the same scale-height for the two components (indicating that only one component is actually present). When two different scale-heights are found, the one of the dominant (most massive) component is often very close to the scale-height obtained for the single fit. Figure  \ref{fig:z0_fits} shows the scale-height of the dominant component in a two-exponential fit as a function of the single exponential scale-height, for all seven galaxies, for all MAPs and at four different radii in the discs (only shown are points for which satisfactory fits were obtained). Only a few points deviate from the one-to-one correlation, and most of them are for old populations. For these outliers, it seems that a single component fit overestimates the scale-height, so that we might artificially measure too much flaring in these old populations.
For most of the populations however, a single exponential is a good fit to the vertical density profiles.

\begin{figure*}
\centering 
\includegraphics[width=0.245\textwidth]{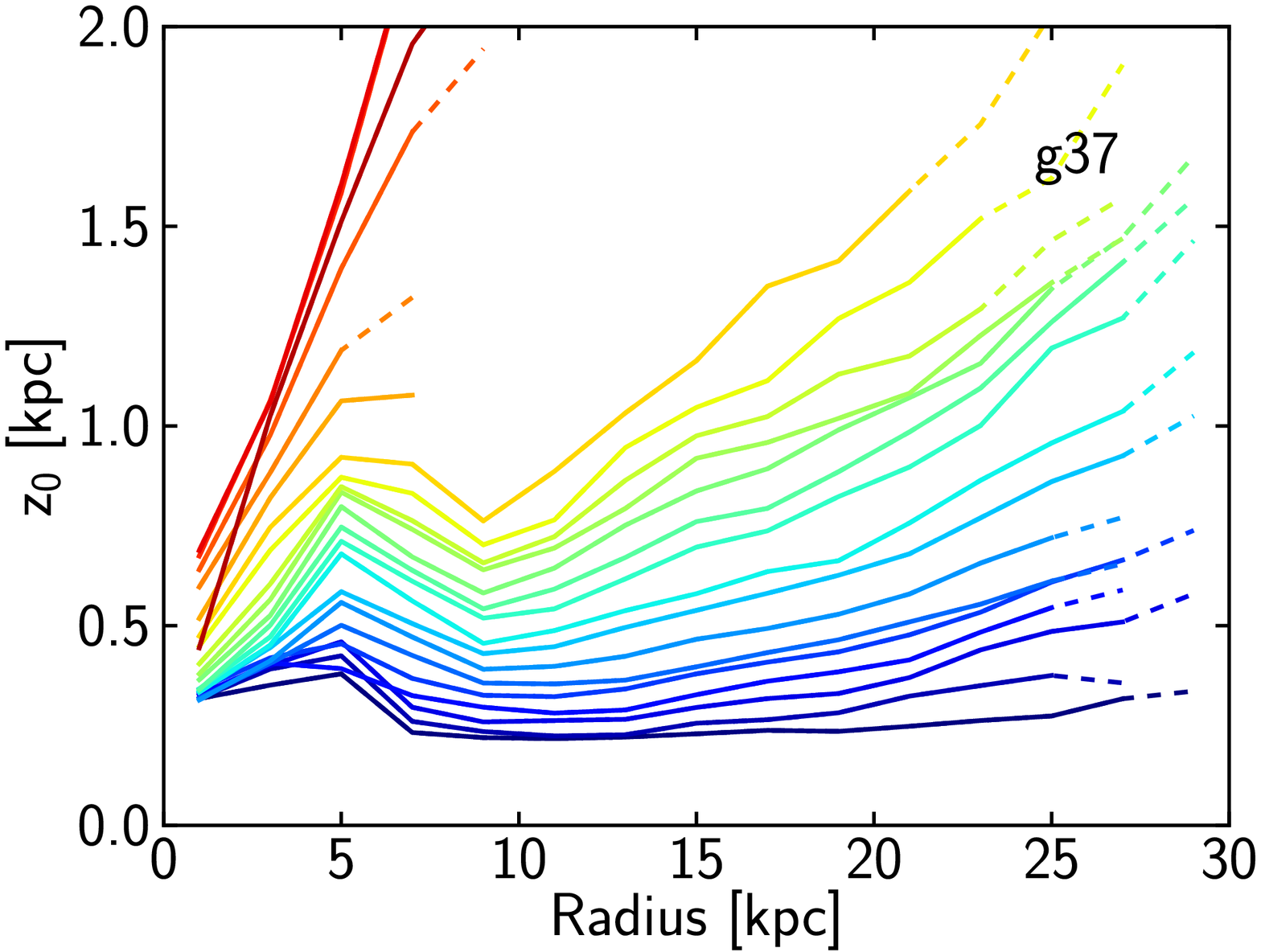}
\includegraphics[width=0.245\textwidth]{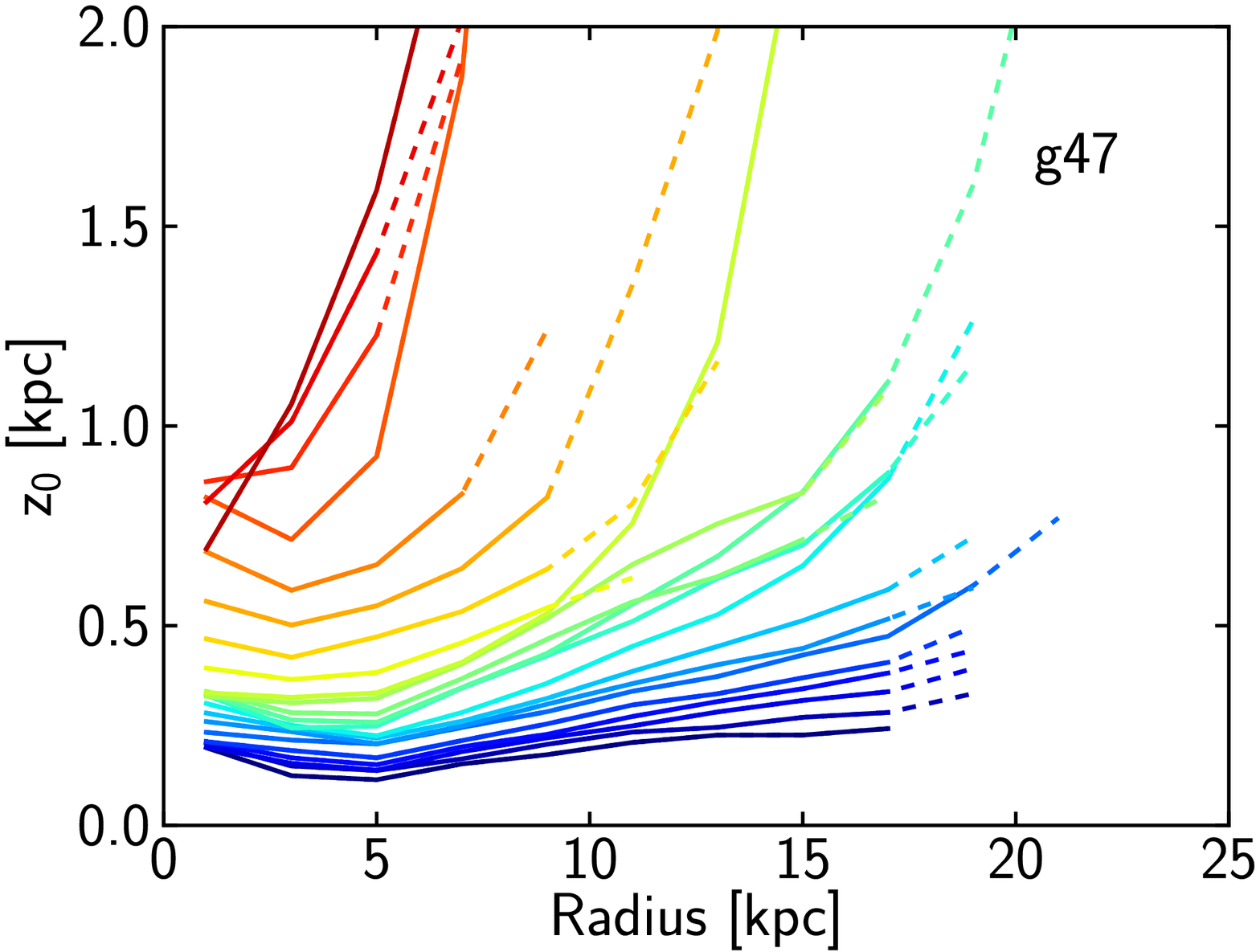}
\includegraphics[width=0.245\textwidth]{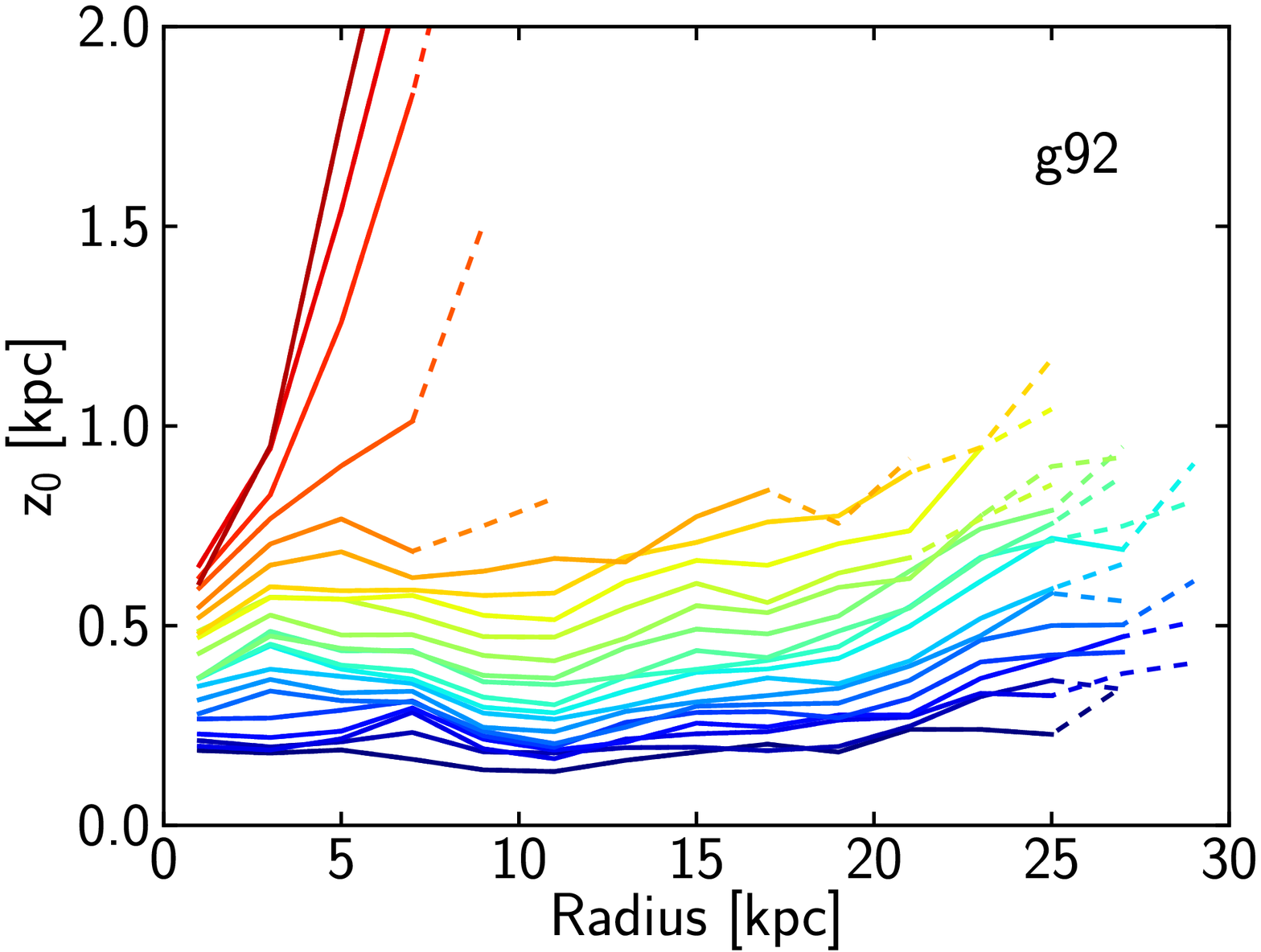}
\includegraphics[width=0.245\textwidth]{colorbar.eps}
\includegraphics[width=0.245\textwidth]{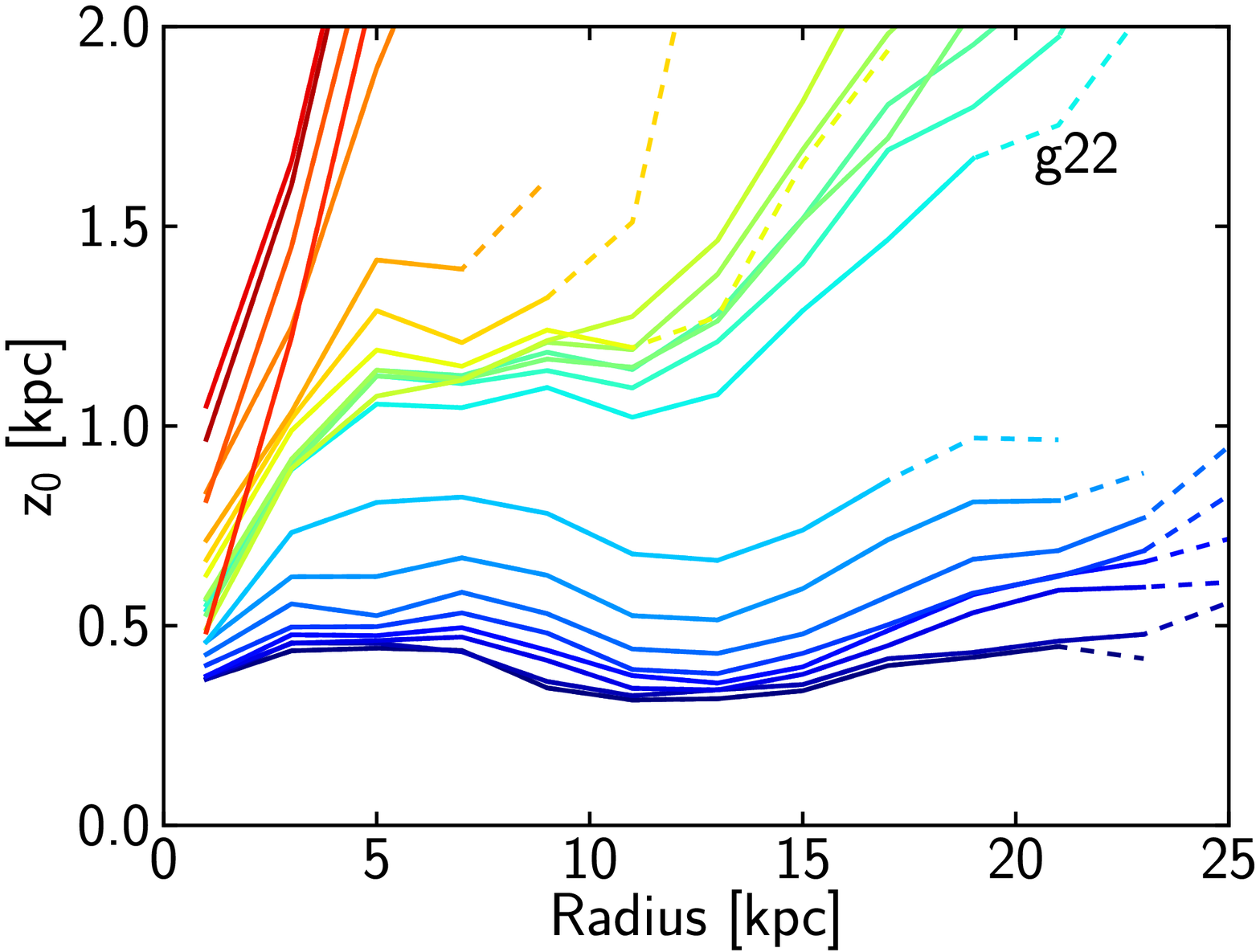}
\includegraphics[width=0.245\textwidth]{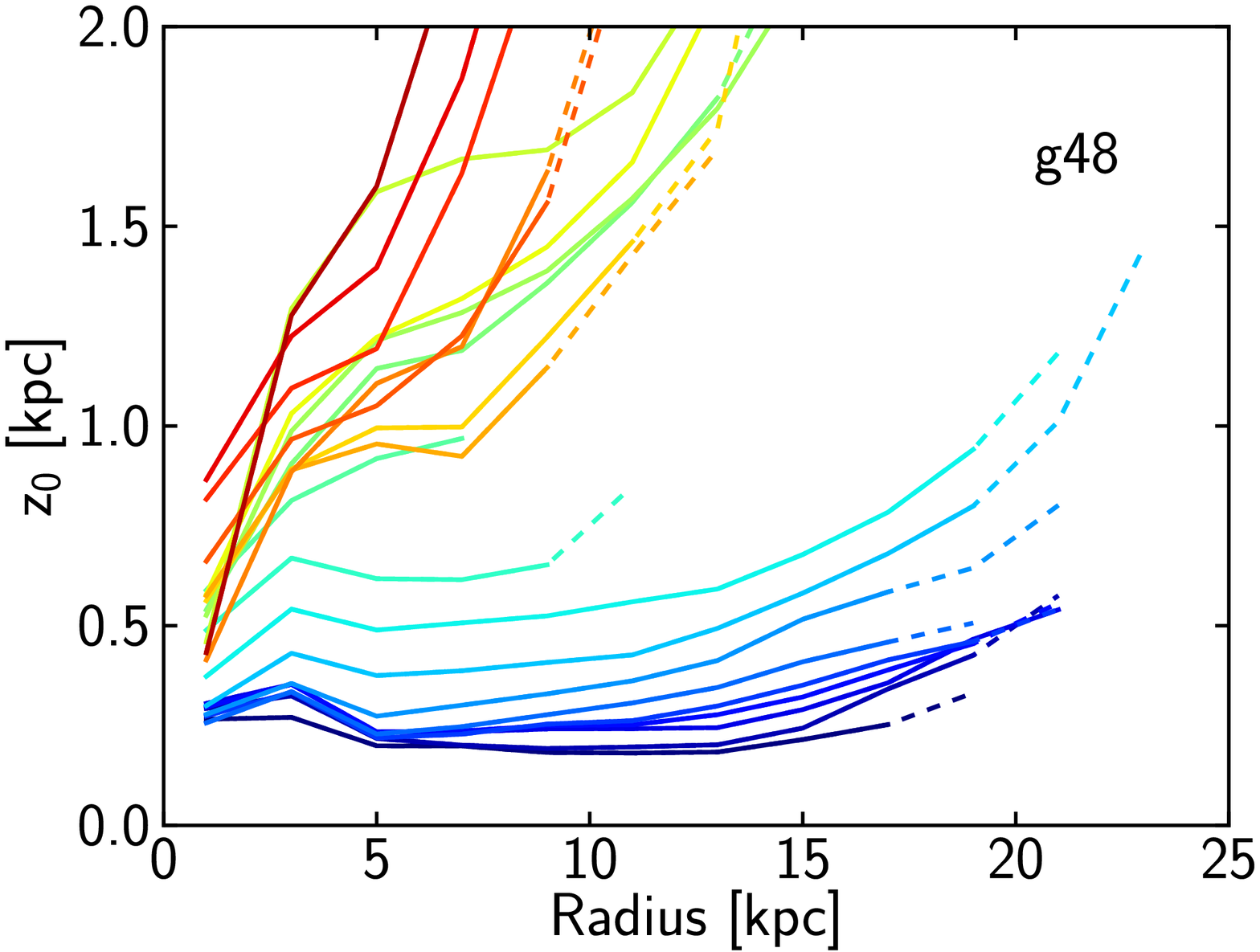}
\includegraphics[width=0.245\textwidth]{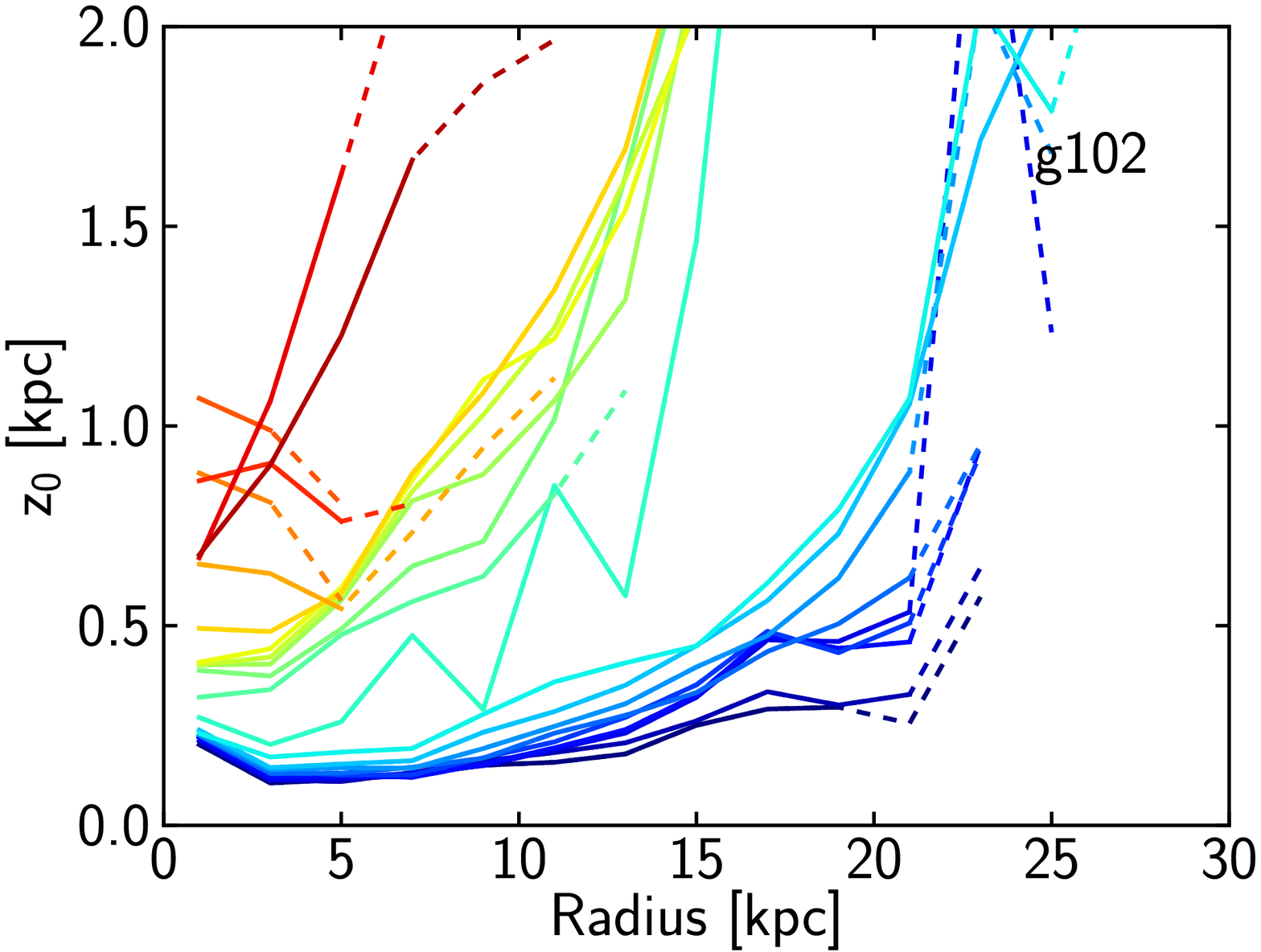}
\includegraphics[width=0.245\textwidth]{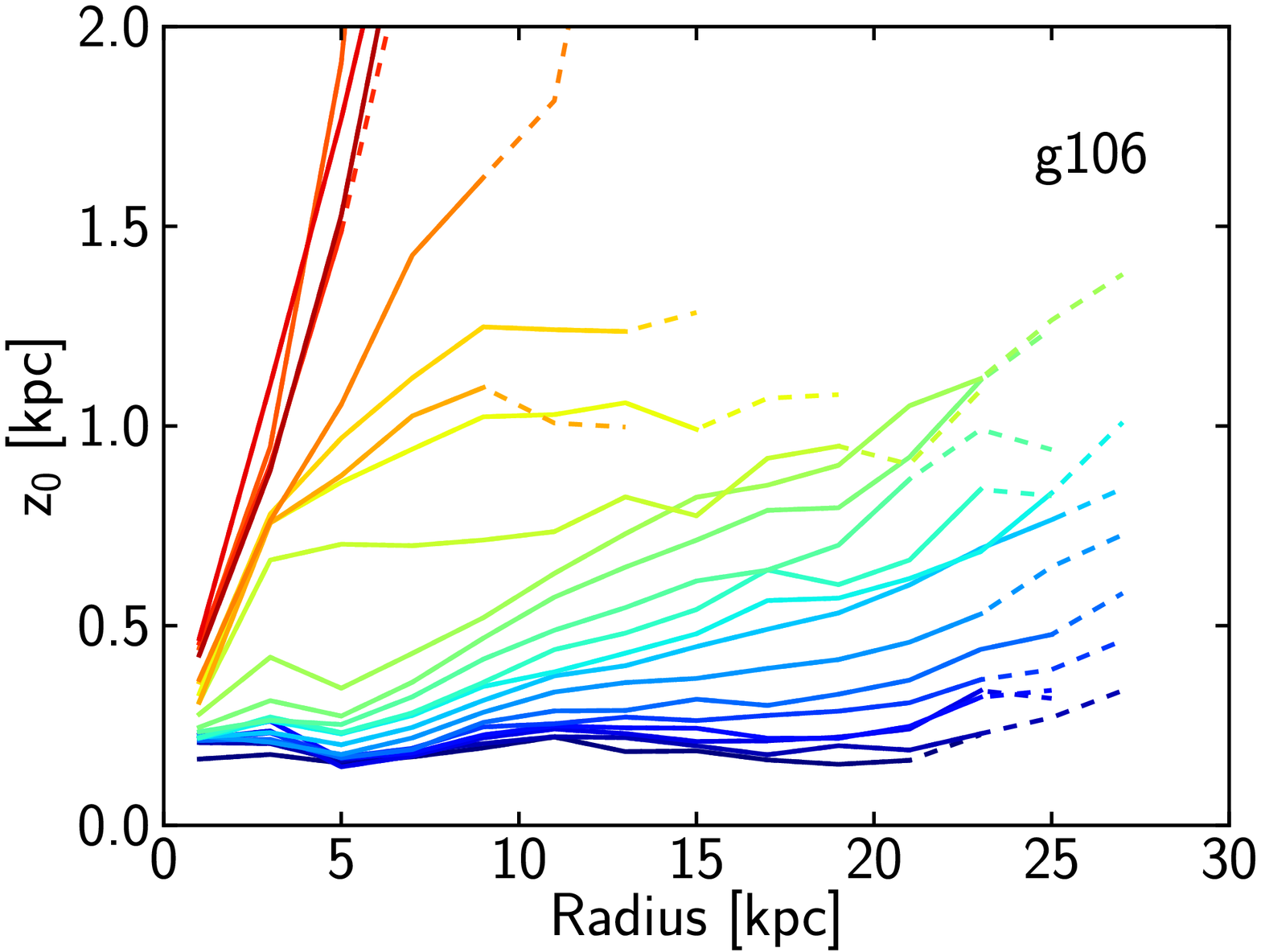}
\caption{Radial profiles of the scale-height for the mono-age populations (the colourcode indicates the age of each population, from 0 in blue to 11 Gyr in red). The solid lines correspond to the radial range up to $R_{95}$ (radius containing 95\% of stars of a given population), the dashed lines extend out to  $R_{98}$. The upper row shows our three quiescent galaxies, while the lower row shows those with an active merger history.}
\label{fig:prof_z0}
\end{figure*}

\begin{figure*}
\centering 
\includegraphics[width=0.245\textwidth]{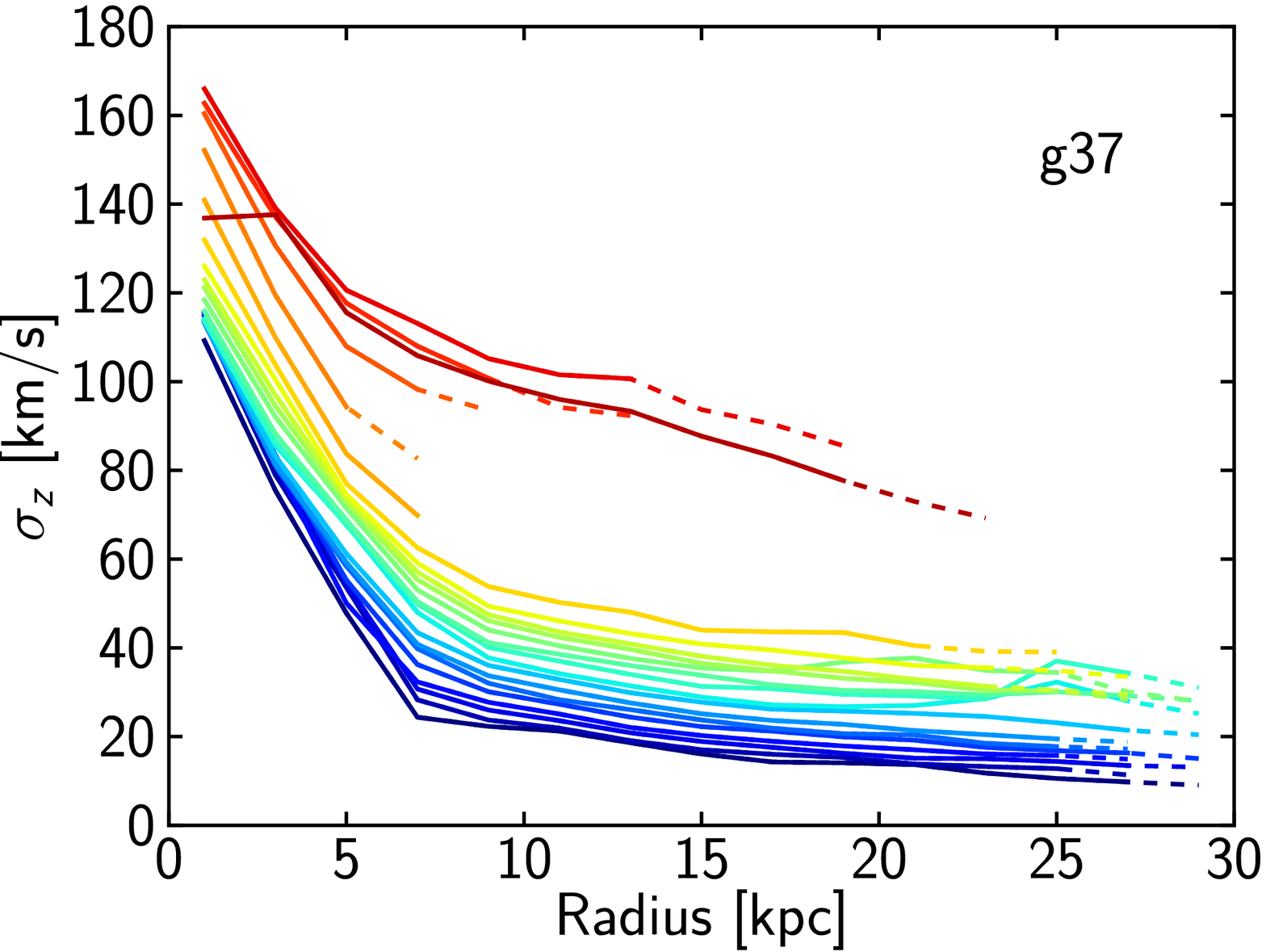}
\includegraphics[width=0.245\textwidth]{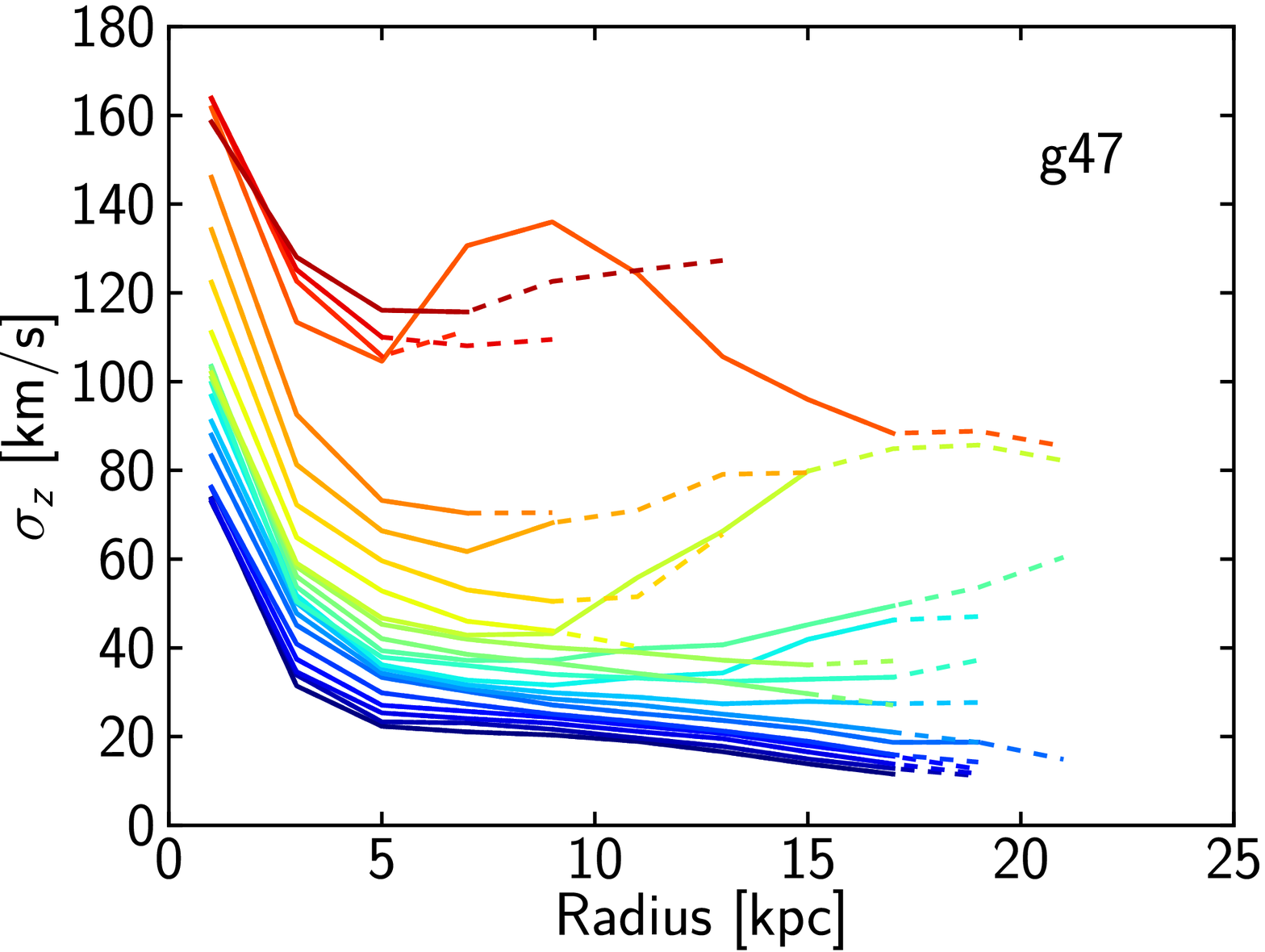}
\includegraphics[width=0.245\textwidth]{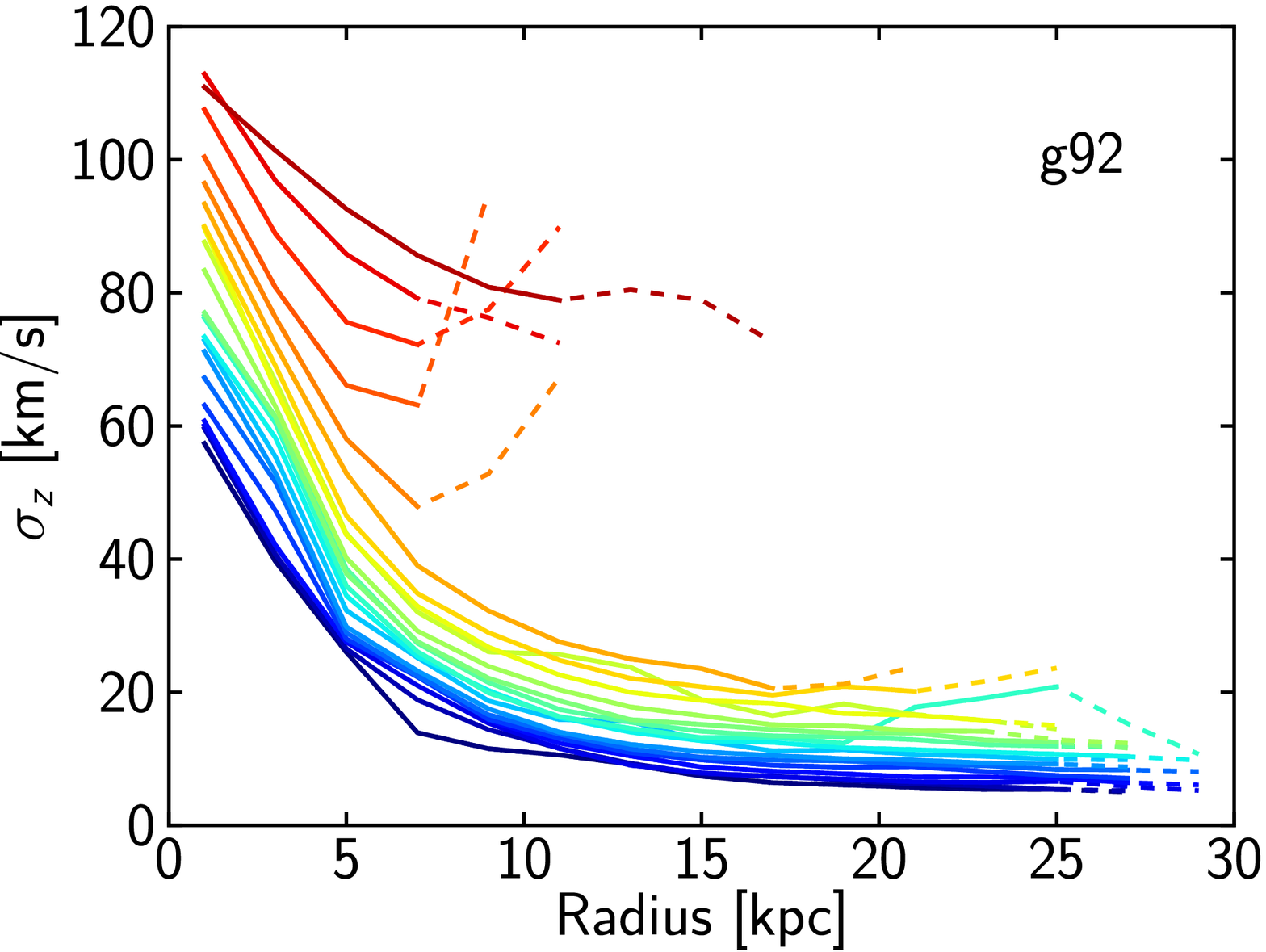}
\includegraphics[width=0.245\textwidth]{colorbar.eps}
\includegraphics[width=0.245\textwidth]{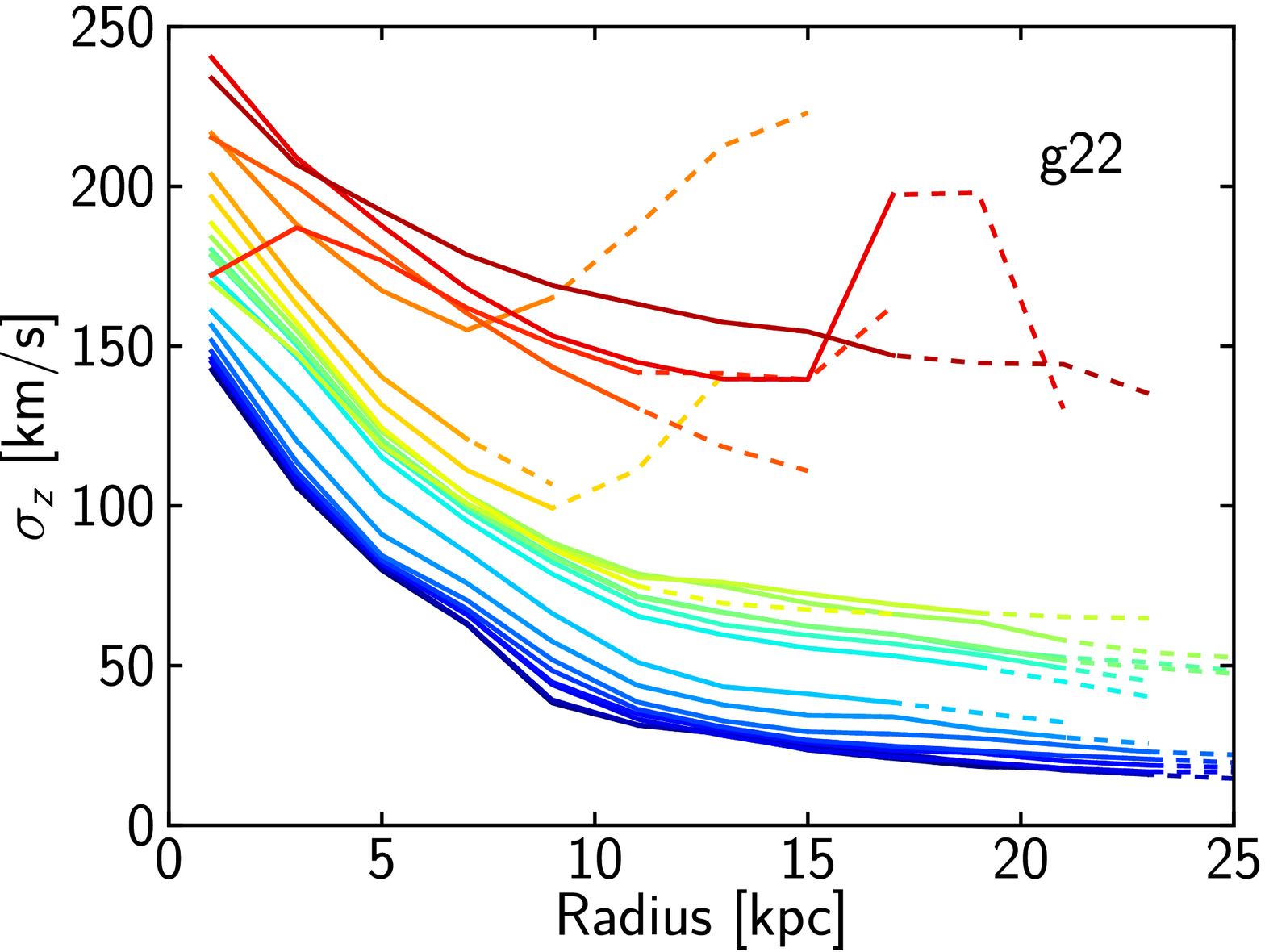}
\includegraphics[width=0.245\textwidth]{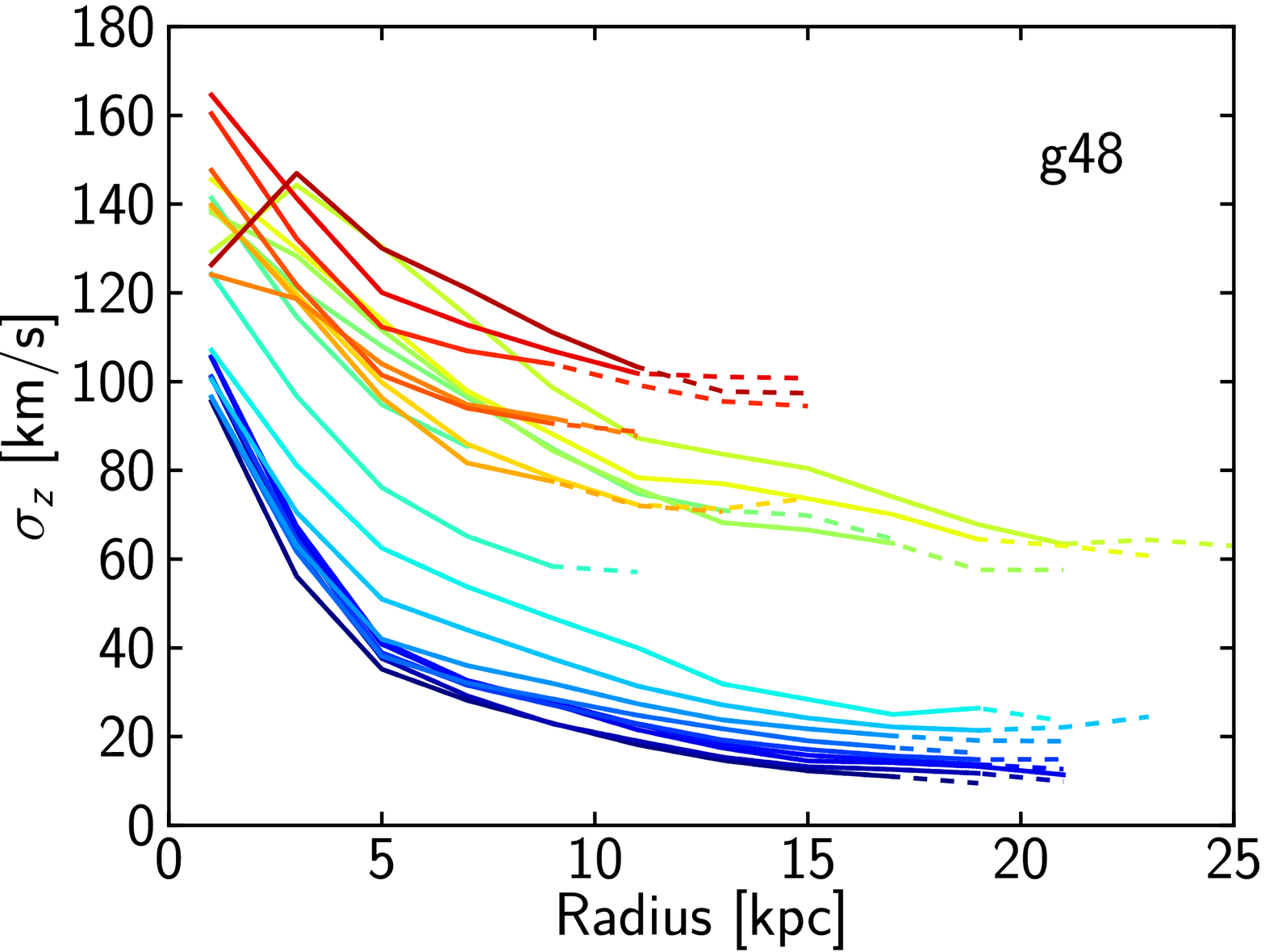}
\includegraphics[width=0.245\textwidth]{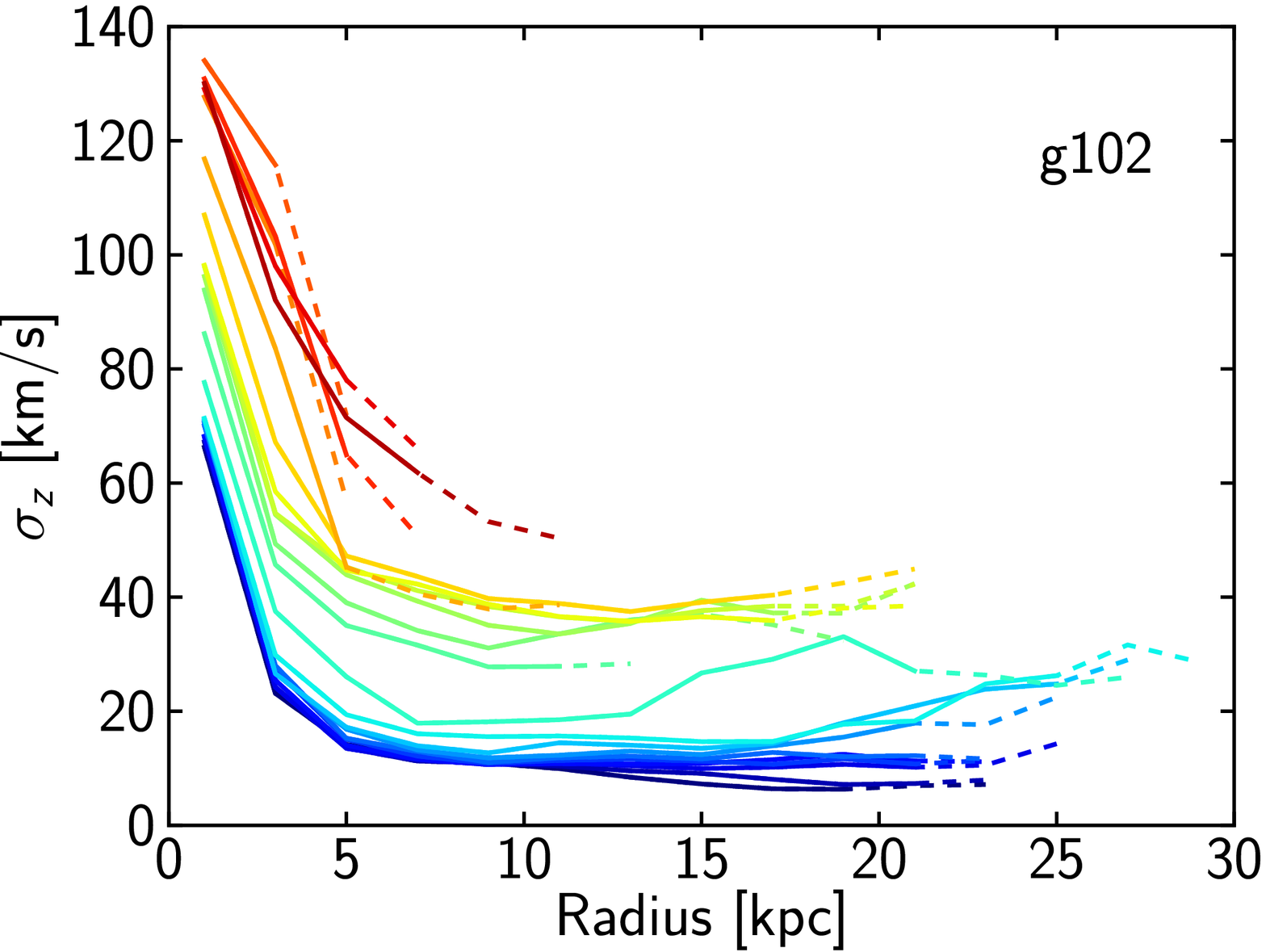}
\includegraphics[width=0.245\textwidth]{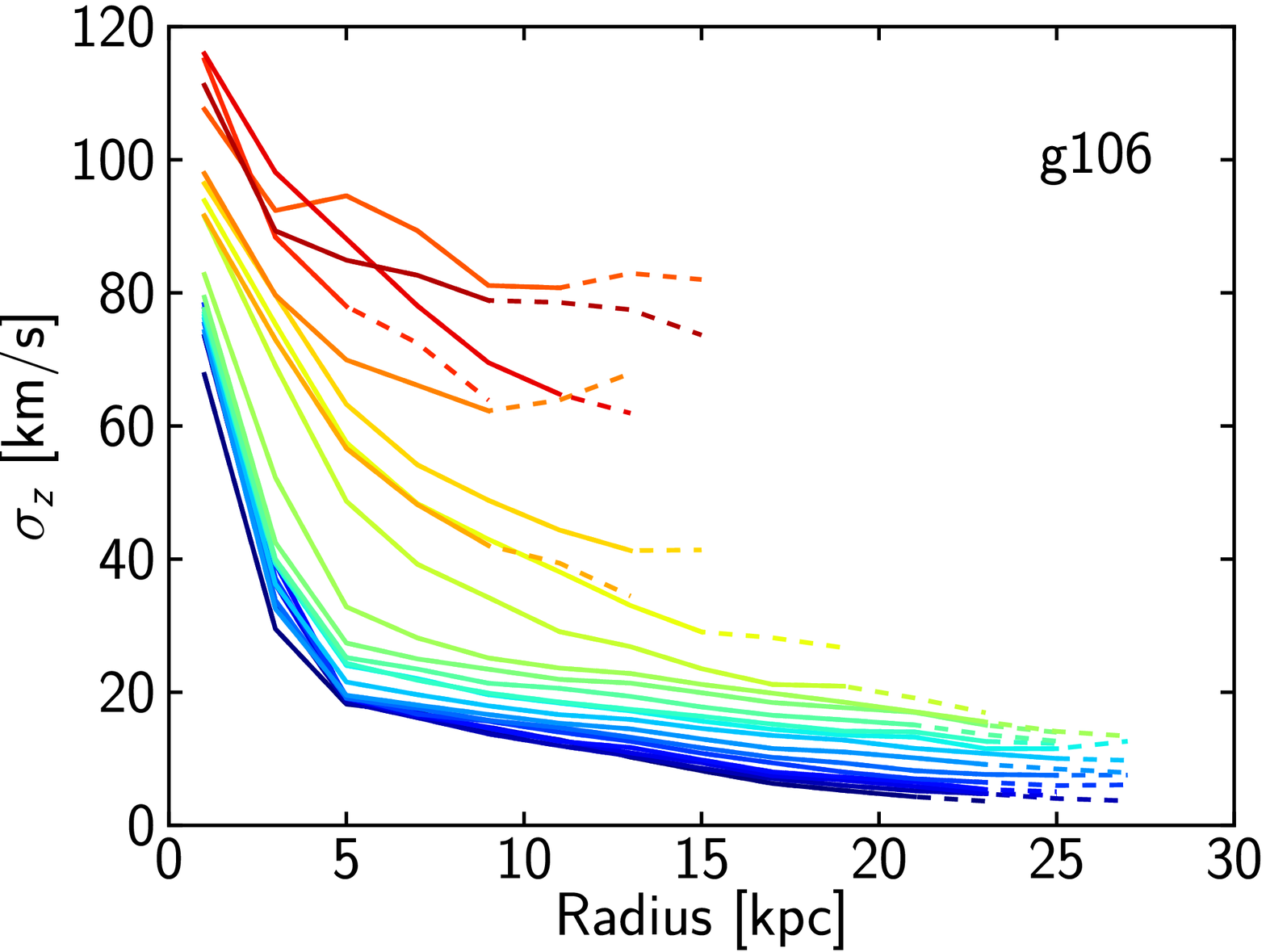}
\caption{Radial profiles of \sz for the mono-age populations. The solid lines show the radial range up to $R_{95}$ for each population, the dashed lines extend out to $R_{98}$. The colourcode and panel order are the same as in Figure \ref{fig:prof_z0}.}
\label{fig:prof_sigma}
\end{figure*}
We show in Figure \ref{fig:prof_z0} the resulting radial profiles of the scale-heights for each population. As already discussed, for all seven galaxies the oldest stars (more than $\sim$9 Gyr old, lines in red colours) are clearly distinct from the rest of the galaxy. They have larger scale-heights and are very flared (i.e their scale-heights increases significantly with radius, although part of that flaring might be artificial as we have just discussed). They are also centrally concentrated, and could be considered as part of the bulge, inner halo or an extreme thick disc. These stellar components reflect a violent early phase in the evolution, with a large fraction of the stars typically accreted from satellites.

For intermediate age and young stars, we find significant differences from one galaxy to another. The three quiescent galaxies show the simplest structures: populations of increasing age have an increasing scale-height, and this is true at all radii. In particular, g92 is unique in having a very flat scale-height profile: each population can truly be described by a single scale-height at least out to $R\sim 20$ kpc (as seen in the Milky Way by \citealt{Bovy2012b}, even though they explore a much more limited radial range). Some flaring is present in the outer disc ($R =$ 20--25 kpc), it is probably a consequence of both heating by small satellites orbiting the galaxy and outwards migration of stars with large vertical actions (see \citealt{Minchev2012b} for a discussion of the effect of radial migration on disc flaring).
Both g37 and g47 show some flaring, with detailed shapes that are quite different, probably due to the fly-by that g47 undergoes at  t$\sim$8.5--9 Gyr (this corresponds to populations shown in green in Figure \ref{fig:prof_z0})

On the other hand, the bottom row of Figure \ref{fig:prof_z0} shows that mergers introduce some complexity. We find in some cases a non-monotonic evolution of scale-height with age (see g48 for clear examples of that situation, with $\sim$5 Gyr old populations thicker than $\sim$8 Gyr old populations), and in other cases a clear bimodality between thick and thin MAPs (as also discussed in Section \ref{sec:dichotomy}). We also find a significant flaring in populations affected by mergers, as already discussed in a number of previous works \citep{Quinn1993,Kazantzidis2008,Read2008,Villalobos2008,Bournaud2009,Qu2011}. This flaring is due to the decreased self-gravity in outer discs, which is less able to counteract the gravitational heating by satellites. Note however that the flaring we observe here could be quite different from the overall thick disc flaring seen in a bimodal thin/thick disc decomposition, depending on the radial distribution of mass in each of the MAPs. 

An interesting case is g106, because order and regularity are preserved in spite of a 1:5 merger at t$\sim$7 Gyr. Populations corresponding to that time are shown in yellow in Figure \ref{fig:prof_z0}. Apart from in the inner disc, these population have a relatively constant scale-height with radius, and represent a good continuity between old and younger stars. The uniqueness of g106 could be due to the timing of that merger, happening earlier than in the other three active galaxies, so that the gas fraction is higher at the time of the merger (the gas fraction at the time of the merger is 0.35 for g106, to be compared to 0.24 and 0.18 for g22 and g102). As shown by \cite{Moster2010b}, a higher gas fraction leads to less disc thickening, because gas can absorb part of the kinetic energy. In addition, the incoming satellite is on a prograde, low-inclination orbit, and triggers a strong spiral structure in the main galaxy, accelerating disc growth. Both the higher gas fraction and the particular orbit for the satellite could contribute to the difference between g106 and the other active galaxies.

\subsection{Vertical velocity dispersions}\label{sec:sigma_profiles}

In the same radial and vertical range used to compute the vertical density profiles, we estimate \sz from the usual second moment of the vertical velocity distribution.
We show in Figure \ref{fig:prof_sigma} the radial profiles of \sz for the MAPs in the seven simulated galaxies.

Similarly to the scale-height analysis in the previous section, we find that the two most quiescent galaxies (g37 and g92) show the most ordered and regular \sz profiles: populations of increasing age have an increasing \sz, at all radii except in the very outer disc. The fly-by experienced by g47 leaves signatures in its \sz profile, with an increased \sz in the outer disc for populations of intermediate age. The inner part of the disc of g47 is unperturbed.

Mergers seem mostly to  produce gaps at all radii in the sequence of increasing \sz with age (i.e., gaps in the age-velocity relations, as discussed in Paper II, to which we refer the reader for a more complete study), as seen for galaxies g22, g48 and g102. Even g106, which has much more regular \sz profiles than the other galaxies with mergers, still shows such a gap, mostly in its inner disc.

\begin{figure}
\centering 
\includegraphics[width=0.45\textwidth]{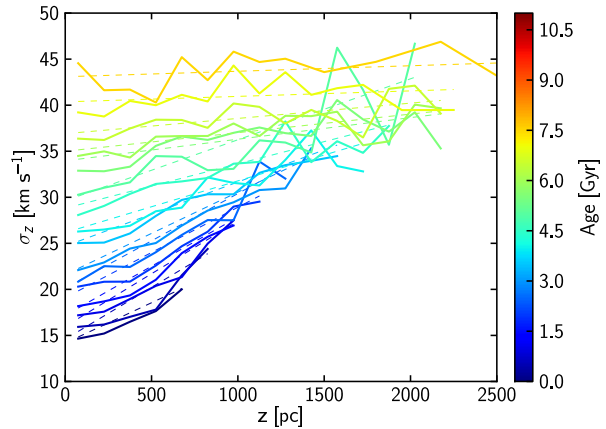}
\caption{Vertical profiles of \sz at a radius of $2 R_d$ for the mono-age populations younger than 8 Gyr in galaxy g37 (solid lines). The dashed lines show linear fits to the profiles, the slopes of  these fits are shown for all galaxies in Figure \ref{fig:slope}.}
\label{fig:vert_prof_sigma}
\end{figure}

\begin{figure*}
\centering 
\includegraphics[width=0.245\textwidth]{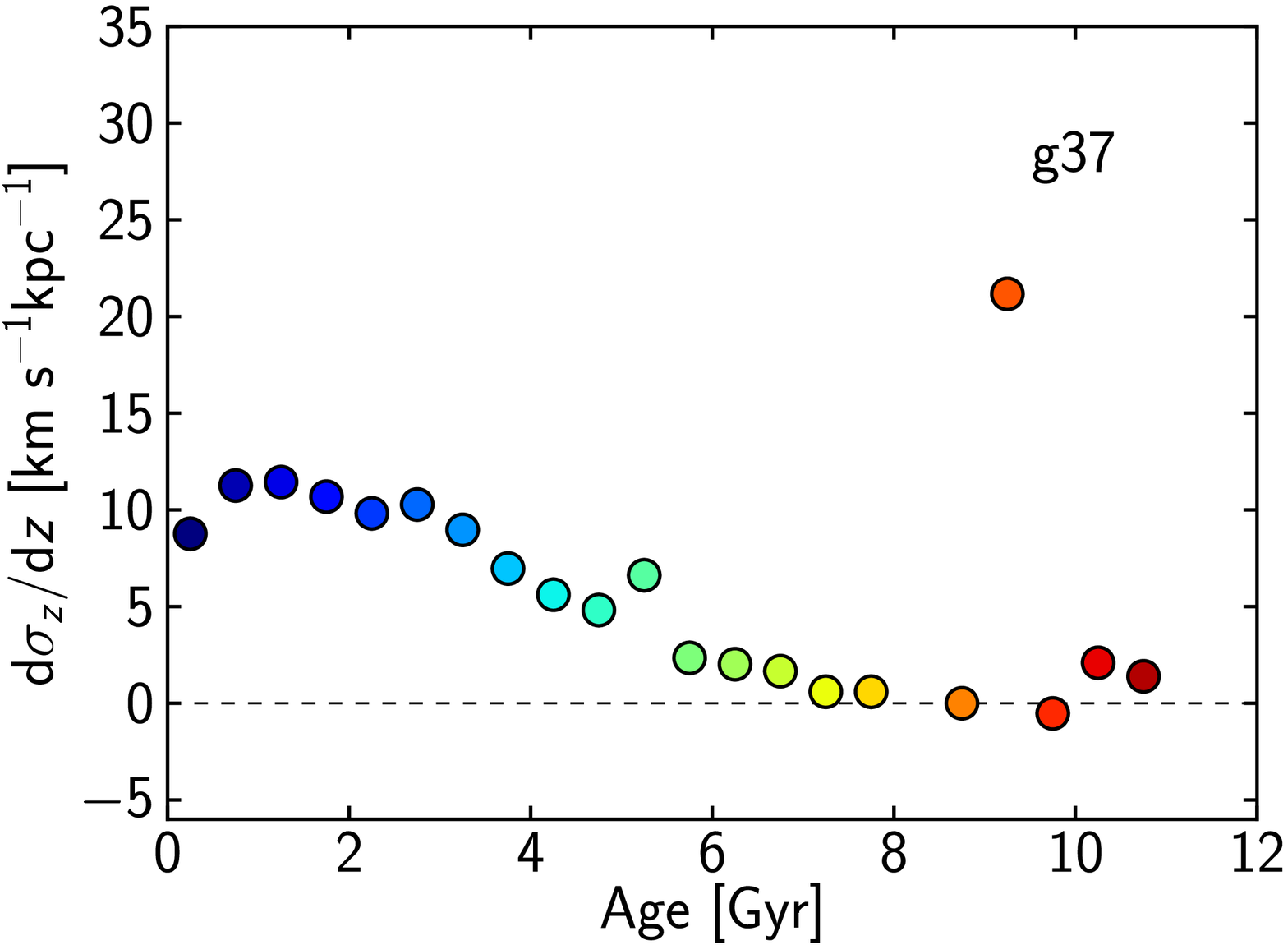}
\includegraphics[width=0.245\textwidth]{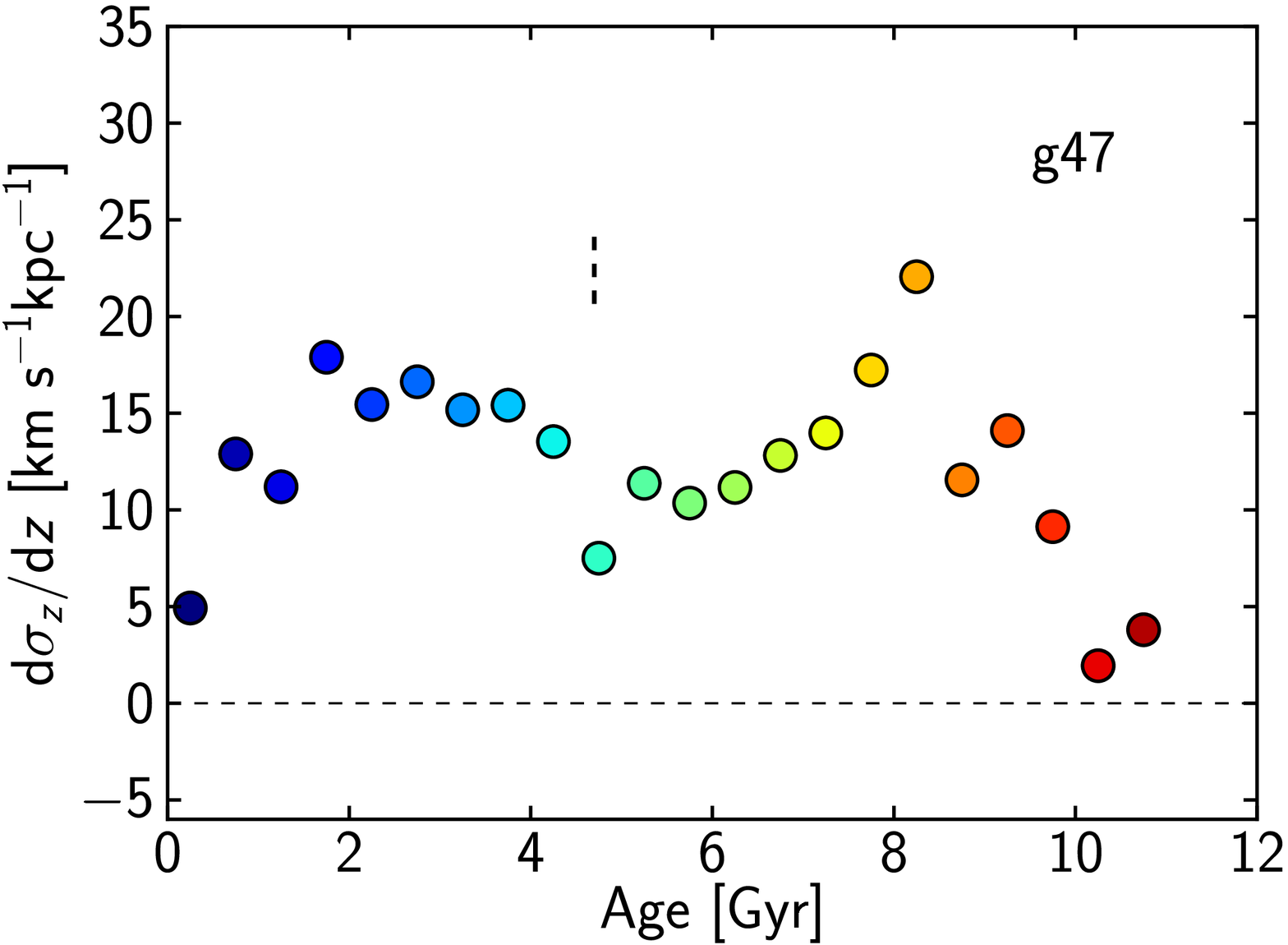}
\includegraphics[width=0.245\textwidth]{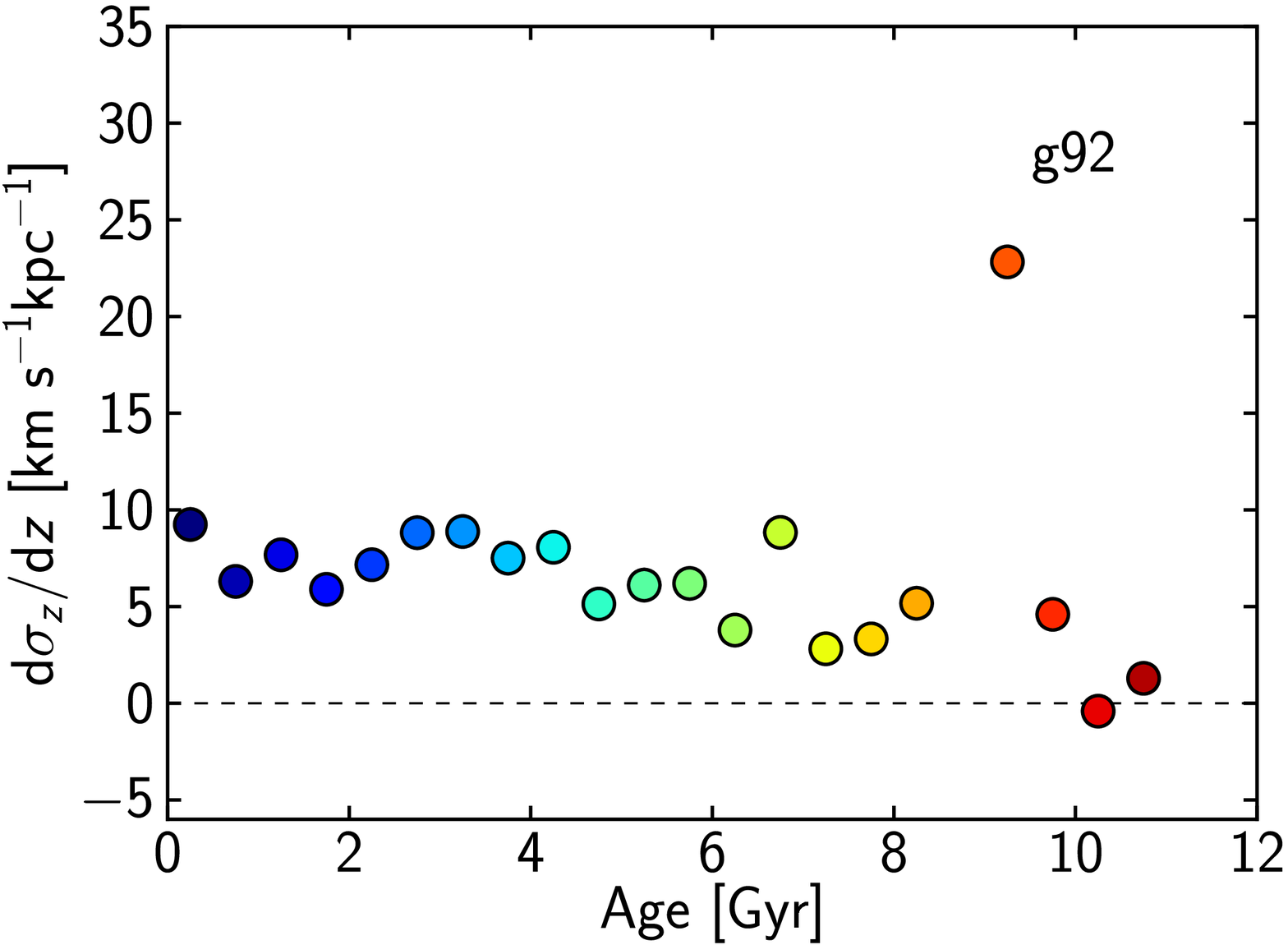}
\includegraphics[width=0.245\textwidth]{colorbar.eps}
\includegraphics[width=0.245\textwidth]{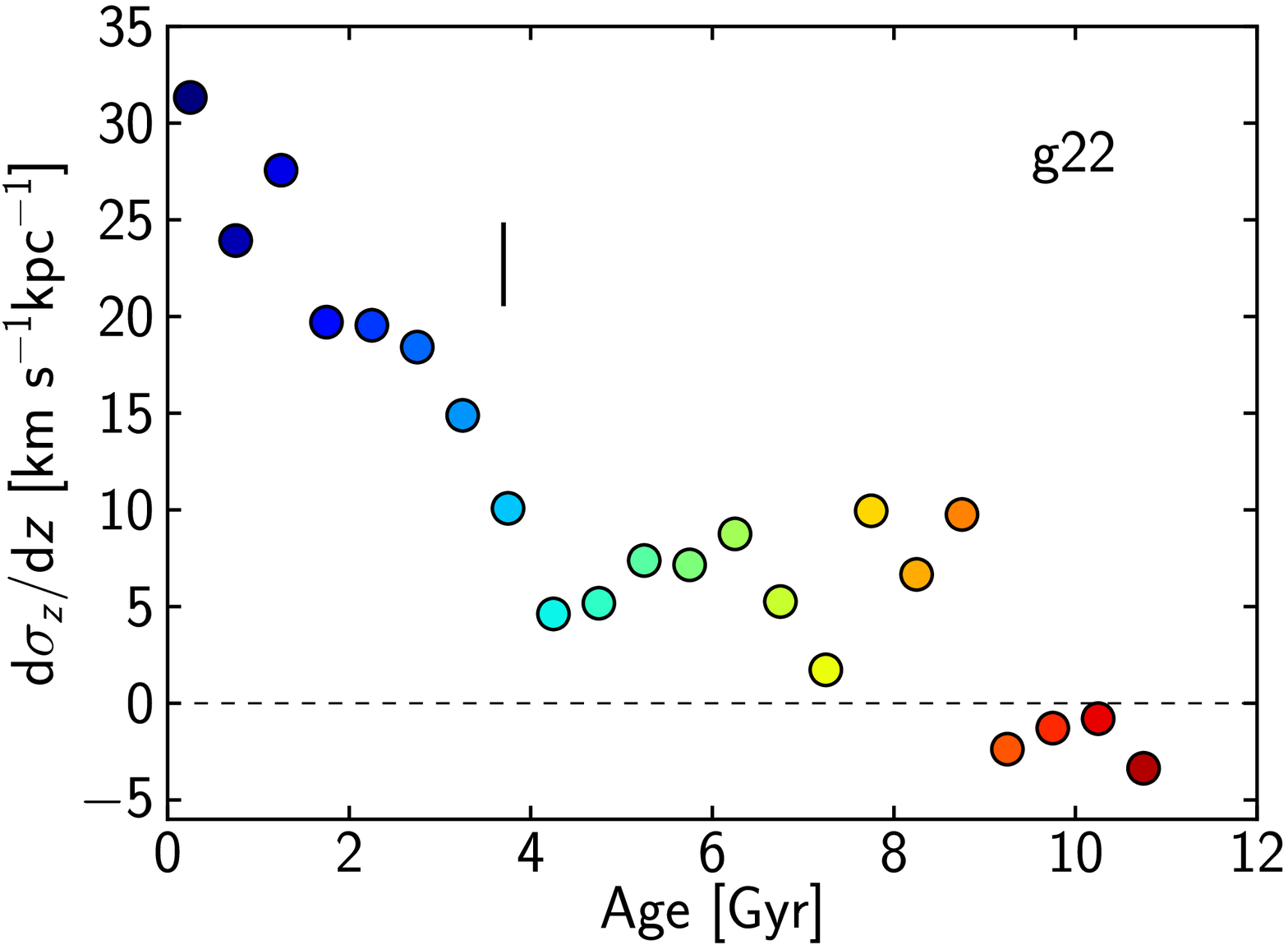}
\includegraphics[width=0.245\textwidth]{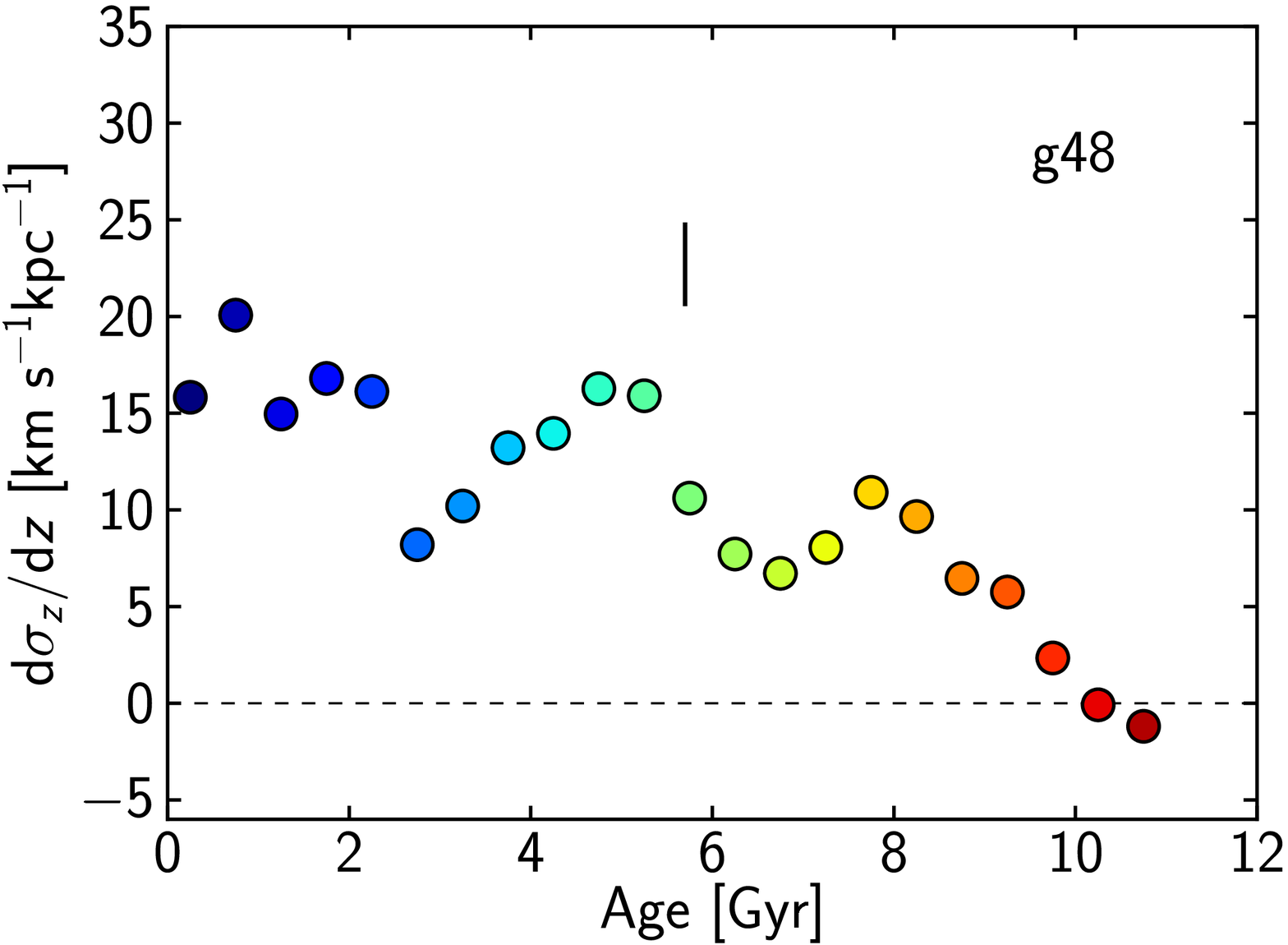}
\includegraphics[width=0.245\textwidth]{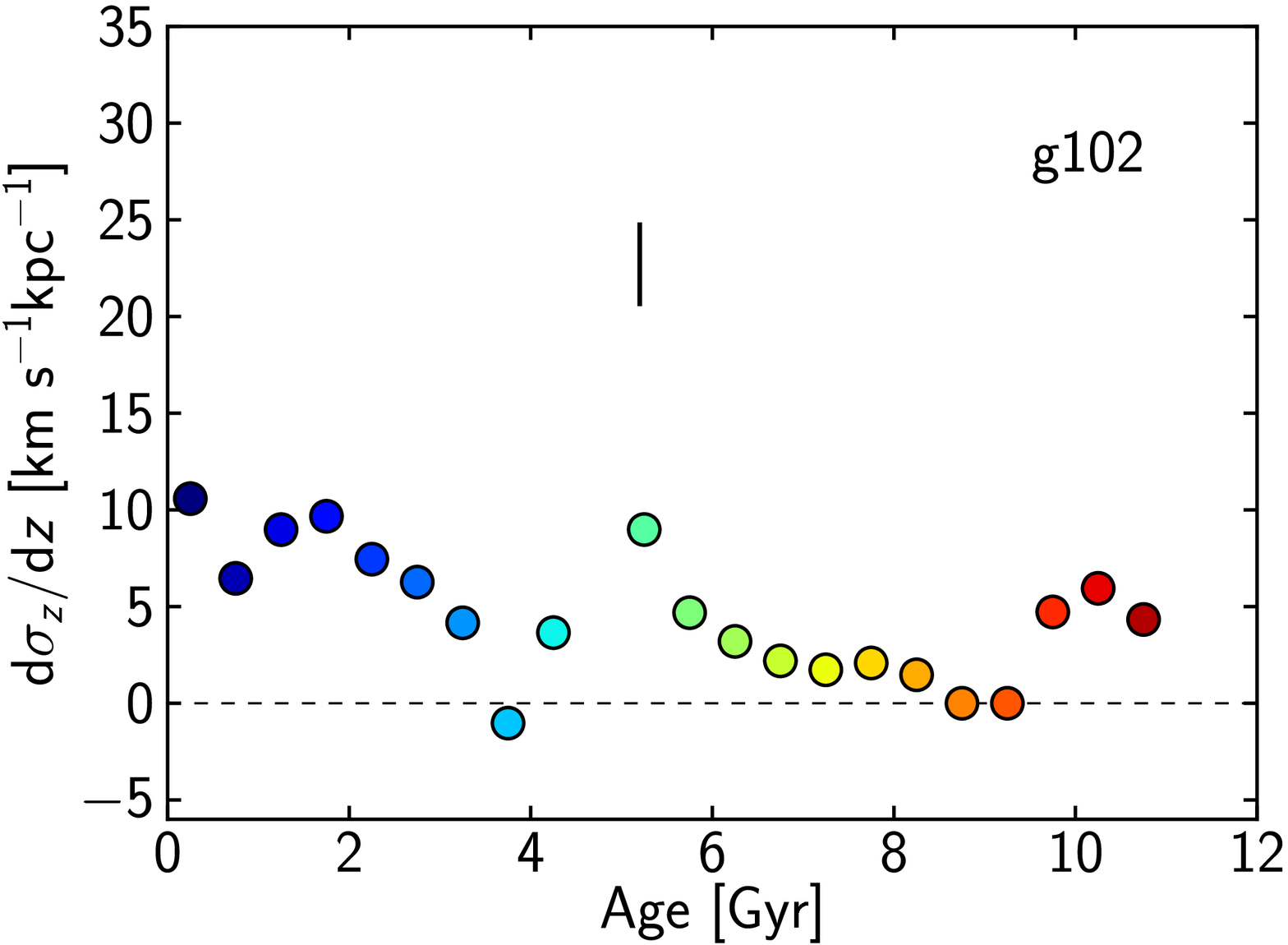}
\includegraphics[width=0.245\textwidth]{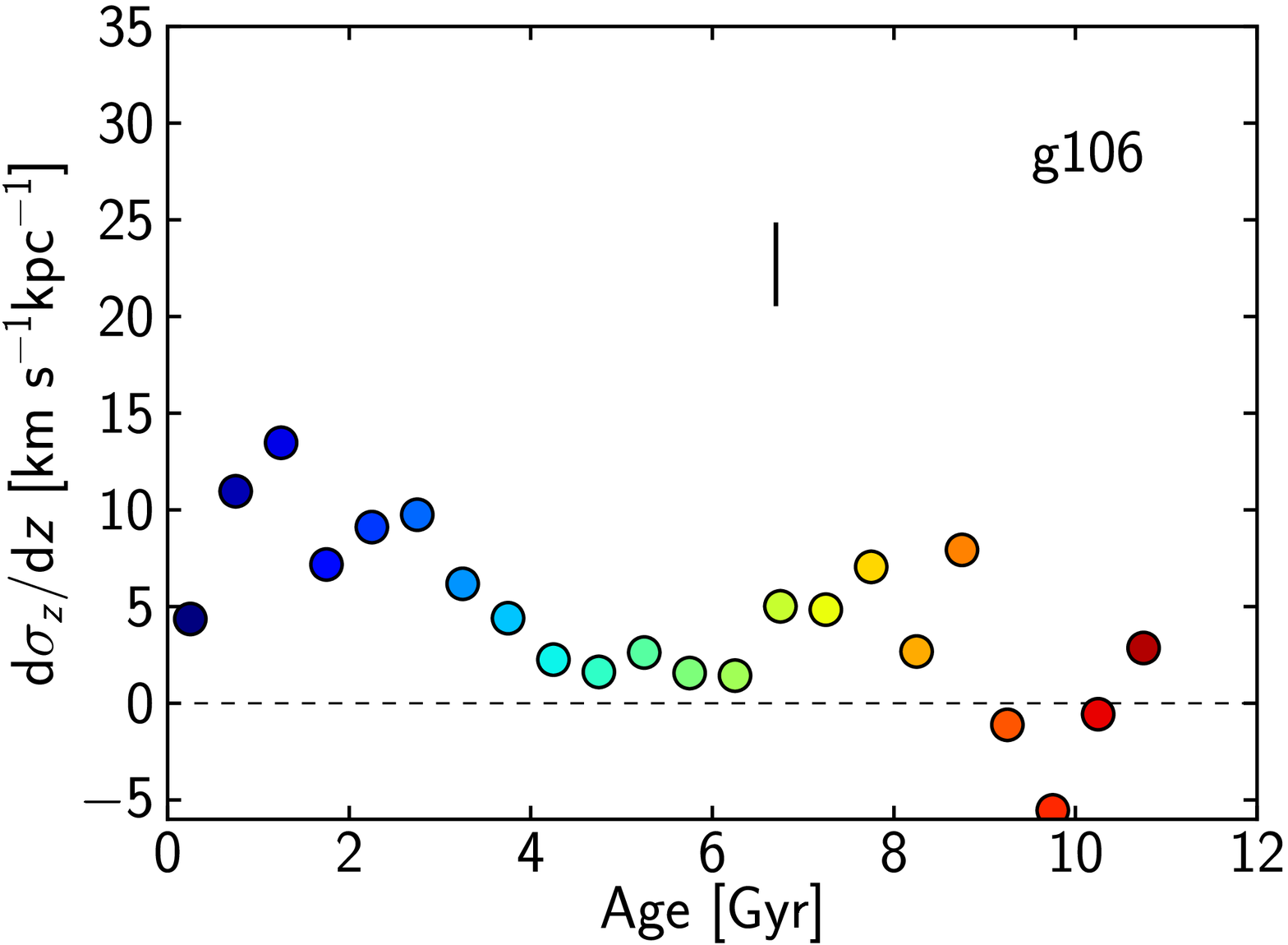}
\caption{Slope of the vertical profiles of \sz at a radius of $2 R_d$ for the mono-age populations. The colourcode and panel order are the same as in Figure \ref{fig:prof_z0}. The small vertical lines mark the time of coalescence for the last merger undergone by each galaxy for $z<1.5$ (the dashed line for g47 marks the end of a fly-by)}
\label{fig:slope}
\end{figure*}

We now examine how \sz varies with height above the disc plane.
We compute vertical profiles of \sz at a radius of $2 R_d$ (more exactly in a 2 kpc-wide annulus around $2 R_d$). Our vertical bins have a minimum height of 150 pc, and if they contain less than 200 particles they are extended (in increments of 150 pc) until they contain more than 200 particles. The vertical profiles are shown in Figure \ref{fig:vert_prof_sigma} for the case of g37. We find that \sz increases with height above the disc, and that the effect is more pronounced for younger populations. We quantify the vertical increase by fitting a line to the vertical \sz profile of each population (dashed lines in Figure \ref{fig:vert_prof_sigma}). We plot the slope of these linear fits, for g37 as well as the other galaxies in Figure \ref{fig:slope}.
 
Nearly all populations (except for the oldest ones) have a positive slope, i.e. \sz increases with height above the disc. Most galaxies also show a clear trend of increasing slope for younger stars. This is probably because young stars are less relaxed, as is the case for stars younger than about 3 Gyr in the solar neighbourhood \citep{Seabroke2007}. The values of the slopes are quite high, mostly in the range 0--15 km/s/kpc, and there are no obvious differences in the magnitudes of the slopes between galaxies with a quiescent or active merger history. The average slope for each simulated galaxy (weighted by the mass in each MAP, and restricted to populations younger than 9 Gyr) is shown in Table \ref{tab-slopes}, the average values are between 5 and 15 km/s/kpc, with again no clear dependence on merger activity.
 
These values are significantly higher than the slopes of 1--3 km/s/kpc found by \cite{Bovy2012c} for the Milky Way. Furthermore, \cite{Bovy2012c} found no dependence of the slope on [$\alpha$/Fe] and [Fe/H], which appears inconsistent with the dependence of the slope on age that we find. To test further the match of our simulations to Milky Way data, we first try to measure the vertical slopes at a radius of $3R_d$ to test for possible  variations in the behaviour with radius. At $3R_d$  we find a less pronounced correlation of the slope with age, as well as smaller values for the slope (Table \ref{tab-slopes}), except for g47 where the slopes increase in the outer disc (this might be a consequence of the fly-by undergone by g47). At $3R_d$, 4 of our galaxies have an average slope between 1.5 and 5 km/s/kpc, somewhat closer to isothermality (i.e., no slope) than at $2R_d$.

A final test is to change the way we measure \sz to be closer to the way it is measured in \cite{Bovy2012c}, by fitting a Gaussian profile to the distribution of vertical velocities. This gives slightly different values of \sz, because the real distributions of vertical velocities are not perfectly Gaussian. We compute the average slopes as previously  (see Table \ref{tab-slopes}). Both at 2 and $3 R_d$, these slopes are lower than with our direct method of measuring \sz. In particular at $3 R_d$, four of the simulated galaxies have an average vertical slope in the range 1.1--4 km/s/kpc, which is marginally consistent with the observed values for the Milky Way \footnote{A possible source of discrepancies between the simulated galaxies and the \cite{Bovy2012c} results is that the SEGUE G-dwarf sample they use only contain stars above 200-300 pc from the mid-plane, probably excluding young stars, which are usually further from isothermality.}. There is still however no clear correlation with the merger histories.
No slope corresponds to the ``isothermal'' solution to the Jeans equation in the case where the mass is distributed in a thin sheet with an exponential vertical profile \citep{Binney1987}. A positive slope indicates that the mass is more vertically extended that in the isothermal case. This suggests that in our simulations the galaxies with the smallest vertical variations of \sz are those with the lowest contribution of halo/bulge/thick disc stars compared to thin disc stars. A visual inspection of Figure \ref{fig:images} indeed suggests that g92 and g102 (the most isothermal galaxies) also have the most concentrated bulges, independently of the total bulge mass (which is quite high for g102).

\begin{figure}
\centering 
\includegraphics[width=0.45\textwidth]{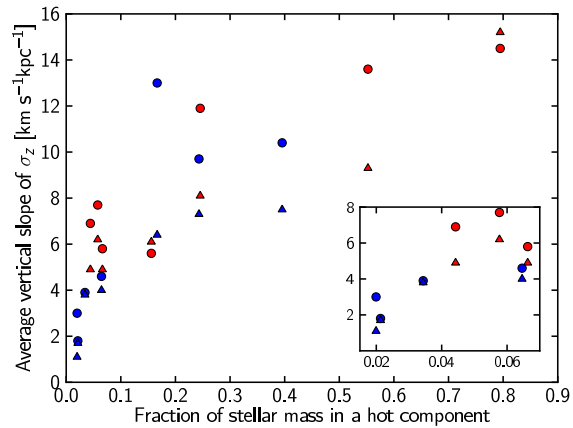}
\caption{Mass-weighted average slope of the vertical profiles of \sz as a function of the fraction of the stellar mass in a hot component (i.e. with \sz$>40$\kms). The data points are shown for all simulated galaxies, at 2 $R_d$ (in red) and 3 $R_d$ (in blue), and with two different ways of measuring \sz as in Table \ref{tab-slopes} (dots: direct measures, triangles: from Gaussian fits). The inset shows a zoom on the bottom left part of the plot. We find that the galaxies closest to isothermality (i.e. an average vertical slope closest to 0) are the ones with the lowest contribution of stars in a  hot component,  although this is not directly related with their merger history (g102 has a very low slope but a 1:15 merger at $t=8.5$ Gyr and a high bulge fraction).}
\label{fig:slope_vs_hot}
\end{figure}

We check this by showing in Figure \ref{fig:slope_vs_hot} the mass-weighted average slope (values shown in Table \ref{tab-slopes}, we use both the direct and fitted values as an indication of the possible range of slopes for each galaxy) as a function of the fraction of stellar mass found in a hot component at 2 and 3 $R_d$ (for stars at $|z|<3$kpc). The fraction of hot stars is derived by measuring \sz for stars in 250 Myr age bins, and computing the mass contribution of populations with \sz$>40$ \kms (this is an arbitrary choice, but different thresholds give consistent results). We confirm that the galaxies closest to isothermality are those with the lowest local fraction of hot stars. This local fraction of hot stars is not only related to the mass of the spheroidal component, but also to its shape. As a consequence, it does not directly reflect the merger history of the galaxy.

\begin{table}
 \caption{Mass-weighted average slopes (in km/s/kpc) of the vertical profiles of \sz measured in 2 different ways (\sz computed directly from the vertical velocities, or from a Gaussian fit to the distribution of vertical velocities), at 2 different radii (2 and 3 $R_d$), restricted to populations younger than 9 Gyr}
 \label{tab-slopes}
 \begin{tabular}{@{}lcccc}
  \hline
  Name &Direct, 2$R_d$&Direct, 3$R_d$&Fit, 2$R_d$&Fit, 3$R_d$\\
  \hline
 g37 &7.7 &4.6 &6.2 & 4.0\\
 g47 &11.9 & 13.0&8.1 & 6.4 \\
 g92 & 6.9 &3.0 & 4.9 & 1.1 \\
 \hline
g22 & 14.5 & 10.4 &15.2 &7.5 \\
g48 & 13.6 &9.7 & 9.3 & 7.3\\
g102 & 5.8 &1.8 & 4.9 &  1.7 \\
g106 & 5.6  & 3.9&6.1 & 3.8\\
  \hline
 \end{tabular}
\end{table}

\subsection{When are MAPs simple?}
From this section, we conclude that:
\begin{itemize}
\item radial profiles are nearly always simple  exponentials for all MAPs, independently of the merger history (although younger stars have more complex profiles)
\item vertical density profiles are also nearly always successfully fitted by simple exponentials. In all cases but one (g92), scale-heights increase with radius. The absence of flaring seems connected with a very quiescent history
\item \sz increases with height above the disc plane for all our galaxies, although the relation flattens strongly (or sometimes even reverses) for old stars. Galaxies for which \sz rises with $z$ slowly have the lowest contribution of bulge/halo stars at the radius where \sz is measured, independently of their total bulge fraction or merger history.
\end{itemize}
\section{Anti-correlation between scale-height and scale-length}

\cite{Bovy2012b} find an anti-correlation of the scale-length and scale-height of mono-abundance populations in the Milky Way. This corresponds to a situation where younger stars are in progressively thinner and more extended components. This configuration was also found in the simulation presented in \cite{Bird2013}.  We examine if this is the case in our seven simulations, and what are the conditions that create such an anti-correlation.

We have already discussed in the previous section the evolution of scale-height and scale-length with age. We found a smooth increase of $z_0$ with age for quiescent galaxies, and more complex dependence for galaxies with mergers (at the exception of g106). The time evolution of scale-lengths was found to be more complex, with quiescent galaxies showing an increase of $R_0$ for younger stars, with the exception of g37 where $R_0$ is mostly constant for all stars younger than 6 Gyr. Mergers influence the radial distribution of MAPs, and they redistribute old stars and gas.

\begin{figure*}
\centering 
\includegraphics[width=0.245\textwidth]{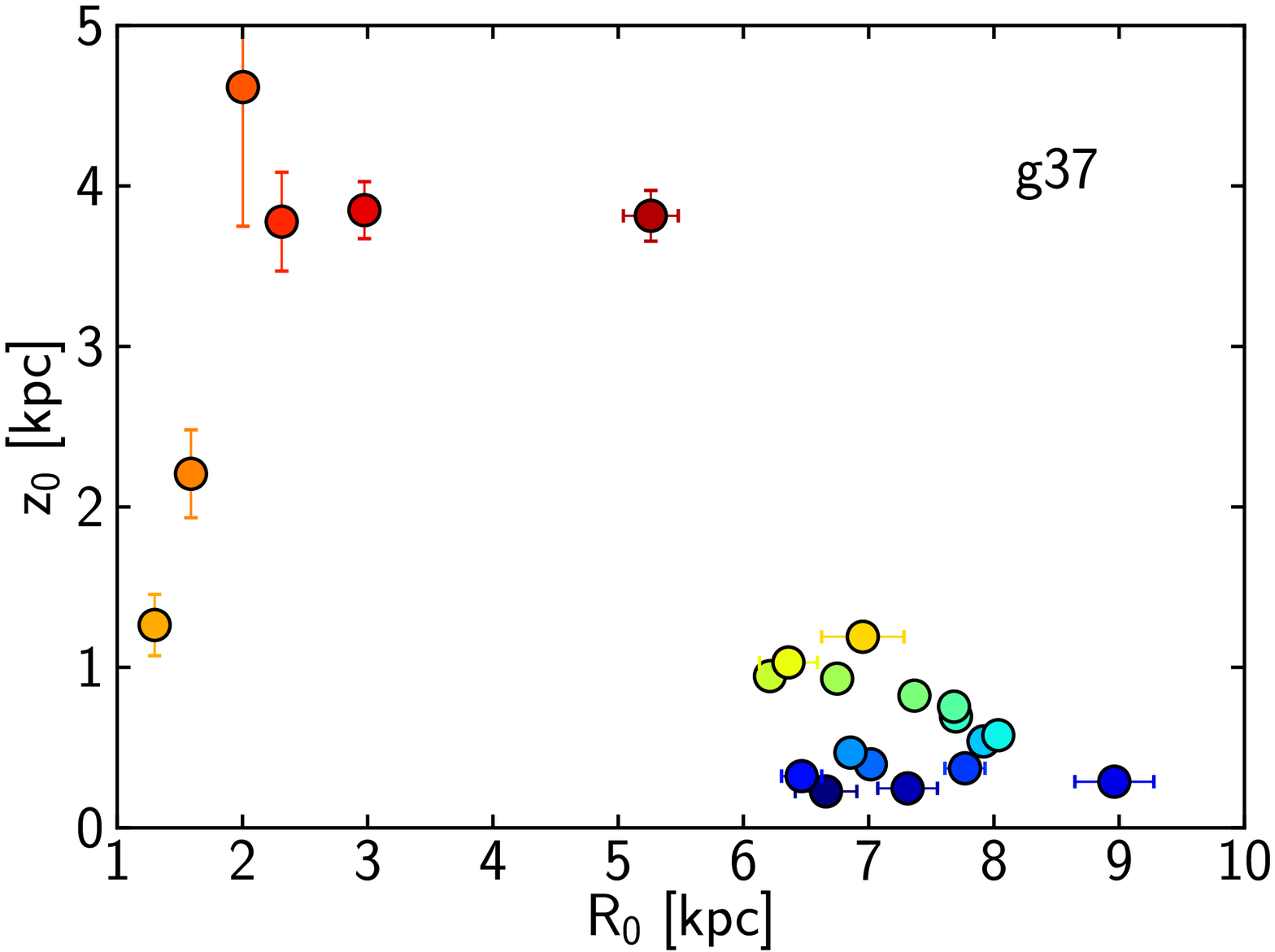}
\includegraphics[width=0.245\textwidth]{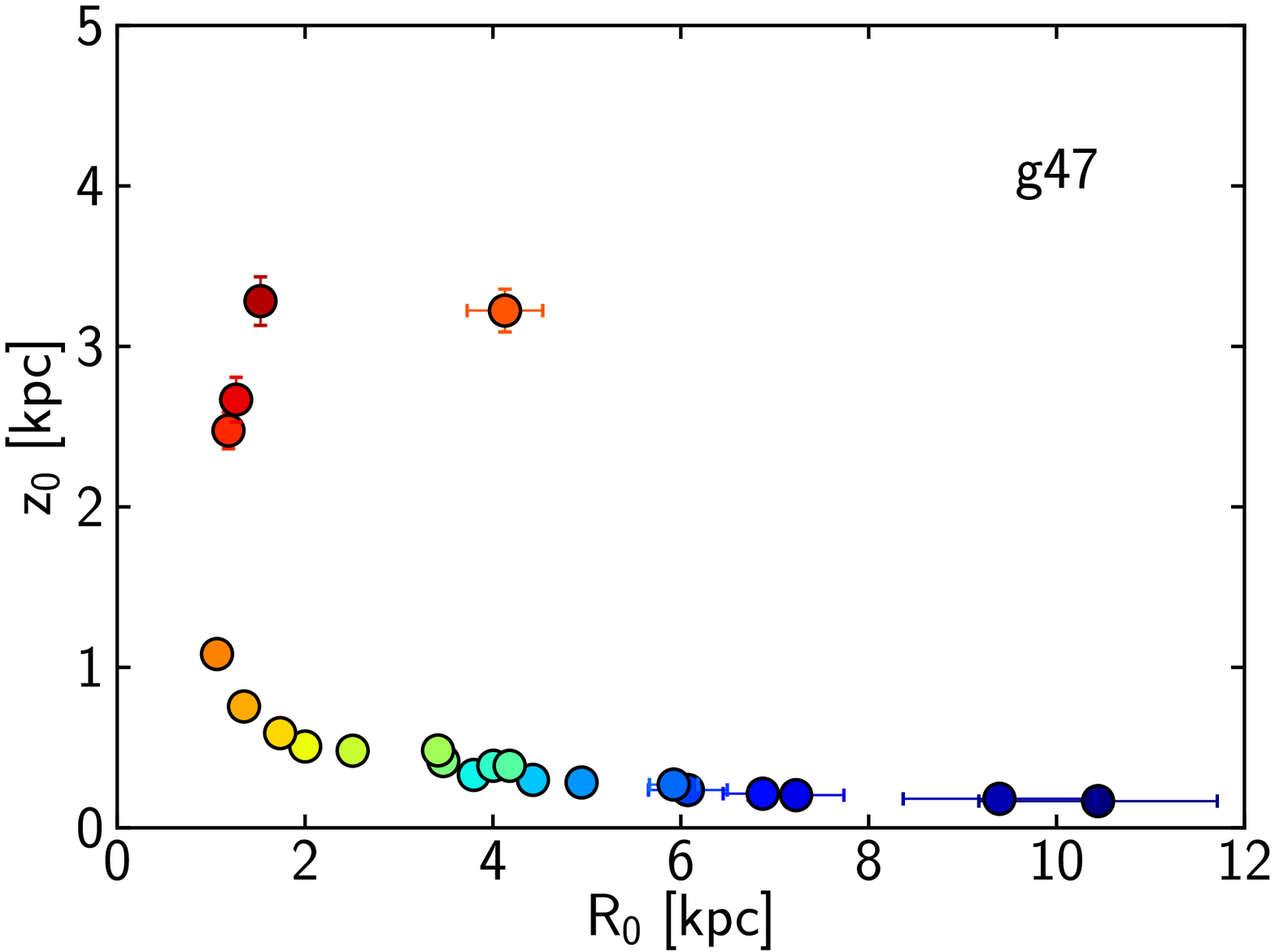}
\includegraphics[width=0.245\textwidth]{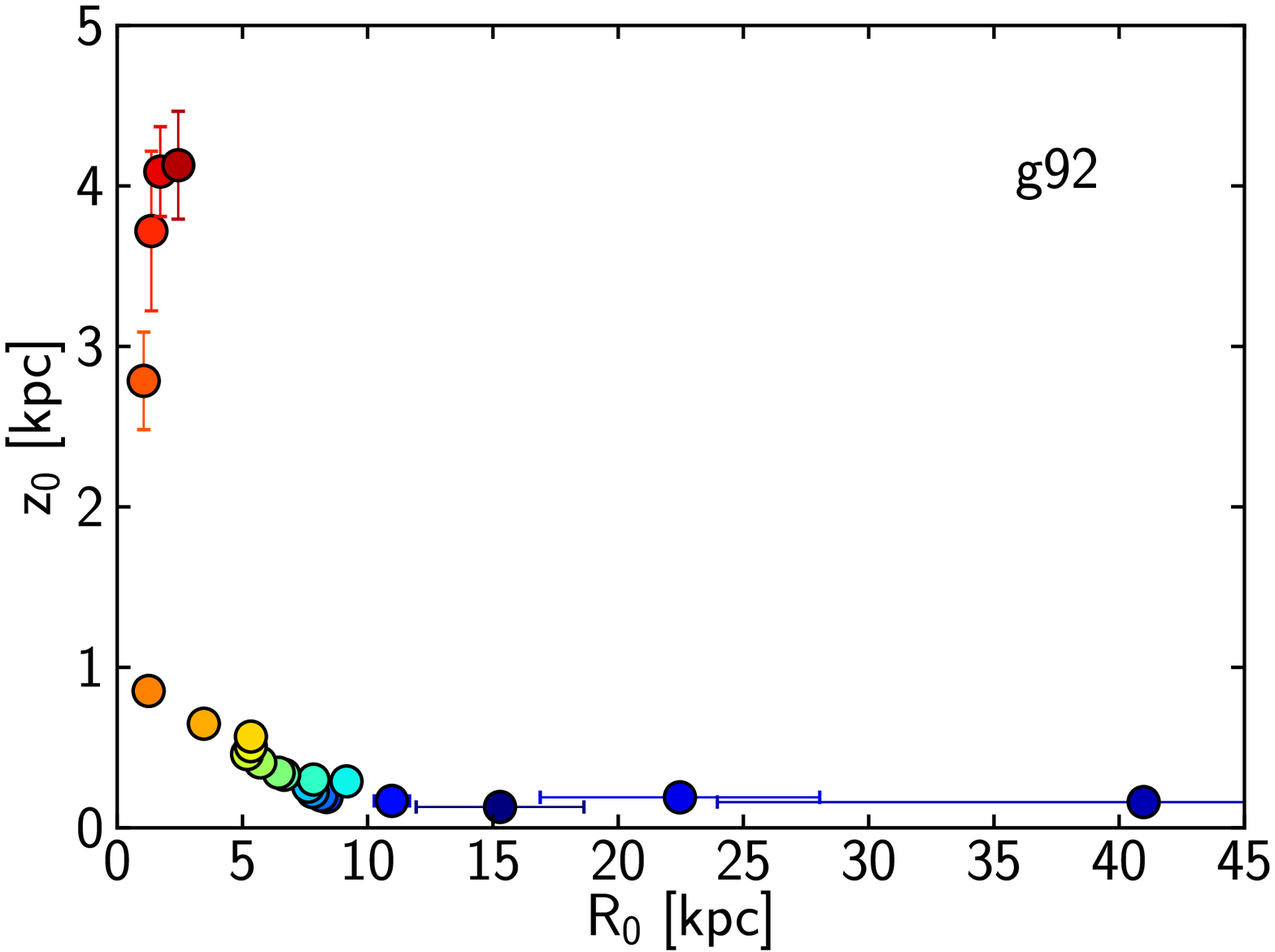}
\includegraphics[width=0.245\textwidth]{colorbar.eps}
\includegraphics[width=0.245\textwidth]{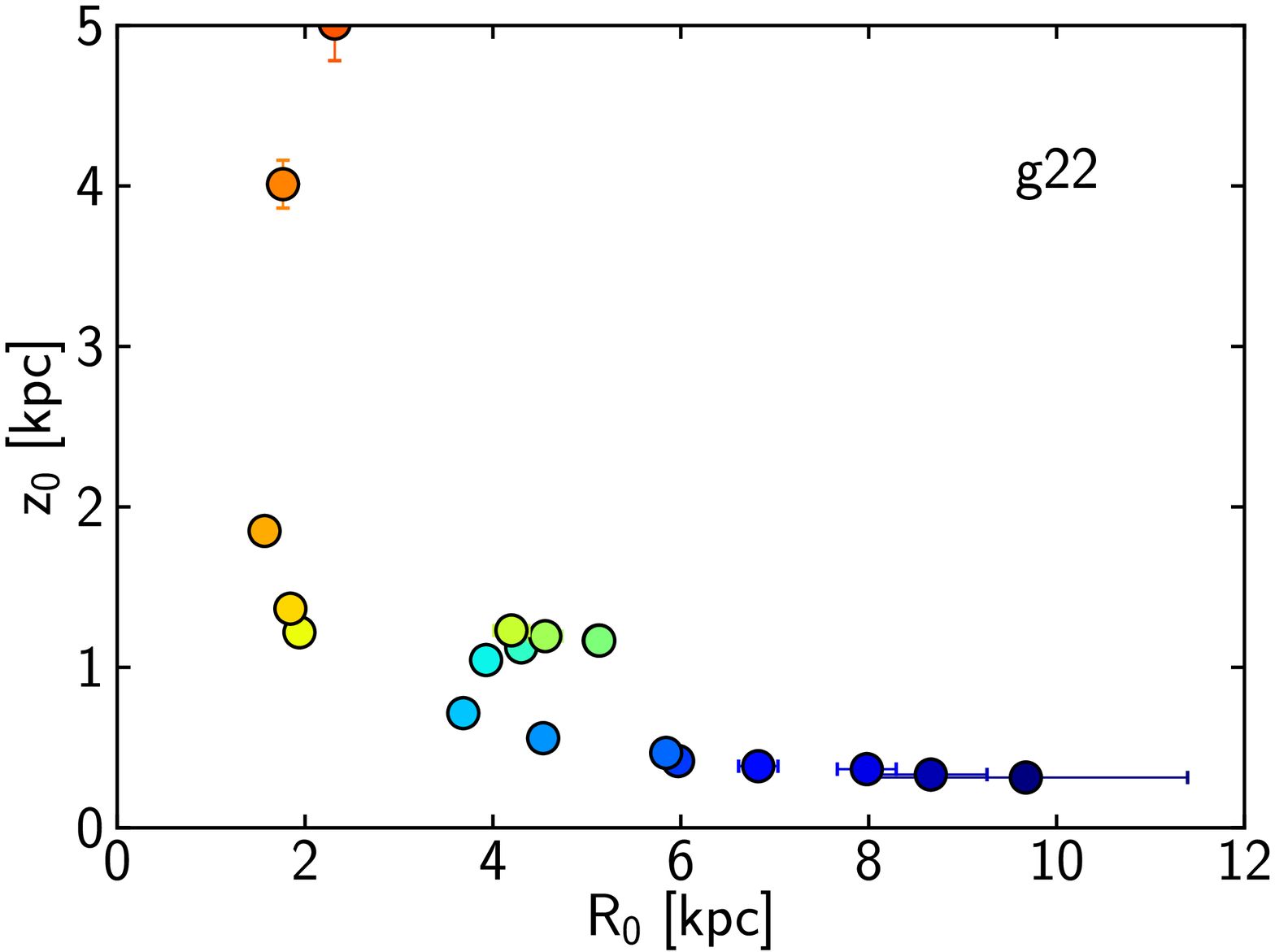}
\includegraphics[width=0.245\textwidth]{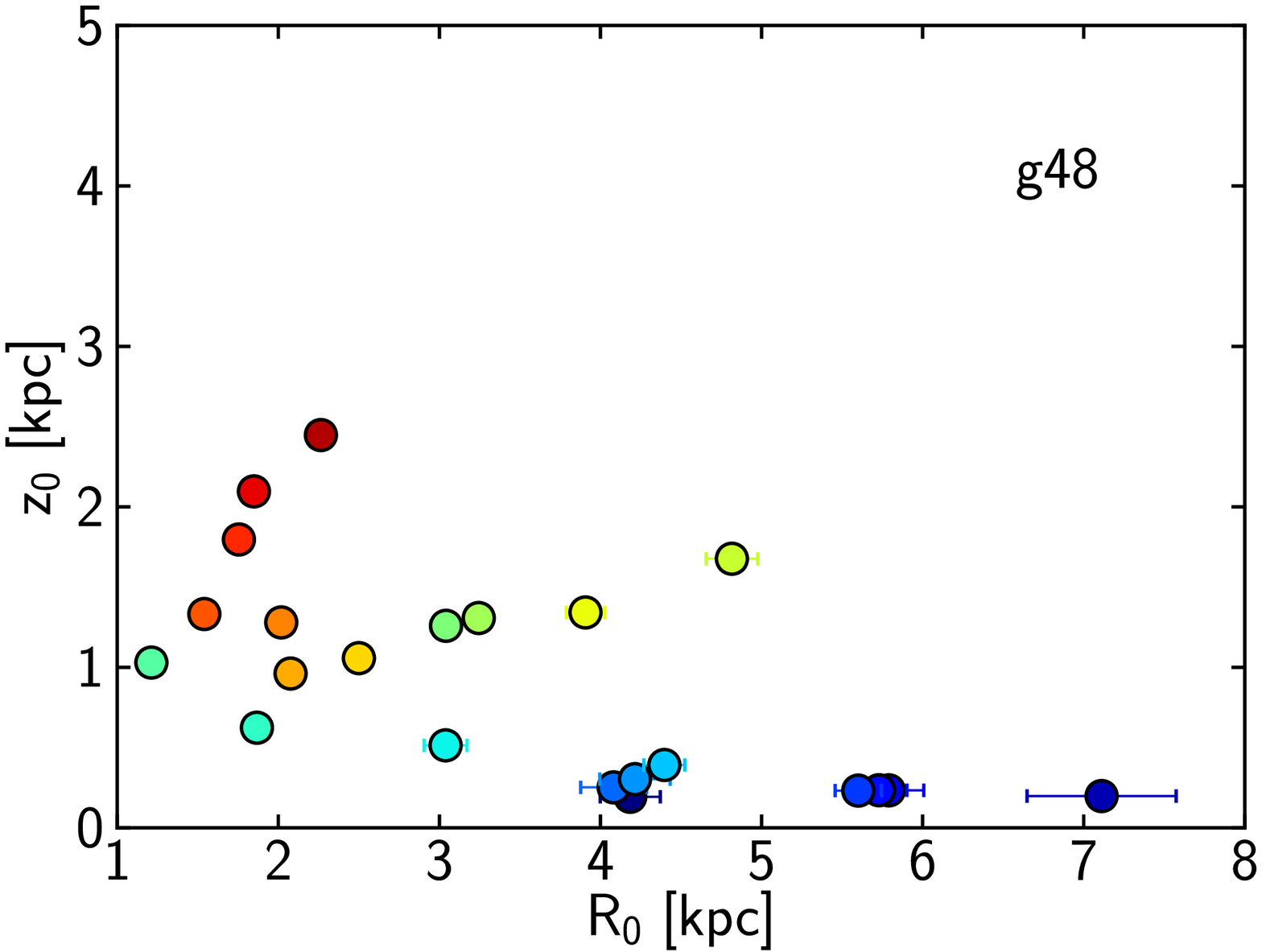}
\includegraphics[width=0.245\textwidth]{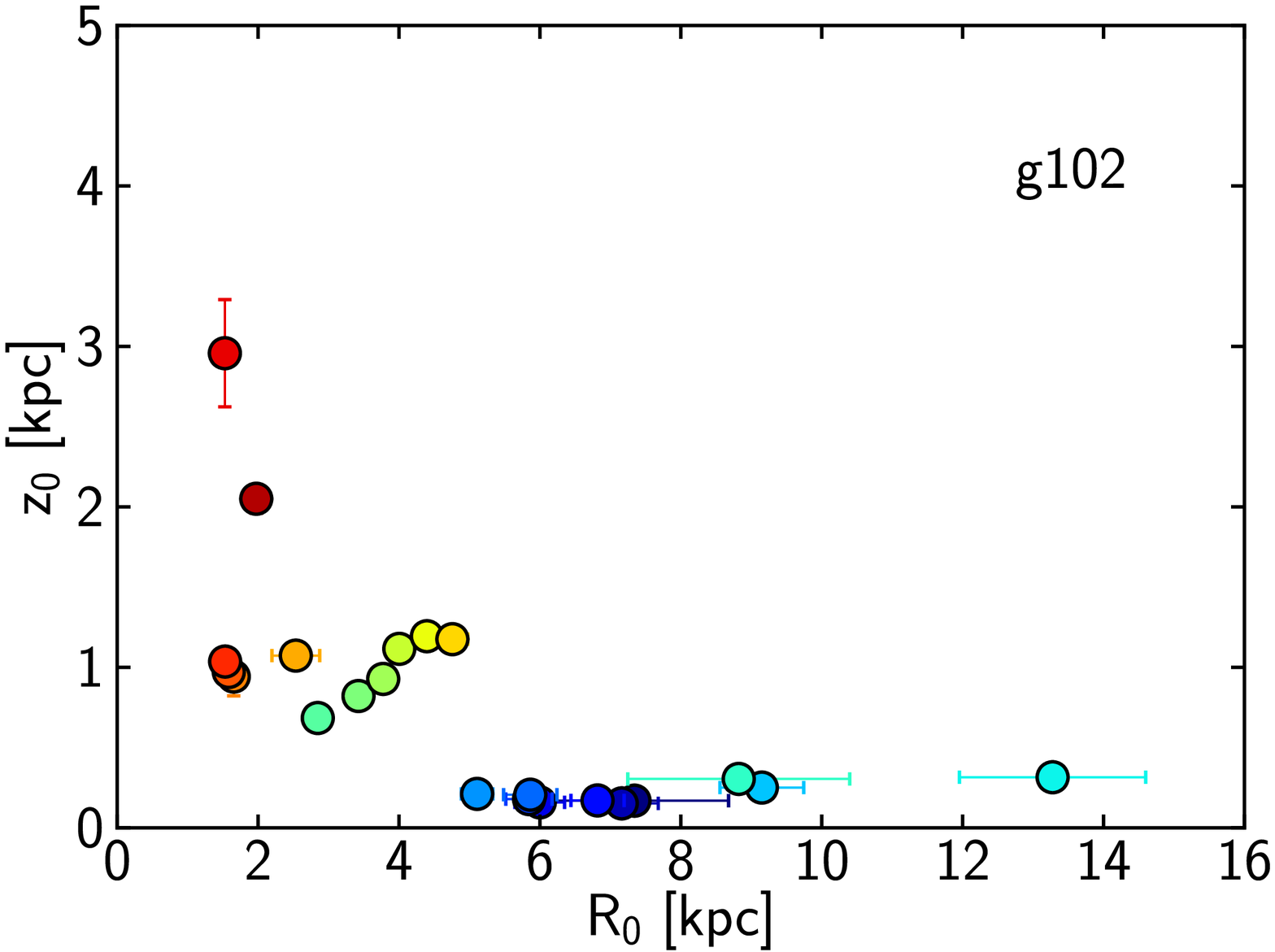}
\includegraphics[width=0.245\textwidth]{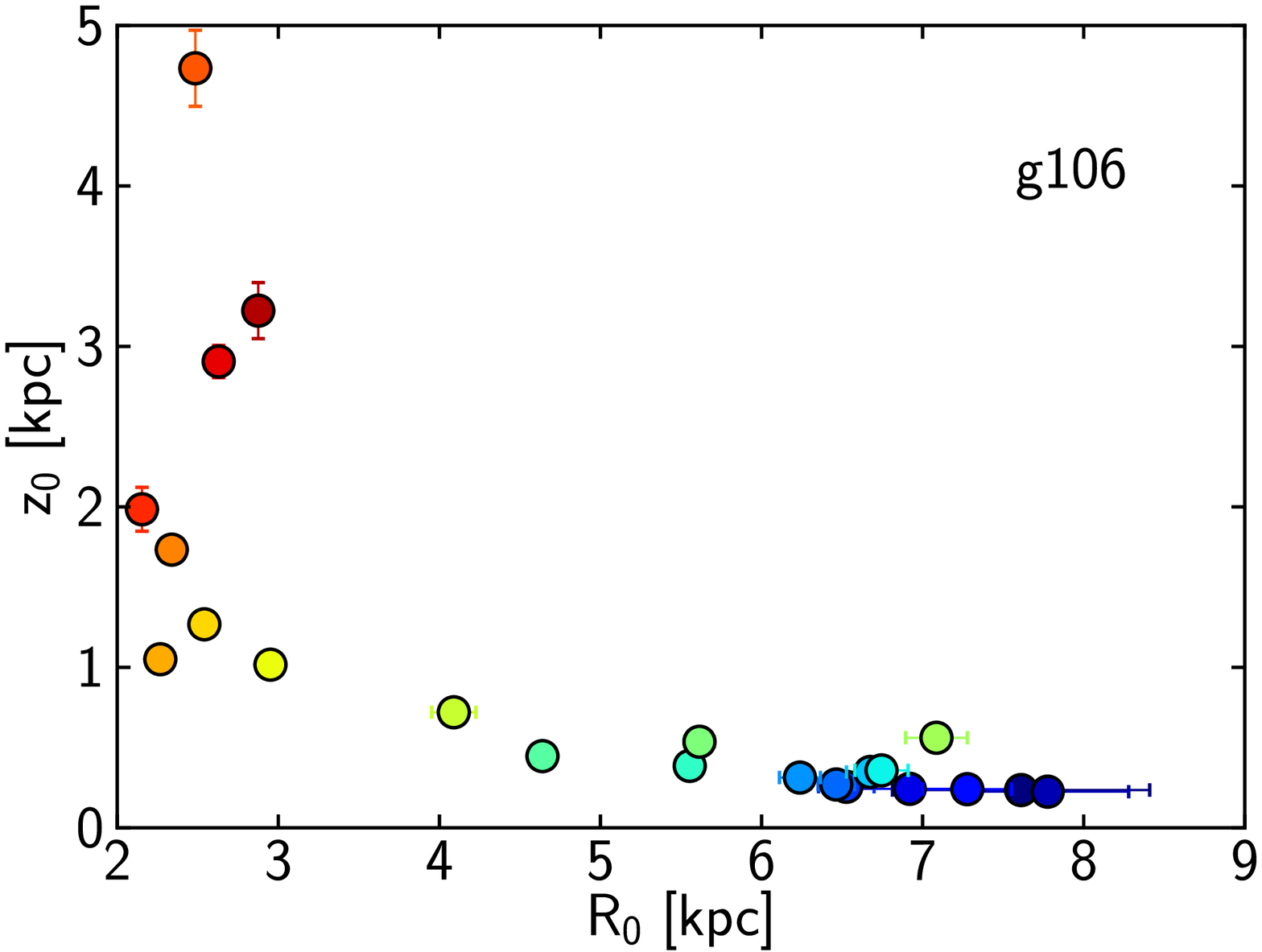}
\caption{Scale-height as a function of scale-length for mono-age populations in the 7 simulated galaxies. The scale-heights are measured at a radius of $2 R_d$. The colourcode and panel order are the same as in Figure \ref{fig:prof_z0}. We find that the observed anti-correlation between scale-height and scale-length can be reproduced in the simulations, and does not necessarily imply an absence of mergers.}
\label{fig:scales}
\end{figure*}

Because of these irregularities in $z_0$ and $R_0$ vs age,  $z_0$ and $R_0$ are not necessarily anti-correlated for all galaxies. In Figure \ref{fig:scales} we show $z_0$ as a function of $R_0$ for MAPs in the seven galaxies (the value of $z_0$ we use here is the one measured at $2 R_d$, but we verified that choosing a radius of $3 R_d$ did not change the results).
We find a variety of structures, even amongst the quiescent galaxies. Both g47 and g92 show a banana-shaped distribution of points, corresponding to an anti-correlation of  $z_0$ and $R_0$. For g37 no such structure is seen, and most points are clustered around $R_0$=6--8 kpc: as described before, $R_0$ does not increase with age for this galaxy.

Three of the four galaxies with a merger show a common interesting feature in Figure \ref{fig:scales}, with populations at the time of the merger gathered along a line of positive slope, departing from the arched locus defined by the rest of the populations (this is really pronounced for g48 and g102, less so for g22). This diagonal feature corresponds to a correlation of $z_0$ and $R_0$ for the populations affected by the merger.

Once again, g106 is different from the other active galaxies: in spite of a merger, the anti-correlation is well preserved. Contrary to g47 and g92, where there is a gap between oldest stars (red colours) and the disc populations, in g106 we find that the populations affected by the merger fill that gap so that there is a continuity of scale-heights from the disc to the inner halo/very thick disc.

We conclude that:
\begin{itemize}
\item the anti-correlation between $R_0$ and $z_0$ can be reproduced by simulations
\item the anti-correlation does not necessarily imply an absence of mergers.
\end{itemize}

\section{Thin-thick disc dichotomy?}\label{sec:dichotomy}

In this Section we explore if our simulated galaxies show a clear thin/thick disc dichotomy, or a continuum of scale-heights for populations of increasing age. We have already discussed that mergers create gaps in the distribution of $z_0$ with age, which suggests that galaxies undergoing mergers would tend to host distinct thin and thick discs. However, even for galaxies with a continuous range of $z_0$ for MAPs, depending on the star formation history of the galaxy and the mass in each population, gaps could appear in otherwise smooth $z_0$ distributions when weighted by the mass in each MAP.

\begin{figure*}
\centering 
\includegraphics[width=0.245\textwidth]{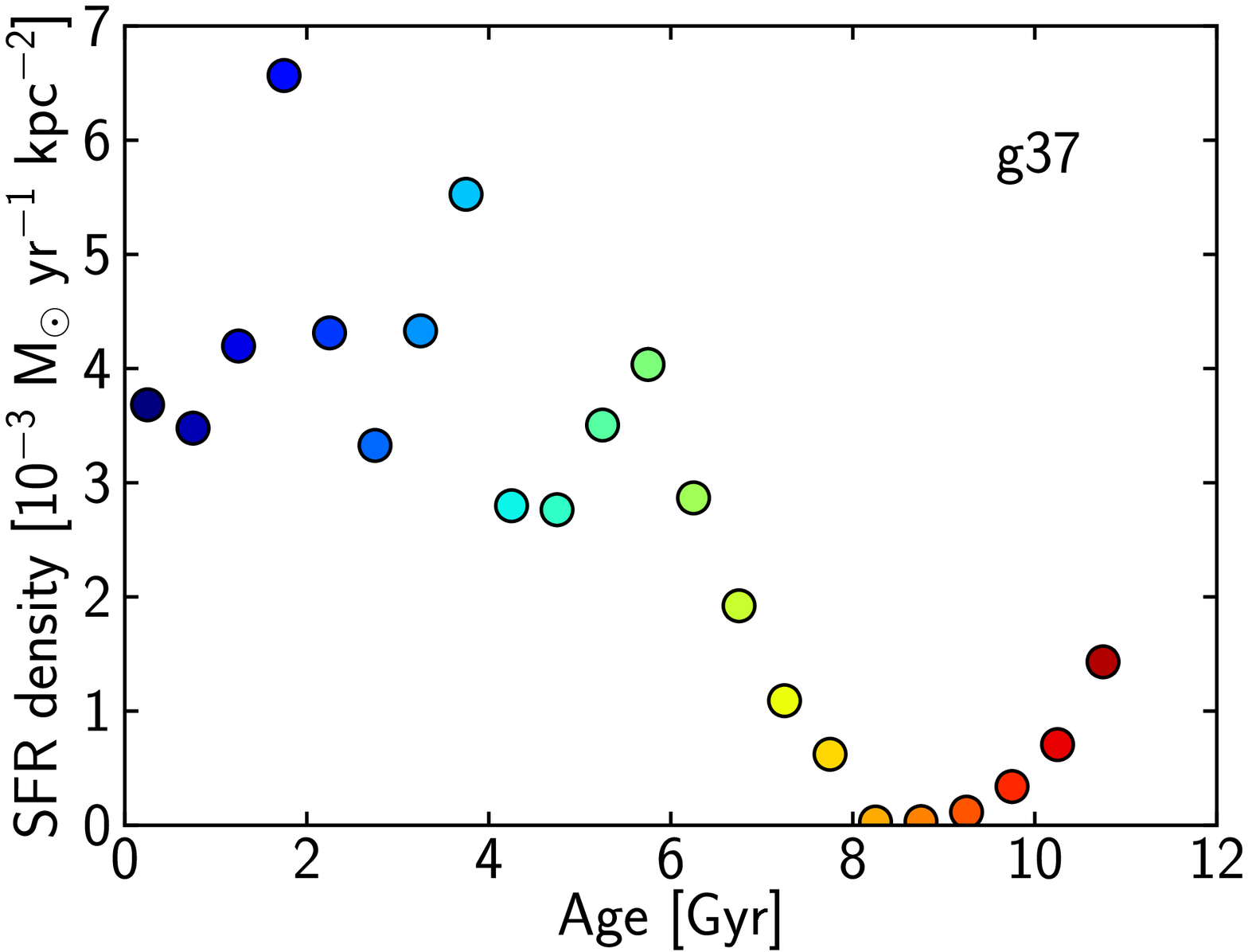}
\includegraphics[width=0.245\textwidth]{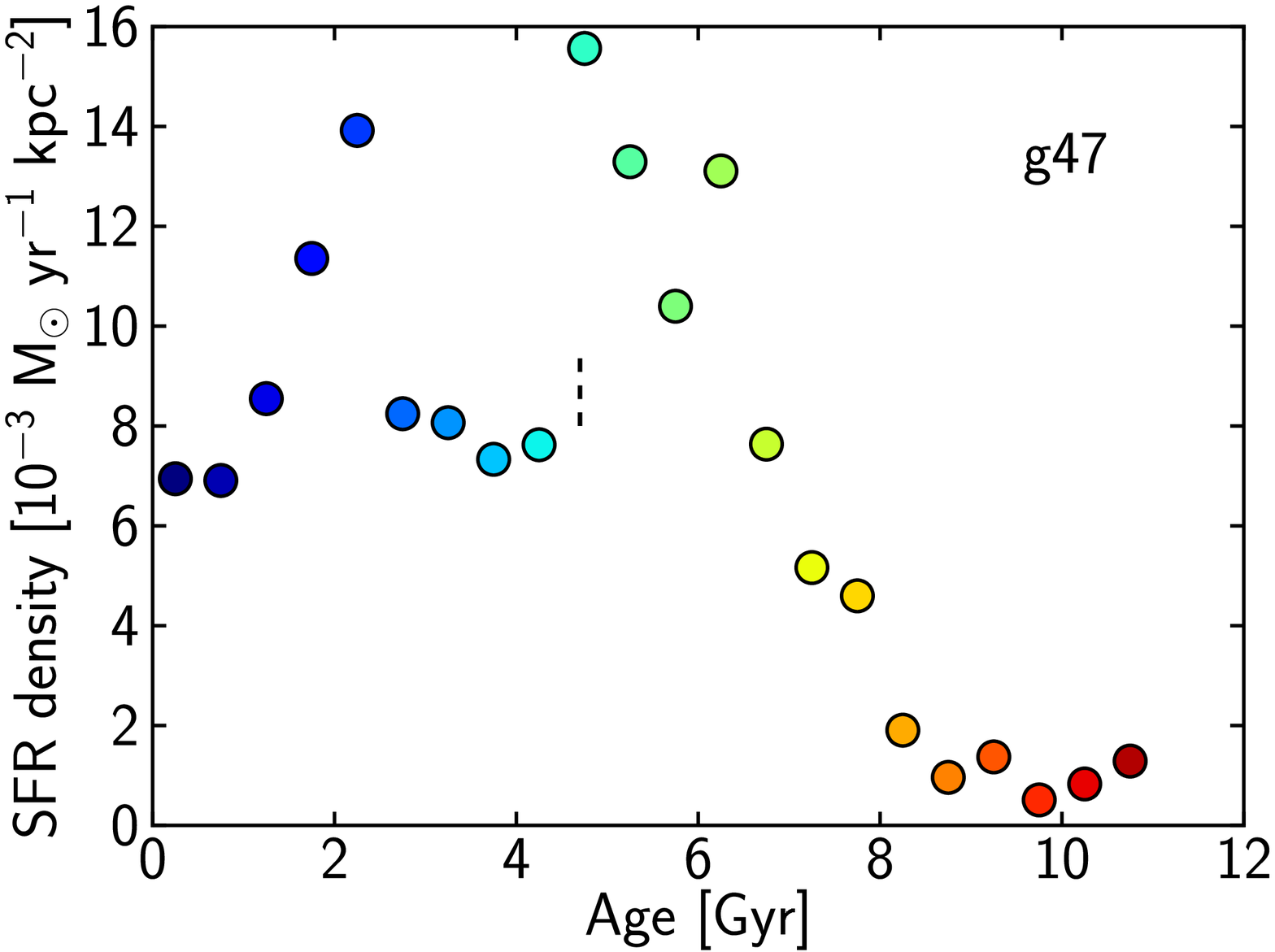}
\includegraphics[width=0.245\textwidth]{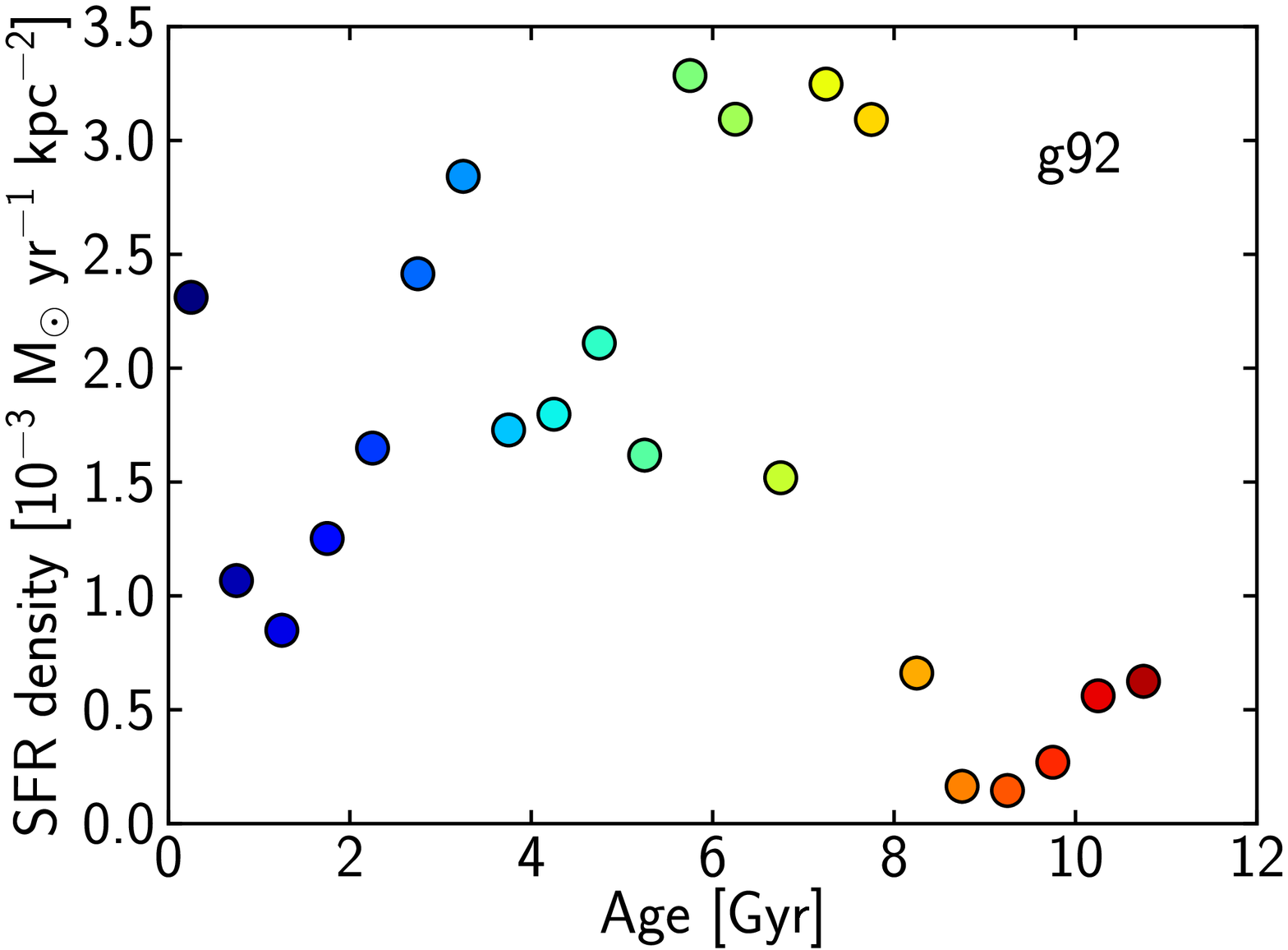}
\includegraphics[width=0.245\textwidth]{colorbar.eps}
\includegraphics[width=0.245\textwidth]{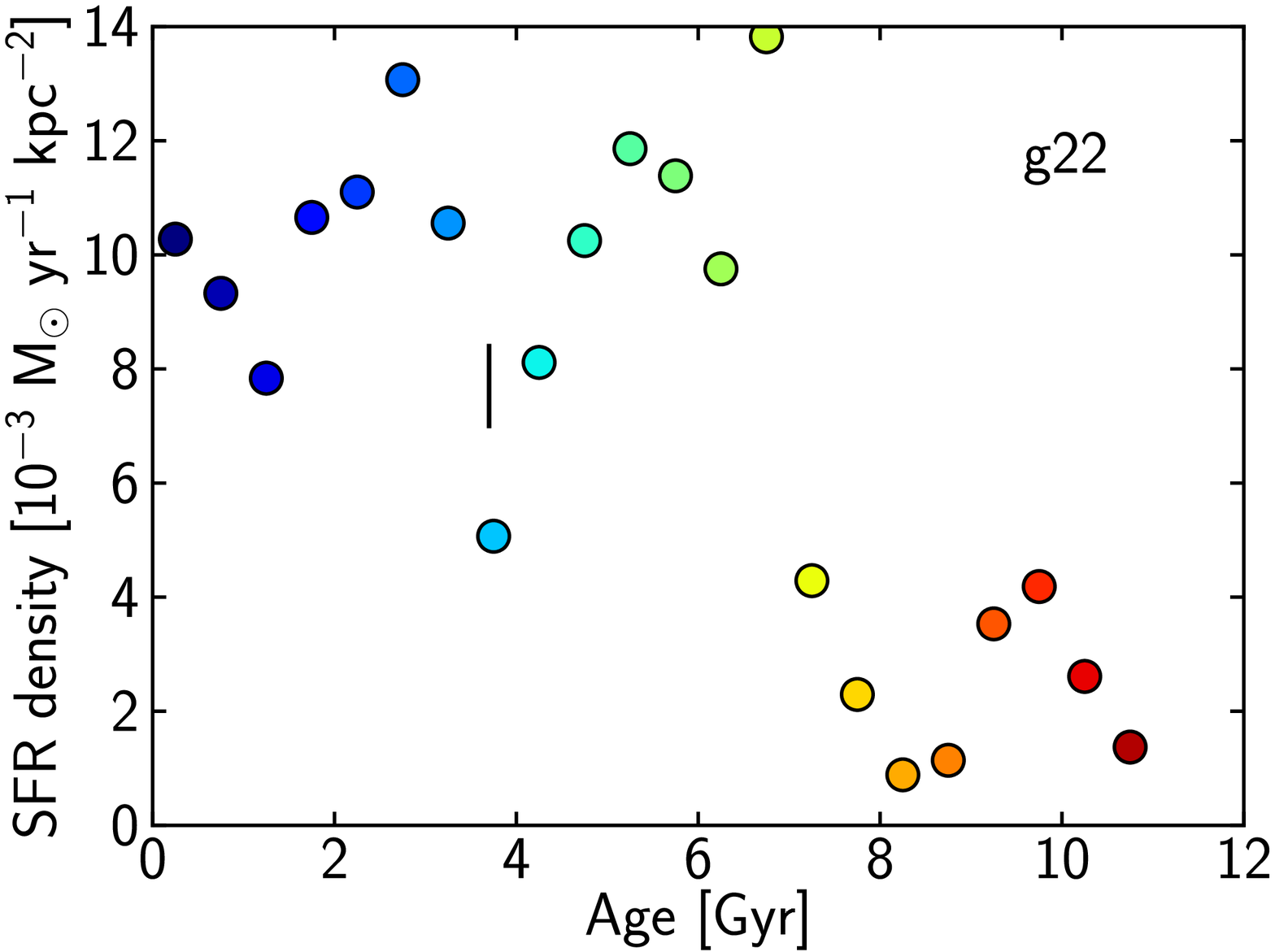}
\includegraphics[width=0.245\textwidth]{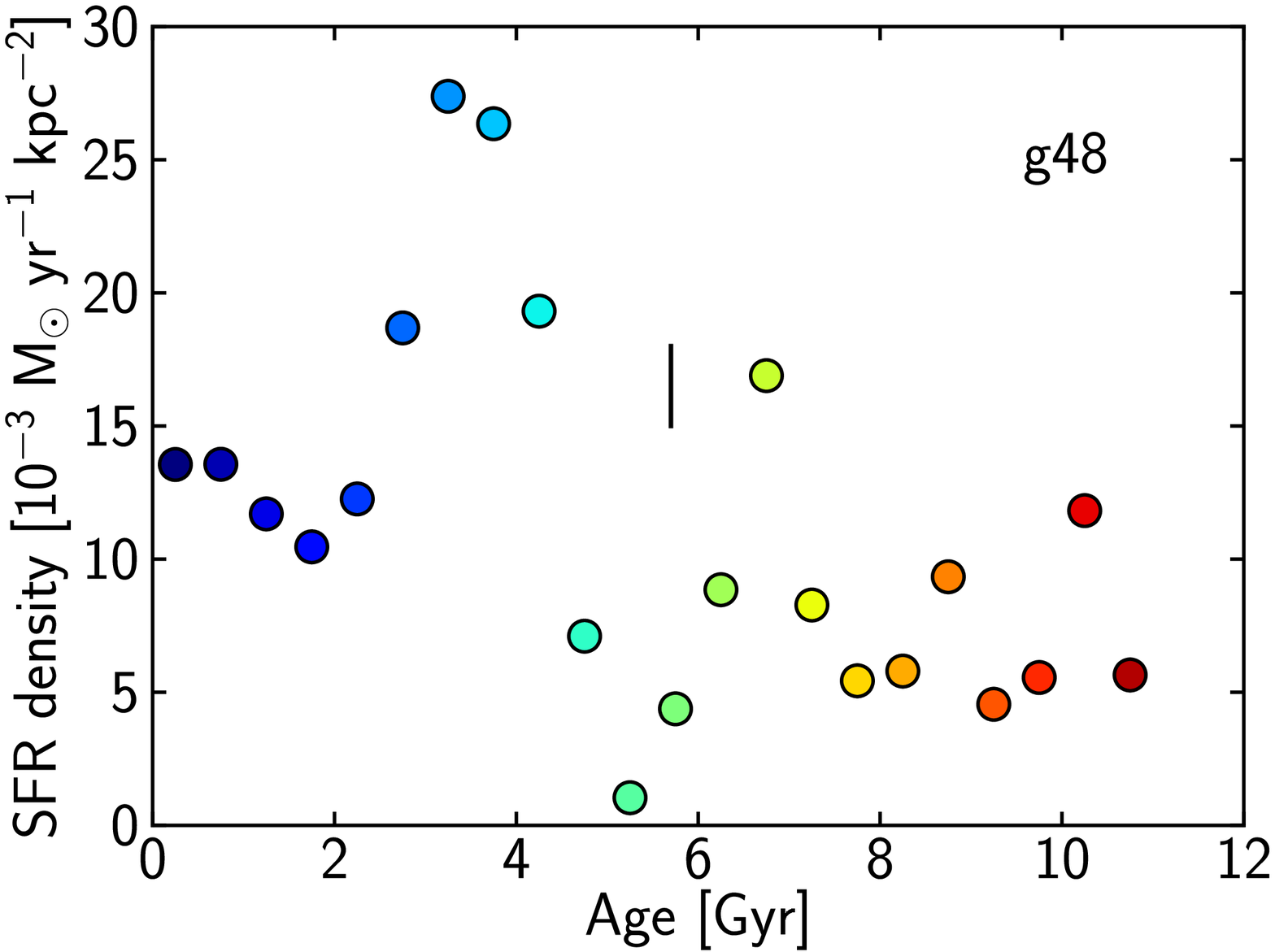}
\includegraphics[width=0.245\textwidth]{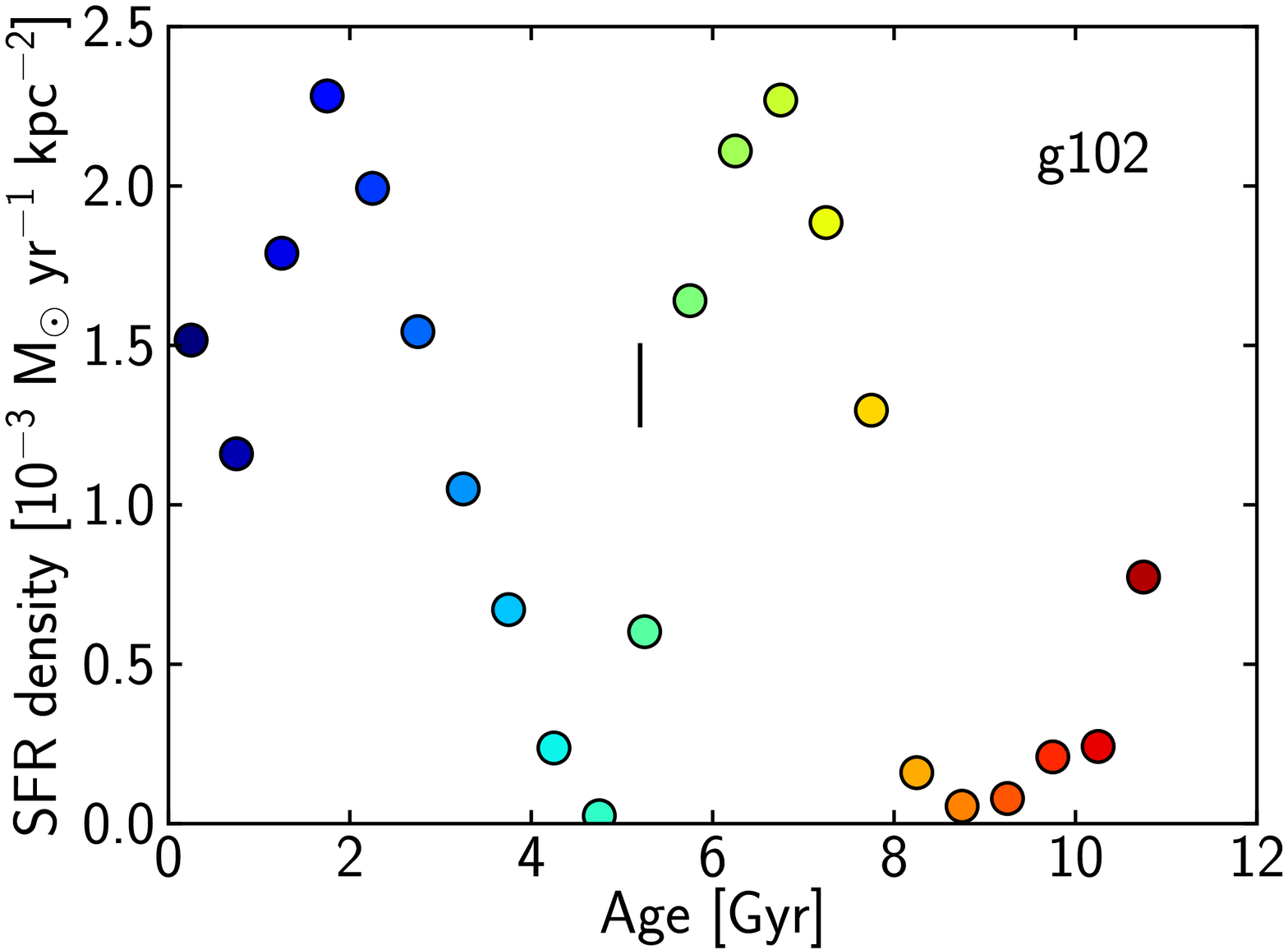}
\includegraphics[width=0.245\textwidth]{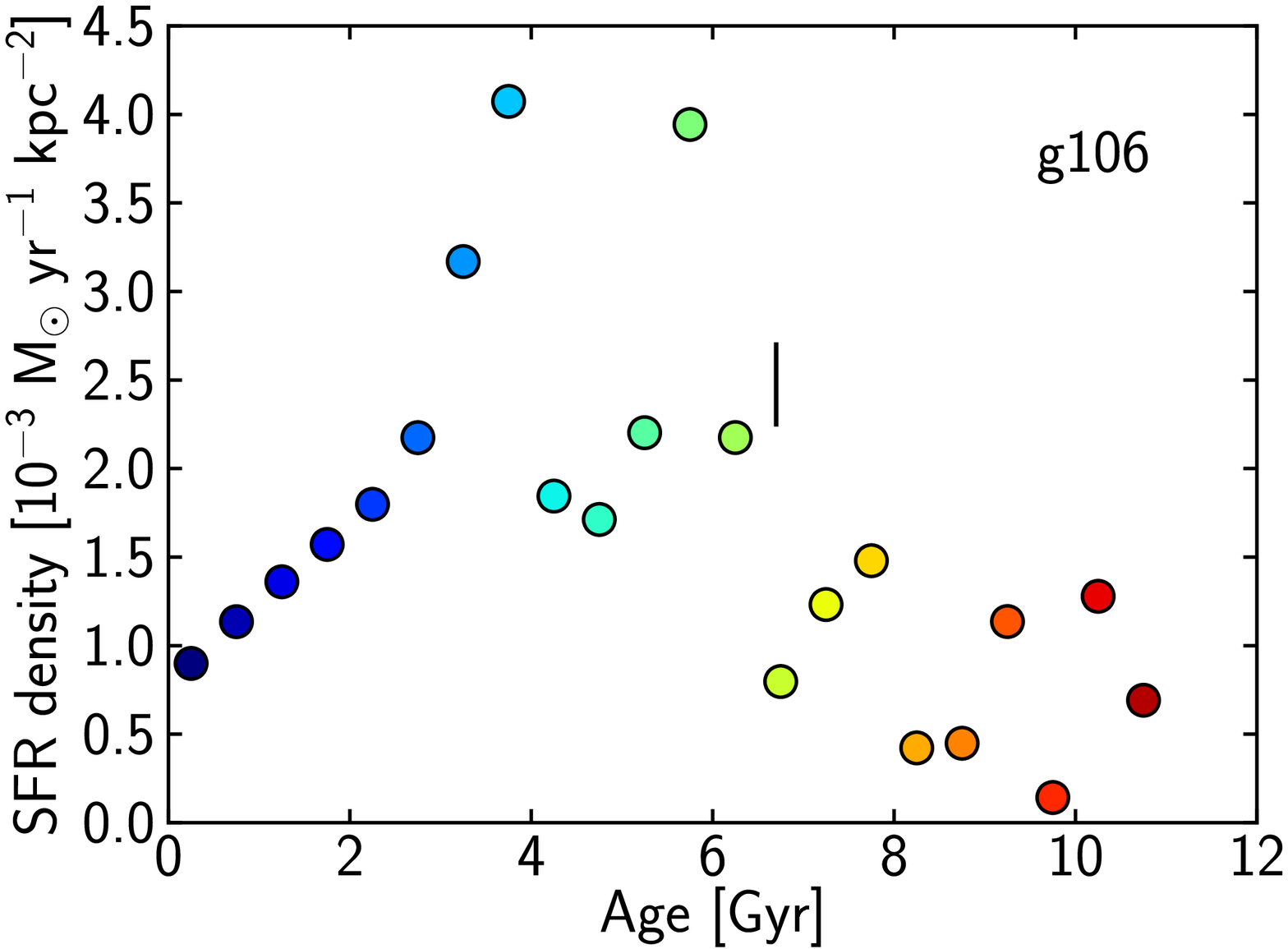}
\caption{Star formation history within a 2 kpc-wide annulus around $2 R_d$. The star formation rates are computed from the stellar mass in each mono-age population, and are thus representative of the stars found at $2 R_d$ at $z=0$ independently of their actual birth location. The small vertical lines mark the time of coalescence for the last merger undergone by each galaxy at $z<1.5$ (the dashed line for g47 marks the end of a fly-by). The colourcode and panel order are the same as in Figure \ref{fig:prof_z0}.}
\label{fig:mass}
\end{figure*}

In Figure \ref{fig:mass} we show for each simulated galaxy the star formation history within a  2 kpc-wide annulus at $2 R_d$ (these values are computed from the stellar mass in each MAP, and are thus representative of the stars found at $2 R_d$ at $z=0$ independently of their birth location). We indeed find that some MAPs contain very little mass. For most galaxies, little mass is in populations older than 8 Gyr. For $t<8$ Gyr, the increasing mass in each MAP corresponds to the start of disc formation, and is due to a mixture of stars born in situ and of stars born in the rest of the galaxy which have migrated to $2 R_d$ by $z=0$. We find that mergers create V-shaped features in this relation (this is particularly obvious for g22, g48 and g102, bottom panels in Figure \ref{fig:mass}). A similar shape was found in Figure \ref{fig:R0} for the evolution of the scale-length of the MAPs. As discussed in Section \ref{sec:Rcarac}, this corresponds to star formation being limited to the central regions of the galaxies after a merger occurred. This also explains why few of these stars are found at a radius of $2 R_d$, and why discontinuities are observed in the time evolution of the mass of MAPs at that radius. Combined with discontinuities in the scale-height of MAPs because of mergers, this has the potential to create a dichotomy between a thin and thick disc.

\begin{figure*}
\centering 
\includegraphics[width=0.245\textwidth]{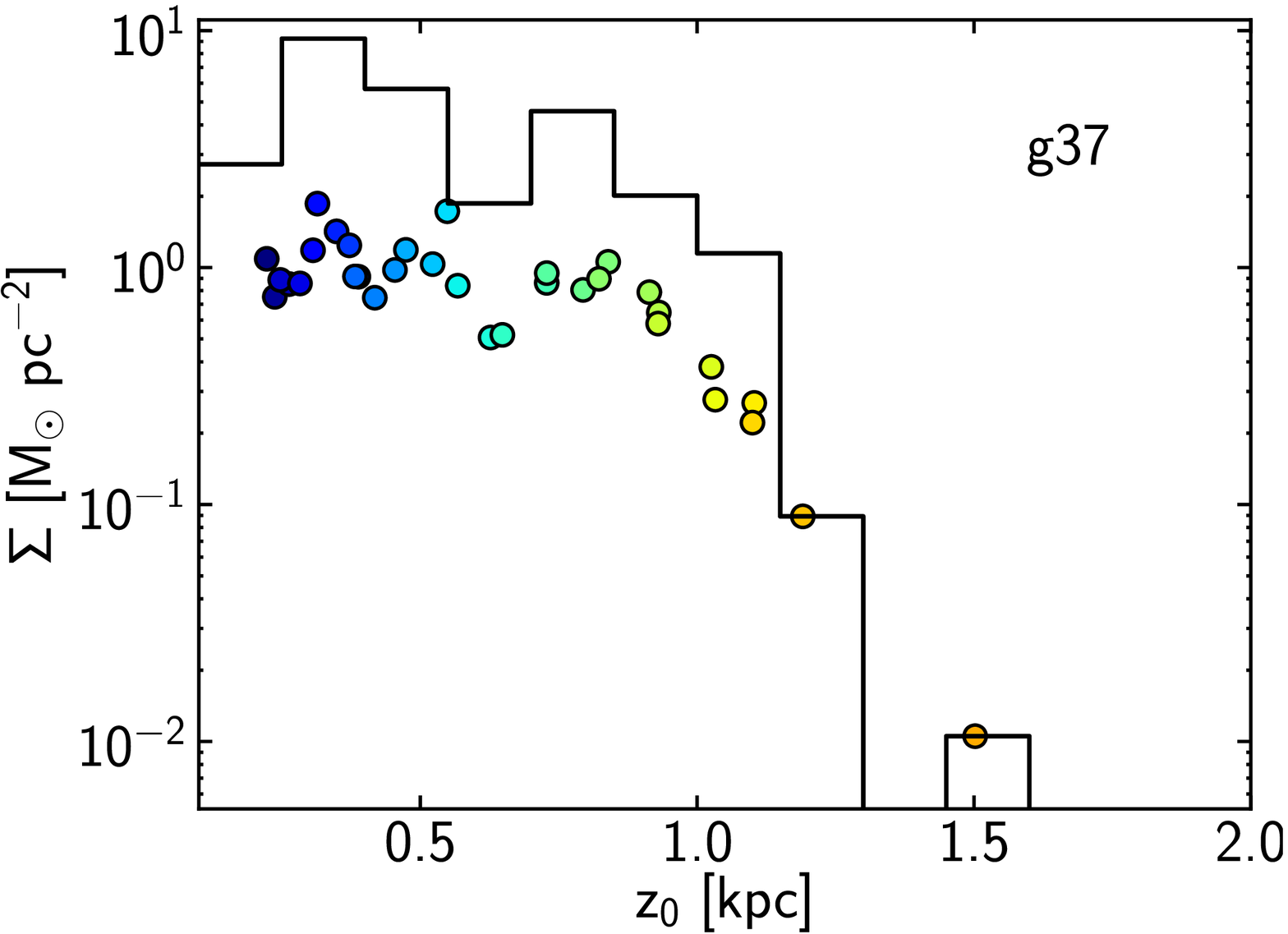}
\includegraphics[width=0.245\textwidth]{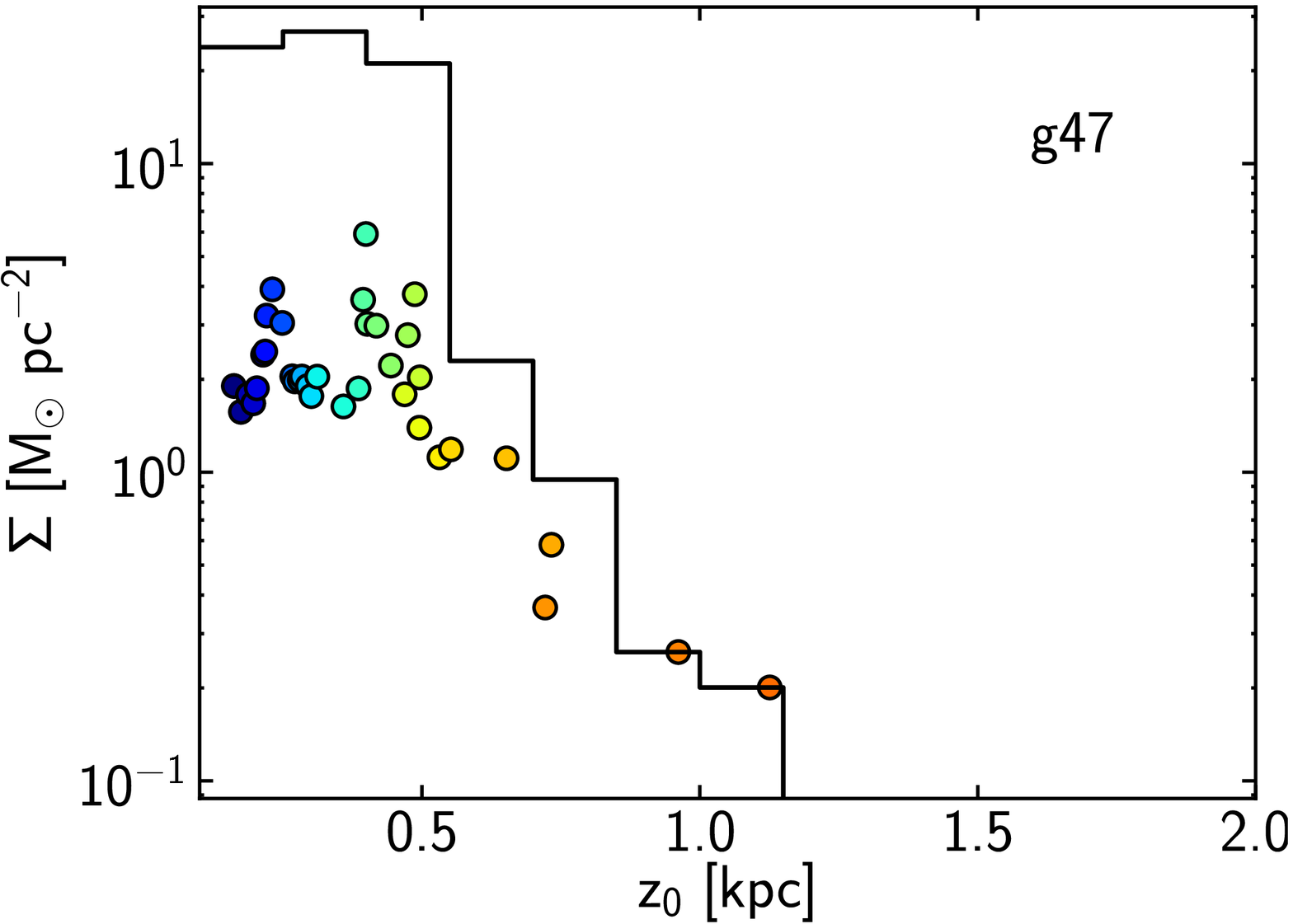}
\includegraphics[width=0.245\textwidth]{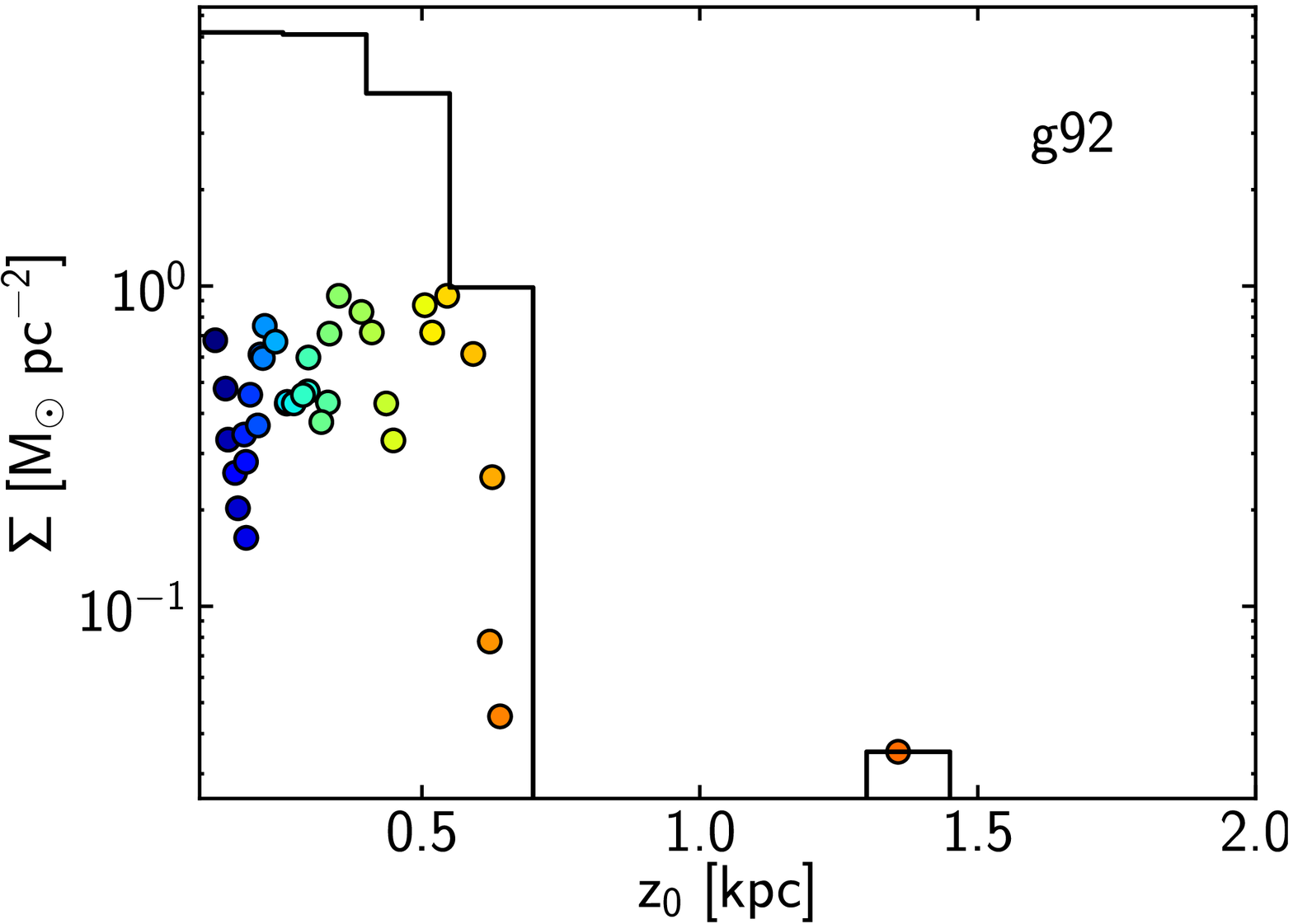}
\includegraphics[width=0.245\textwidth]{colorbar.eps}
\includegraphics[width=0.245\textwidth]{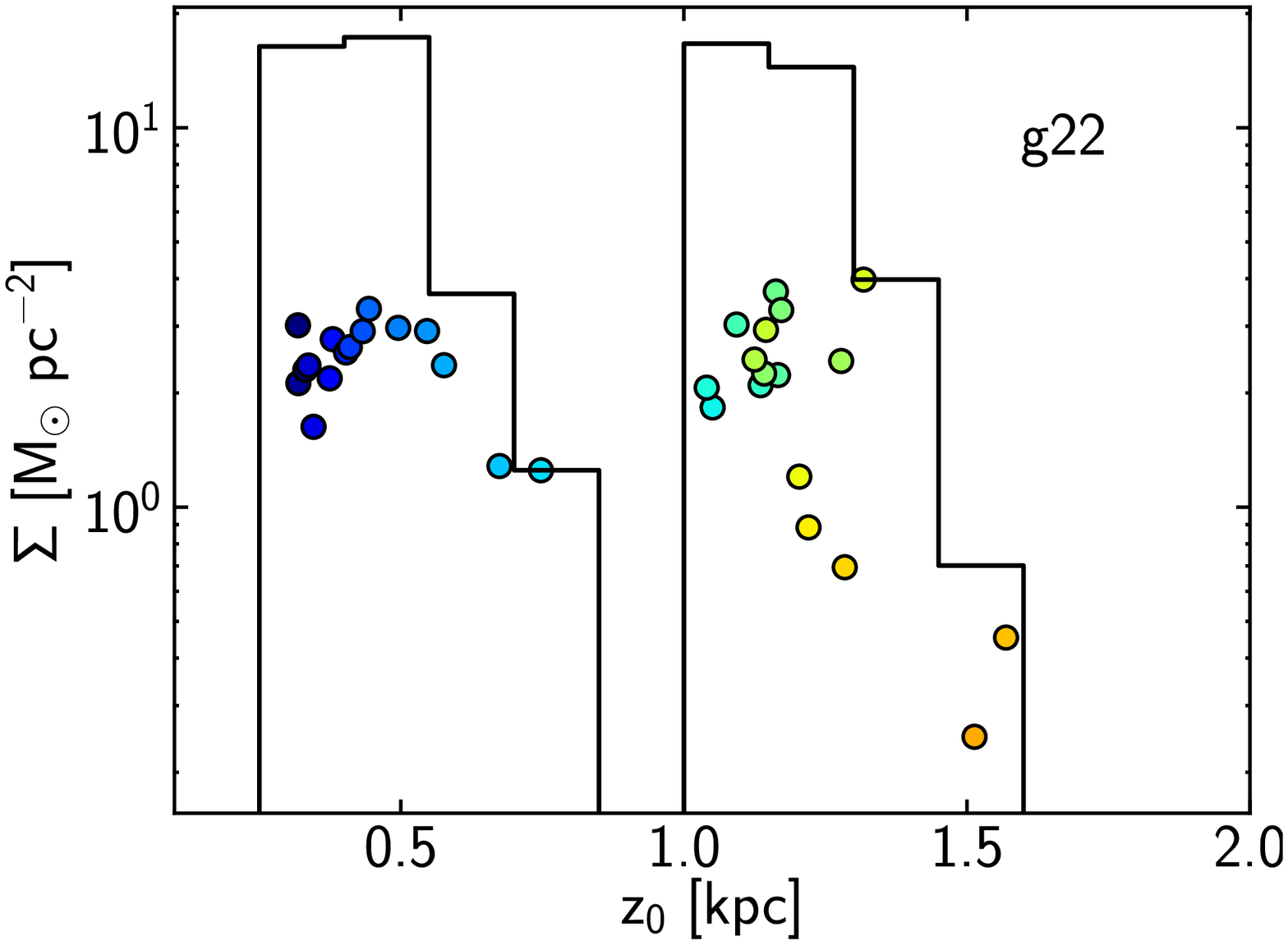}
\includegraphics[width=0.245\textwidth]{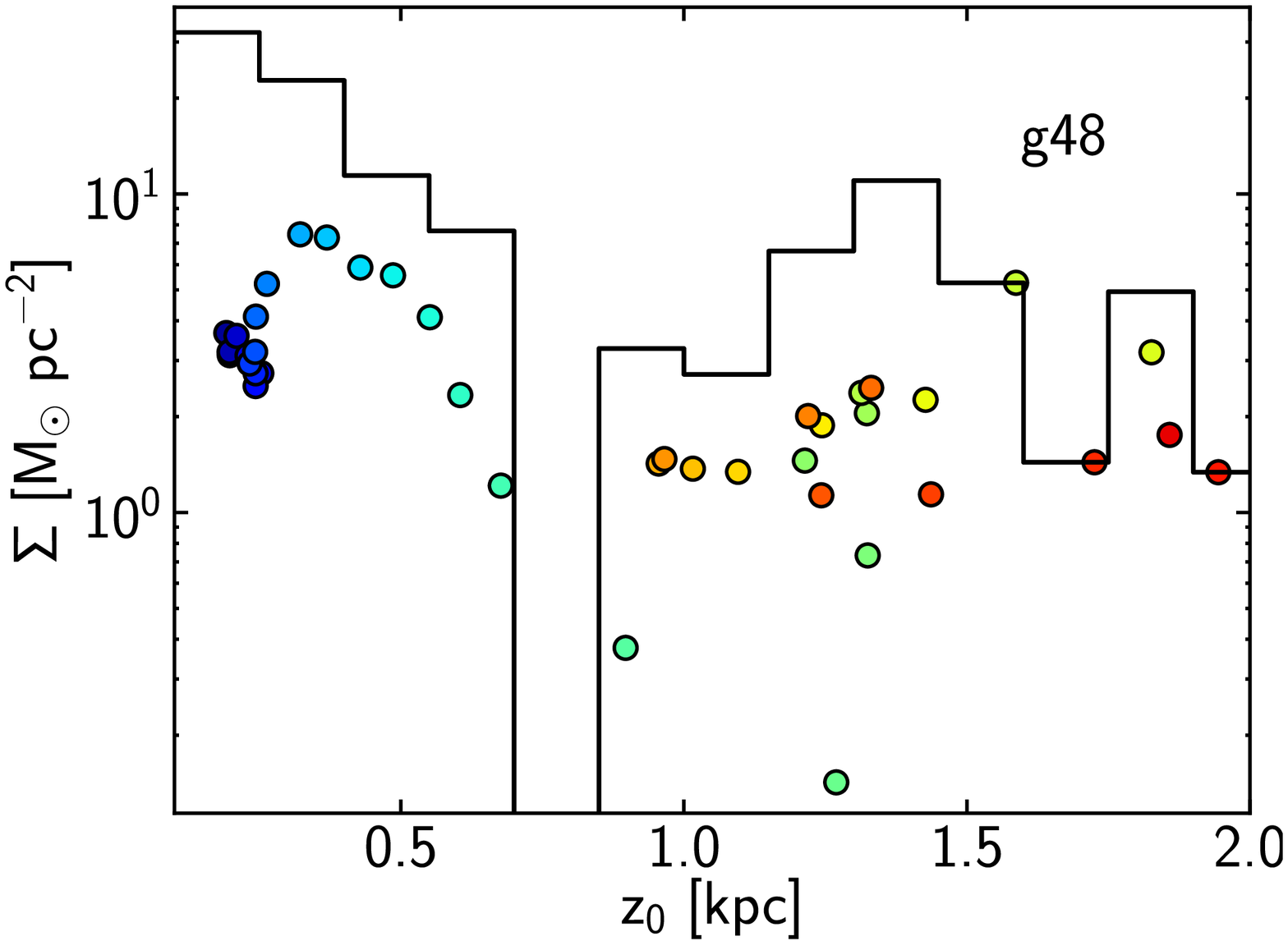}
\includegraphics[width=0.245\textwidth]{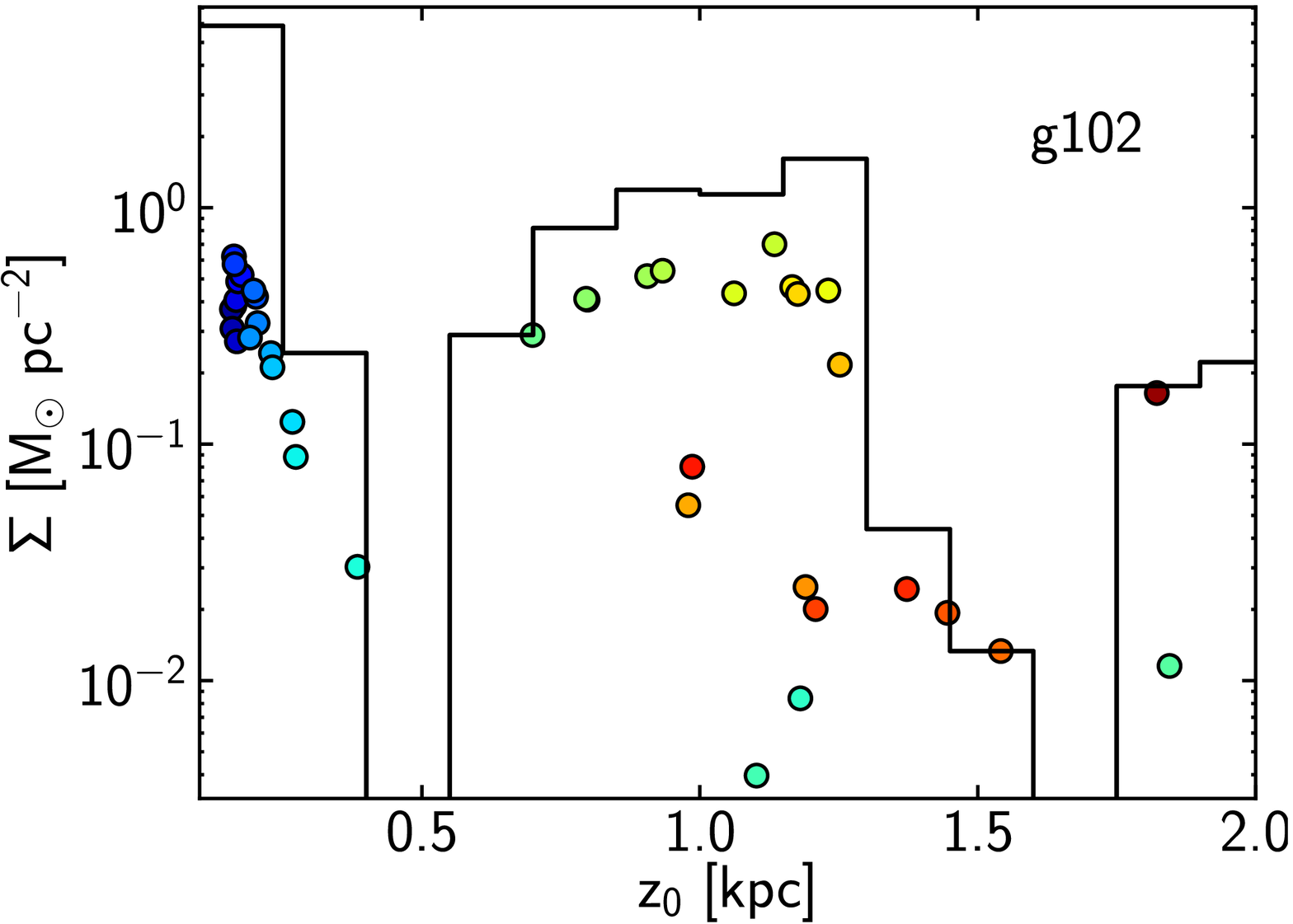}
\includegraphics[width=0.245\textwidth]{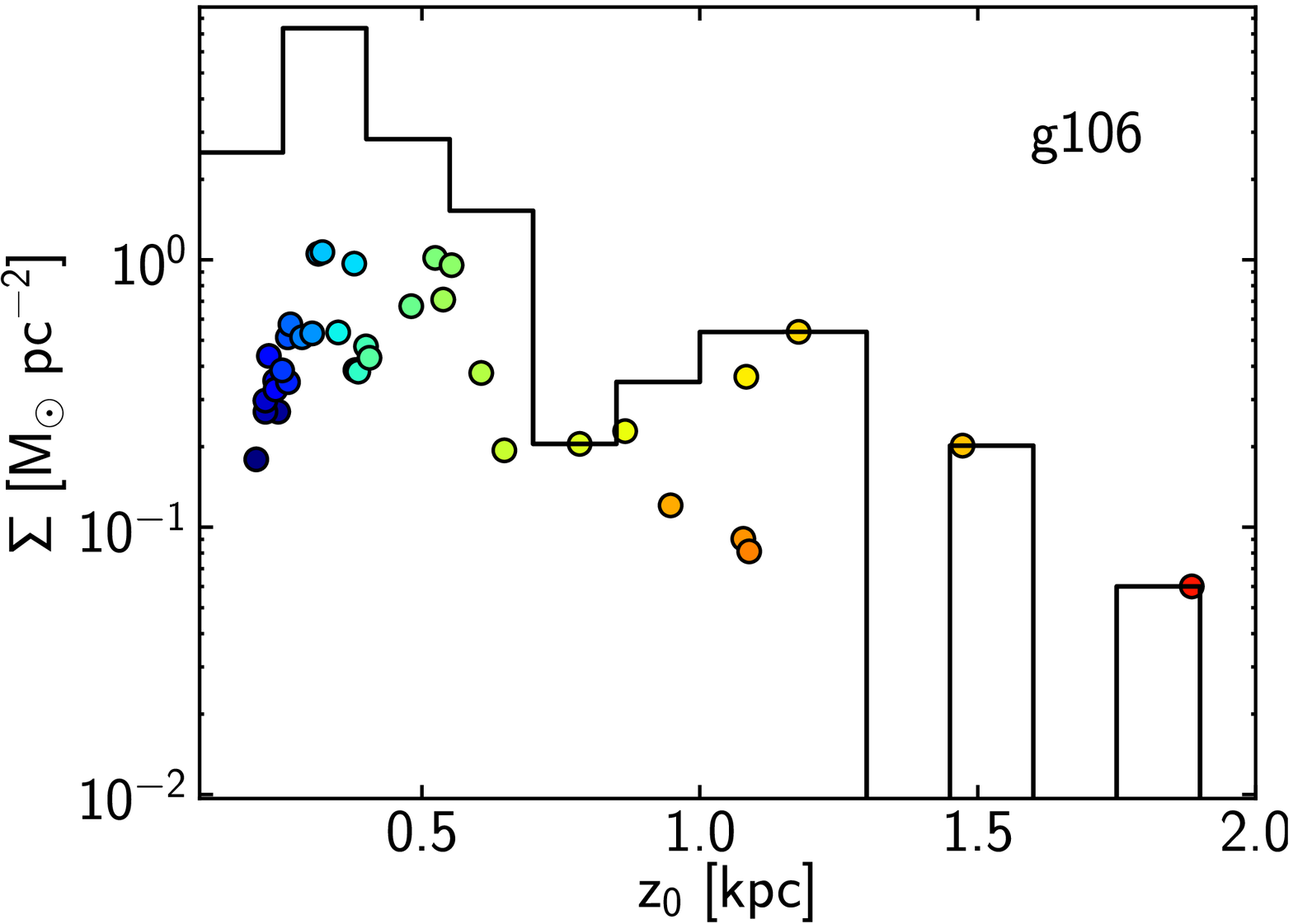}
\caption{Contribution of populations with a given scale-height to the total stellar surface mass density at a radius of $2 R_d$ (mono-age populations are here taken in 250 Myr age bins). The histogram is built by summing the contributions in bins of 150~pc. The colourcode and panel order are the same as in Figure \ref{fig:prof_z0}.}
\label{fig:bimodality_250Myr}
\end{figure*}

To explore that issue, we use a similar plot as the one introduced by Bovy et al (2012a). We study how much each MAP contributes to the total stellar mass surface density at a given radius as a function of their $z_0$. In Figure \ref{fig:bimodality_250Myr} we show the contribution of each MAP, as well as the contributions binned as a function of $z_0$ (in bins of 150 pc). To improve the statistics, we here consider MAPs in 250 Myr age bins instead of 500 Myr as previously (the large number of particles in each MAP allows us to reduce the width of each age bin while still keeping enough star particles to determine $z_0$).

This Figure shows that there is a clear difference between galaxies with quiescent and active merger histories. A first difference is that the scale-height for the thickest population is greater in galaxies with mergers relative to quiescent galaxies. In addition, the three quiescent galaxies show a continuous distribution of scale-heights (the shapes of these distributions being quite different from one galaxy to another), while the galaxies with mergers show some bimodal distributions, with clear gaps for g22, g48 and g102.  These gaps are a result of jumps in $z_0$ for populations influenced by the mergers, and match the time of coalescence of the last merger so that the ``thick'' populations are the pre-merger populations (note that mergers also create jumps in the age-velocity relation as discussed in Paper II).
We find that g106 does not have such a clear gap, but still shows more bimodality than the quiescent galaxies.

We also note that even in the case of a continuous distribution of scale-heights, a simple two-component fit to the vertical density profile of stars could give the impression of a bimodality. For instance, for g92, which has a very regular and continuous structure, a decomposition into a thin and thick disc is possible (but perhaps not physically meaningful), and gives scale-heights of $\sim$300~pc and $\sim$800~pc, relatively constant with radius out to $R=4R_d$. This means that the standard decompositions into thin and thick discs are entirely compatible with a continuum of scale-heights for MAPs (as also shown by \citealp{Rix2013} for the Milky Way). 

Assuming that mono-age populations are similar to mono-abundance populations, we conclude that, indeed, a continuous distribution of $z_0$ as shown in the Milky Way is a strong indicator of a very quiescent merger history. It is interesting to note that \cite{Stinson2013} have performed a similar test for one of their galaxies undergoing mergers and they find a bimodality in the distribution of $z_0$ (their Figure 10), supporting our basic picture.

\section{Discussion}
\subsection{Properties imprinted at birth vs subsequent evolution}

\subsubsection{Radial profiles}

\begin{figure}
\centering 
\includegraphics[width=0.45\textwidth]{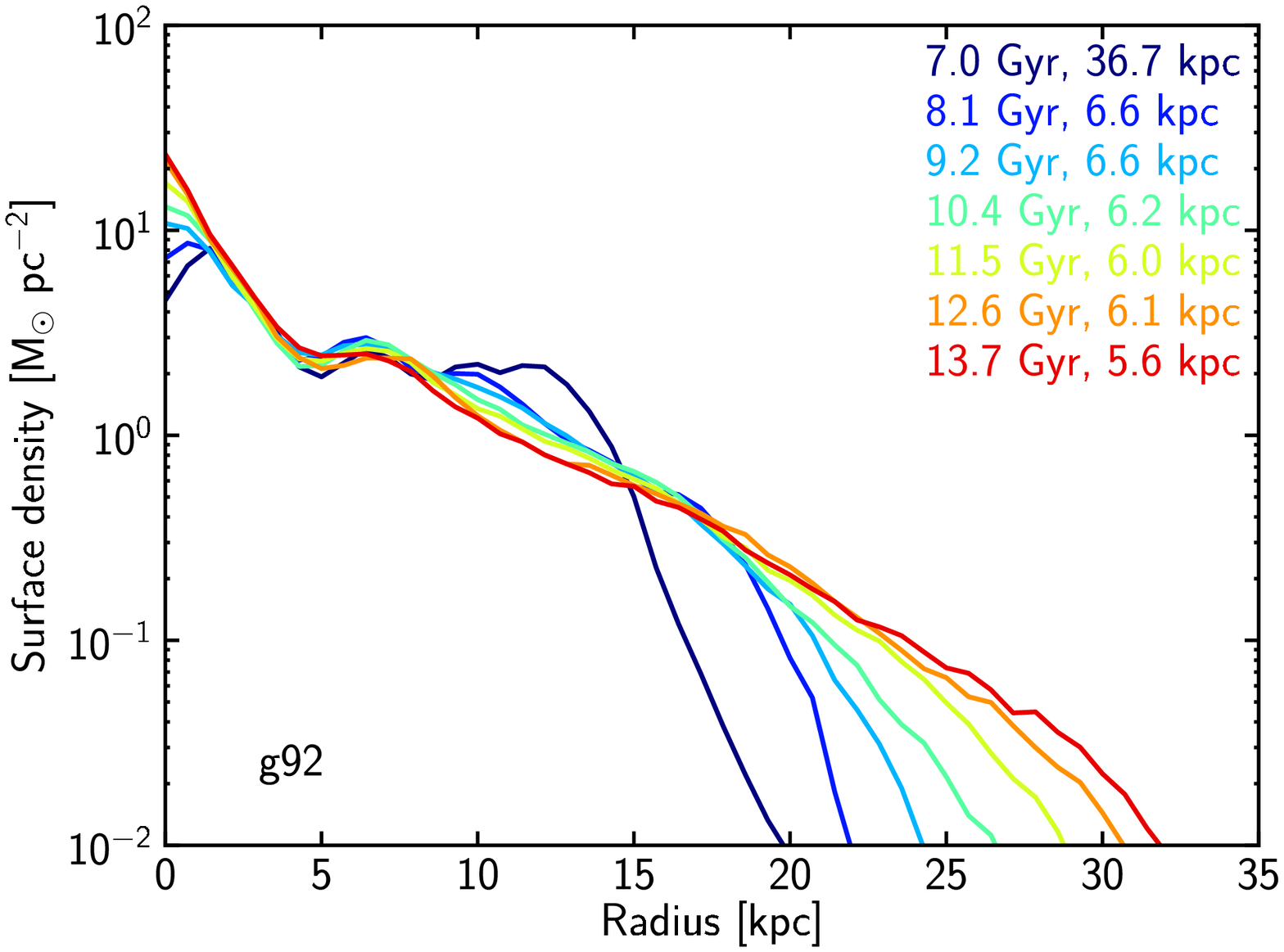}
\includegraphics[width=0.45\textwidth]{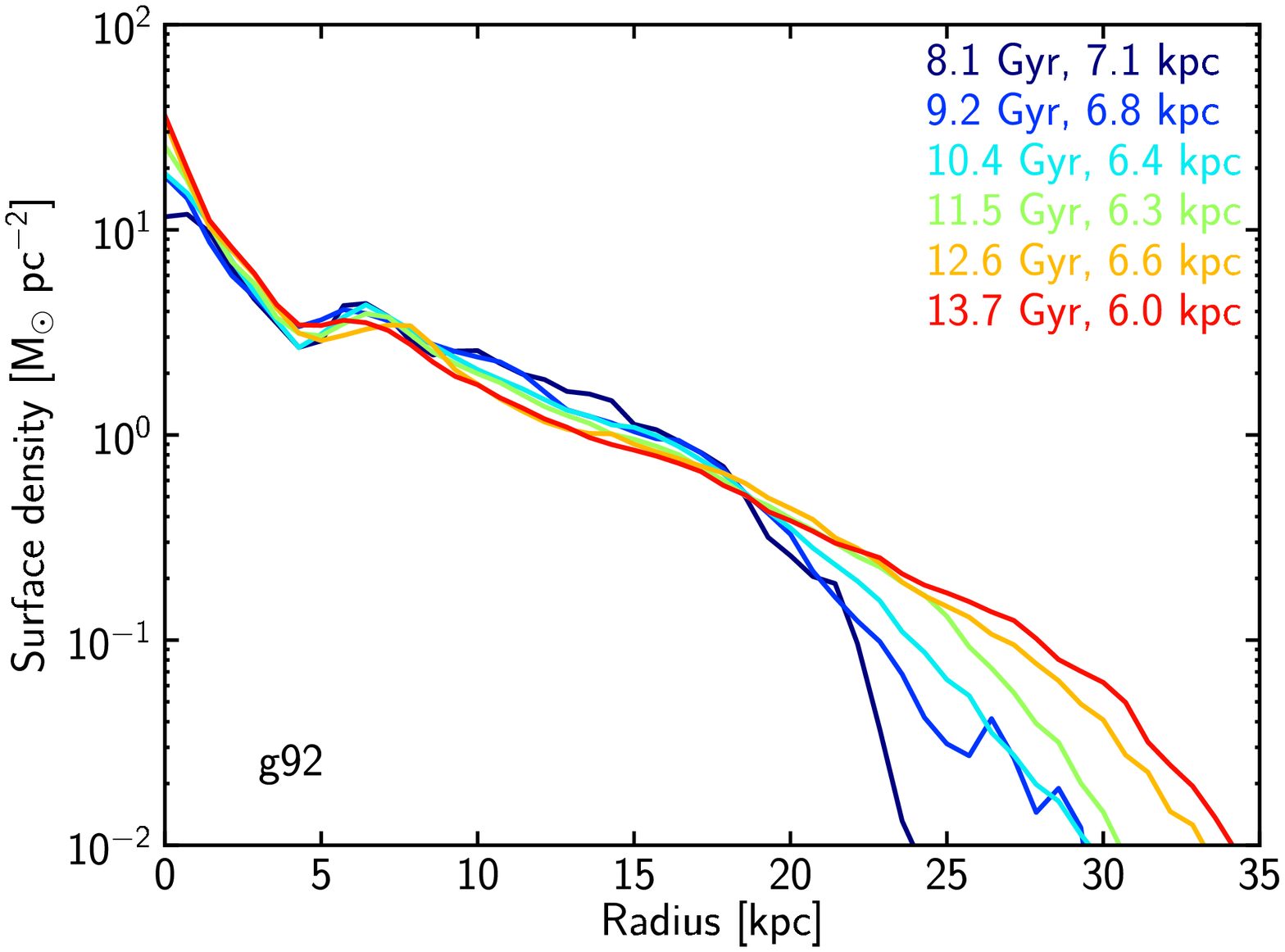}
\caption{Time evolution of the radial surface density profile for two population of stars in g92. The top panel shows the evolution of stars born between 6.5 and 7 Gyr after the Big Bang, while the bottom panel shows stars born between $t=7.6$ and 8.1 Gyr. The evolution of each population is shown at regular times between birth and 13.7 Gyr (from dark blue to red curves). Also shown are the scale-length from an exponential fit of each profile. In both cases, soon after birth, the density profile is irregular and truncated at a small radius. With time, the profiles become smoother and the truncation moves out to larger radii. While both populations end up with a very similar profile at $t=13.7$ Gyr, their starting points are very different: in one case (top panel) a very flat profile with a large scale-length at birth, and in the other case a birth scale-length which is very close to the final one.  }
\label{fig:radial_evol}
\end{figure}
We now study how stars in MAPs are re-distributed radially as a function of time. As an example, Figure \ref{fig:radial_evol} shows the time evolution of the radial surface density profile for two MAPs in galaxy g92. It illustrates some general features seen in our simulated galaxies.

Soon after birth, the surface density profile of a MAP is irregular, following overdensities in the gas disc (typically spiral arms and rings). As stars migrate radially, the irregularities are blurred and the profile smoothens significantly. This redistribution of mass occurs for all populations in all simulated galaxies. Another effect of migration is that, while the profile for young stars often shows a break in the outer regions, as stars move outwards, this break is either erased or moved to larger radii \citep[see also][]{Minchev2012a, Aumer2013}. This populates the outer disc with stars that were born in the inner regions (this is the case for both populations shown in Figure \ref{fig:radial_evol}). Note that this expansion in the outer disc is matched by an increase of mass in the inner regions, to conserve the total angular momentum.

To quantify the radial extension of the profiles, we measure how $R_{95}$ (the radius enclosing 95\% of stars) changes with time for a given MAP. The strongest variations are seen for g92; for instance for the population shown in the top panel in Figure \ref{fig:radial_evol}, $R_{95}$ grows from 13.5~kpc at birth ($t=7$ Gyr) to 22 kpc at $z=0$. Other galaxies usually show smaller variations of $R_{95}$ with time, but if $R_{95}$ evolves, it always increases with time. We also find that the expansion of the stars is not limited to the first Gyr after birth, but is a continuous process: for both MAPs shown in Figure \ref{fig:radial_evol}, the outer parts of the profile change continuously.

While exponential fits are not always the most informative way to describe the profiles (particularly when they are irregular soon after birth), we also use them to characterize the profiles (each profile is fitted from $R_d$ to  $R_{95}$). We find that the measured scale-lengths are either constant with time, or decline (which corresponds to a steepening of the profiles). The MAP shown in the bottom panel in Figure \ref{fig:radial_evol} corresponds to a situation where there is only little evolution of the scale-length (from 7.1 kpc at birth to 6 kpc at $z=0$). By contrast, the MAP in the top panel of that Figure has a very flat density profile at birth, with a sharp truncation. As the truncation moves outwards with time, the overall profile also steepens.

For MAPs where the scale-length declines with time, the strongest changes are often seen in the first Gyr of evolution, as it is the case in the top panel of Figure \ref{fig:radial_evol}: for this sub-population of g92, there is a significant steepening of the profile in the first Gyr, and the scale-length remains nearly constant for the remaining  $\sim$ 5 Gyr of evolution.

We thus find that radial migration has a strong influence on the density profiles of MAPs, and makes it hard to reconstruct birth properties from the present-day state: both MAPs shown in Figure \ref{fig:radial_evol} have similar profiles at $z=0$, but in one case the stars were born with a very flat profile and a sharp truncation, while in the other case the scale-length did not change much with time. In both cases, though, migration had a strong influence on the outer disc, populating it with stars born in the inner disc.

\subsubsection{Vertical profiles}

The time evolution of the vertical velocity dispersion, and the sources of heating, are the focus of Paper II, to which we refer the reader for more details. We here briefly summarize the findings of that paper.

We find that the vertical structure of the oldest stars (ages greater than 8--9 Gyr) is mostly imprinted at birth: these stars are born hot in a violent phase at high redshift. After that, though, stars are mostly born with a constant (or slightly decreasing) velocity dispersion, at least in galaxies with a quiescent merger history. We find that the shape of the age-velocity relation is then due to heating after birth. 

We find that radial migration is not the source of that heating, at least at the radius of 2$R_d$ that we studied in Paper II: at that radius, the stars born in situ mostly determine the total \sz. Stars coming from the inner disc have a slightly larger \sz, and stars from the outer disc a smaller \sz, so that the net effect of migration on the total \sz is small (see Figure 7 in Paper II).

We discuss in Paper II that the sources of heating vary from one galaxy to another, but include heating due to disc growth, and heating due to a combination of spiral arms and bars coupled with overdensities in the disc and vertical bending waves.

\subsection{The merger history of the Milky Way}

Mergers with a ratio of 1:10 should be quite common for Milky-Way mass galaxies in the last 10 Gyr \citep{Stewart2008}, and are usually thought to be quite damaging to the discs \citep{Quinn1993,Velazquez1999}, even if their effect might have been overestimated \citep{Moster2010b}. Indeed, \cite{Moster2010b} argue that a recent 1:10 merger would not be in conflict with the present-day structure of the Milky Way, provided that the Milky Way contained at least 20\% of gas at the time of the merger.

Our simulations suggest that the structure of MAPs is very sensitive to these mergers (and is thus a better tracer of the merger history than the global bulge fraction). We showed that mergers in the range 1:10--1:15 happening in the last 7 Gyr leave some clear jumps in the AVR and create bimodal distributions of scale-heights. These results, in combination with the observations presented in \cite{Bovy2012a}, would mean that the Milky Way has experienced no such merger since $z=1$ (as already speculated for instance by \citealp{Quinn1993}).

It is however very likely that mergers played an important role in shaping the density distribution of older stars. Indeed, we find that internal evolution is not enough to produce the hottest thick disc stars, such as are found in the Milky Way. This is illustrated by g92, which shows a very regular and smooth structure for most MAPs, and density profiles that do not flare vertically. In that, g92 is probably similar to the Milky Way. However, the range of scale-heights found for the disc, from 200 to 800 pc for young to old stars, is slightly smaller than in the Milky Way, where the thickest stars have a scale-height of 1000~pc \citep{Bovy2012a}. We also find that in g92 the oldest disc stars only reach a vertical velocity dispersion of $\sim$ 30 \kms (for a galactocentric distance of 2$R_d$, see Figure \ref{fig:prof_sigma} and Paper II), while \cite{Bovy2012c} find that the oldest local disc populations have a \sz of about 50 \kms.

Some violent early event seems necessary to produce the thickest and hottest disc populations (in addition to building the halo), as is the case for g106, and as is demonstrated in \cite{Minchev2014} by comparing g106 with RAVE data. This idea is consistent with the results of \cite{Liu2012}, where the thick disc of the Milky Way is found to be made of two components, the oldest one likely coming from mergers, and the rest being formed by internal mechanisms.

We have not discussed thick disc formation by clump instabilities, since this is not a mechanism that significantly affects the evolution of our selected simulated galaxies. It is however a possible mechanism for forming thick discs. It remains to be seen if it is consistent with the complex chemical signatures discussed in \cite{Minchev2014}.

\section{Summary}

We study a sample of seven simulated galaxies, which we slice into stellar populations in age bins of 500 Myr (``mono-age populations'', or MAPs). We measure the spatial and kinematical properties of these populations to study if and how they retain traces of the formation history of their host galaxies. In particular, we have studied the influence of mergers on the structure of MAPs by comparing galaxies with no significant merger in the last $\sim 9$ Gyr (3 galaxies in our sample) to galaxies with various types of mergers at low redshift (4 other galaxies).

All the studied galaxies undergo a phase of merging activity at $z>1.5$, so that stars older than 9 Gyr are always found in a spheroidal and centrally concentrated component. A fraction of these stars are accreted from the merging satellite galaxies. Younger stars are mostly found in the discs, and the properties of these discs are the main focus of this paper. We do not discuss the properties of bulges and bars, and often present quantities measured at a galactocentric radius of 2 or 3 $R_d$, where $R_d$ is the global disc exponential scale-length.

We first find that most MAPs have radial surface density profiles that can be fitted by an exponential (Section 3.1), with the exception of the youngest stars (which have complex profiles, following spiral arms and rings found in the gas discs). The complexity of radial profiles for young stars is actually common at all redshifts, and the final profiles result from gradual smoothing of the initial distribution because of radial migration of stars. In most cases, most of the radial evolution occurs in the first Gyr after birth, but with ongoing changes down to $z=0$. An important feature of migration is that it is very efficient at populating the outer disc with stars born in the inner regions, and it has a strong impact on the overall shape of the density profiles (Section 6.1).
In the end, we find that younger stars tend to have larger scale-lengths, both as a result of their properties at birth and of a steepening of the profiles with time. Mergers have a noticeable effect on radial density profiles, by both redistributing the old stars and changing the spatial distribution of newly formed stars.

We have also studied the vertical  density profile and vertical velocity dispersion of MAPs (Sections 3.2 and 3.3). We find that the vertical density profiles can be fitted by simple exponentials. Looking at the variation of scale-height with radius, we often see an increase of the scale-height at larger radius, giving a flared geometry to the MAPs.
The three quiescent galaxies show the simplest structures: populations of increasing age have an increasing scale-height, and this is true at all radii. Mergers introduce some complexity, and also seem to increase the flaring of populations affected by mergers (as expected from previous studies).

We similarly find that \sz is an increasing function of age (see details in Paper II). The increased \sz for older stars comes from two different origins: oldest stars are truly born kinematically hot (following the mergers happening at high redshift), while for stars younger than $\sim 7$ Gyr we find that \sz at birth is small and nearly constant with time. These stars are subsequently heated to their present day \sz. We find that radial migration is only a minor source of heating (if any). We also show in Paper II that \sz is very sensitive to mergers, which create jumps in the age-velocity relation.
Finally, we quantify the isothermality of each MAP by measuring \sz at a fixed radius, but as a function of height above the disc. We find that \sz increases with height above the disc plane for all our galaxies. Galaxies closest to isothermality have the lowest contribution of bulge/halo stars at the radius where \sz is measured, independently of their total bulge fraction or merger history.

As a result of the properties at birth, radial redistribution and vertical heating, we find that the relation between scale-length and scale-height is complex (Section 4). Assuming the equivalence between mono-age and mono-abundance populations, the anti-correlation observed in the Milky Way \citep{Bovy2012b} is reproduced by some of our simulations (\citealp{Stinson2013} show however that even if such an anti-correlation is found for mono-age populations, it might not be seen for mono-abundance populations).

In any case, the presence of an anti-correlation does not necessarily imply an absence of mergers: we both find a quiescent galaxy with no anti-correlation (g37) and a galaxy with mergers but a really nice anti-correlation (g106). We suggest that a better indicator of past mergers is the repartition of stellar mass as a function of scale-height (Section 5). We find that our quiescent galaxies show a continuous distribution of scale-heights, while a bimodality between a thick and thin component is found when mergers occur. The continuous distribution found in the Milky Way by \cite{Bovy2012a} points then towards an evolution dominated by secular processes. 
We also note that even in the case of a continuous distribution of scale-heights, a simple two-component fit to the vertical distribution of stars could give the impression of a bimodality. This means that the standard decompositions into thin and thick discs are totally compatible with a continuum of scale-heights for mono-age populations.
However, our simulations do not produce the thickest, hottest components in the absence of mergers. That would suggest that the old component of the Milky Way has a mixed origin, also in agreement with \cite{Liu2012}.

\section*{Acknowledgments}
We thank the referee for a constructive report.
We thank Hans-Walter Rix, Jo Bovy and Alex B\"{u}denbender for useful comments. MM is supported by a Humboldt Research Fellowship.
CF acknowledges financial support by the Beckwith Trust.

{}

\end{document}